\documentclass[notitlepage,openbib,12pt]{article}
\usepackage{amsfonts}
\usepackage{amsmath}
\usepackage{amssymb}
\usepackage{graphicx}
\usepackage{color}
\usepackage[colorlinks,citecolor=blue]{hyperref}
\usepackage{natbib}
\usepackage{mathpazo}
\usepackage{amssymb}

\newtheorem{theo}{Theorem}
\newtheorem{prop}{Proposition}

\newtheorem{define}{Definition}

\newtheorem{lemma}{Lemma}

\makeindex
\input epsf
\input{tcilatex}
\begin{document}

\title{\textbf{Rationalizable Implementation of Social Choice Functions:
Complete Characterization}\thanks{%
I thank Federico Echenique, Anirudh Iyer, Ritesh Jain, Ville Korpela,
Michele Lombardi and anonymous referees for insightful comments.}}
\author{Siyang Xiong\thanks{
Department of Economics, University of California, Riverside, United States,
siyang.xiong@ucr.edu}}
\maketitle

\begin{abstract}
We provide a necessary and sufficient condition for rationalizable
implementation of social choice functions, i.e., we offer a \emph{complete
answer} regarding what social choice functions can be rationalizably
implemented.
\end{abstract}

\newpage

\baselineskip= 18pt

\section{Introduction}

\label{sec:intro}

What social choice functions can be fully implemented \emph{\`{a} la} \cite%
{em}? This question has been studied extensively in the literature (see,
e.g., \cite{moj2001} and \cite{msj2002} for surveys), and most papers adopt
the solution concept of Nash equilibria. However, we adopt a different
solution concept in this paper, and study what social choice functions can
be fully implemented in rationalizable strategies.

Nash equilibrium imposes two requirements: (1) (common knowledge of) players
taking best strategies to their beliefs regarding other players' strategies,
and (2) players' beliefs being correct. Clearly, the second requirement is
too demanding. If we impose only the first requirement, we get the solution
concept of rationalizability. Compared to Nash equilibrium,
rationalizability has two advantages.\footnote{%
In Appendix \ref{sec:Nash-rationalizability}, we use examples to illustrate
their difference.} First, though Nash equilibrium has a simpler definition
than rationalizability, the epistemic foundation of the former is much more
complicated than the latter (see \cite{raab1995}). As a result, from an
epistemic view, the interpretation of the results of rationalizable
implementation is more clear than that of Nash implementation. Second, if
players do not have common knowledge of primitives, a mechanism designer
should require \emph{robust mechanism design}.\footnote{%
Conceptually, robust mechanism design means that we implement a social
choice function not only at each state, but also on a neighbourhood of each
state.} Recent papers (e.g., \cite{dbsm2009}, \cite{dbsm2011}, \cite%
{moot2012}) have shown that robust mechanism design usually leads to
requiring rationalizable implementation. Thus, compared to Nash
implementation, rationalizable implementation better helps us understand
robust mechanism design.

Rationalizable implementation is first studied in \cite{bmt} (hereafter,
BMT). Focusing on social choice functions, BMT show that strict Maskin
monotonicity is necessary for rationalizable implementation, and
furthermore, given two additional technical additional conditions, strict
Maskin monotonicity is also sufficient. In three subsequent papers, \cite{ks}%
, \cite{rjain} and \cite{xiong2} study rationalizable implementation of
social choice correspondences.\footnote{\cite{cksx2} also study
rationalizable implementation, but they allow for monetary transfer.} In
this paper, we focus on social choice functions, and our goal is to fully
characterize rationalizable implementation when no technical condition is
imposed.

Previous characterization of rationalizable implementation (BMT, \cite{rjain}%
, \cite{ks}) hinges critically on the following two assumptions.

\begin{itemize}
\item No Worst Alternatives\ (Definition \ref{def:NWA}, hereafter NWA);%
\footnote{%
NWA is first introduced by \cite{acrs}.}

\item Responsiveness (Definition \ref{def:responsive}).
\end{itemize}

NWA means that the targeted social choice function cannot choose a worst
outcome of any agent at any state. Responsiveness means that the targeted
social choice function is injective. There are plenty of examples in which
either responsiveness or NWA fails. For instance, when the number of states
is larger than the number of social outcomes, responsiveness fails
automatically.\footnote{%
When $\left\vert \Theta \right\vert >\left\vert Z\right\vert $, a social
choice function $f:\Theta \longrightarrow Z$ can never be injective, i.e.,
responsiveness fails.} Technically, this paper develop tools to dispense
with responsiveness and NWA, and use these tools to fully characterize
rationalizable implementation when neither assumption is imposed.

Besides the technical contribution, our results also clarify two conceptual
puzzles implied by the results in BMT. \ Even though Nash equilibria and
rationalizability are two very different solution concepts, the
characterization of full implementation in these two solution concepts are
surprisingly similar: Maskin monotonicity for the former and strict Maskin
monotonicity for the latter.\footnote{%
Given strict preference, Maskin monotonicity is equivalent to strict Maskin
monotonicity.} Our results identify the source of such a coincidence: NWA.
When NWA is relaxed, rationalizable implementation is fully characterized by 
\emph{strict event monotonicity} (Definition \ref%
{def:strict-group-monotonicity}), which embeds an argument of iterated
deletion of never-best replies---a feature that distinguishes
rationalizability from Nash equilibria. When NWA\ holds, we still have
iterated deletion, but the order of deletion does not matter, and hence,
strict event monotonicity reduces to strict Maskin monotonicity.

Second, given NWA, BMT also aim to fully characterize rationalizable
implementation when responsiveness is relaxed: they show that strict Maskin
monotonicity$^{\ast }$ (Definition \ref{def:monotonicity*}) suffices for
rationalizable implementation, and given an additional assumption called
"the best-response property," strict Maskin monotonicity$^{\ast }$ is also
necessary. However, "the best-response property" is not defined on
primitives, and hence, it is not clear whether "the best-response property"
suffers loss of generality.

Given NWA, we prove that strict Maskin monotonicity$^{\ast \ast }$
(Definition \ref{def:monotonicity**}) fully characterizes rationalizable
implementation. In \cite{sx2022}, we provide an example in which NWA and
strict Maskin monotonicity$^{\ast \ast }$ hold, while strict Maskin
monotonicity$^{\ast }$ does not. Therefore, it suffers loss of generality to
impose "the best-response property."

The remainder of the paper proceeds as follows: we describe the model in
Section \ref{sec:model}; we provide motivating examples in Section \ref%
{sec:example}; we illustrate canonical mechanisms in Section \ref%
{sec:illustration}; we deal with responsiveness and NWA in Sections \ref%
{sec:responsive} and \ref{sec:NWA}, respectively; we provide the full
characterization in Section \ref{sec:no-condition}.

\section{Model}

\label{sec:model}

The model consists of 
\begin{equation*}
\left\langle \mathcal{I}=\left\{ i_{1},..,i_{I}\right\} \text{, \ }\Theta
=\left\{ \theta _{1},..,\theta _{n}\right\} \text{, \ }Z\text{, }f:\Theta
\longrightarrow Z\text{, \ }\left( u_{i}:\triangle \left( Z\right) \times
\Theta \longrightarrow 
%TCIMACRO{\U{211d} }%
%BeginExpansion
\mathbb{R}
%EndExpansion
\right) _{i\in I}\right\rangle \text{,}
\end{equation*}%
where $\mathcal{I}$ is a finite set of $I$ agents with $I\geq 3$, $\Theta $
a finite set of $n$ states, $Z$ a countable set of pure social outcomes, $f$
a social choice function (hereafter, SCF) which maps each state in $\Theta $
to an outcome in $Z$, $u_{i}$ the expected utility function of agent $i$.

Throughout the paper, we adopt $Y\equiv \triangle \left( Z\right) $. With
abuse of notation, we also let $z\in Z$ denote a degenerate lottery in $Y$
which assigns probability $1$ to $z$. Then, the function $u_{i}$ with the
restricted domain $Z$ (i.e., $u_{i}|_{Z}:Z\times \Theta \longrightarrow 
%TCIMACRO{\U{211d} }%
%BeginExpansion
\mathbb{R}
%EndExpansion
$) is the Bernoulli utility function of agent $i$, and $u_{i}$ is the
utility function extended to $Y$, i.e.,%
\begin{equation*}
u_{i}\left( y,\theta \right) =\dsum\limits_{z\in Z}y_{z}u_{i}\left( z,\theta
\right) \text{,}
\end{equation*}%
where $y_{z}$ denotes the probability of $z$ under $y$. Throughout the
paper, we we use $-i$ to denote $\mathcal{I}\backslash \left\{ i\right\} $
and assume $\left\vert f\left( \Theta \right) \right\vert \geq 2$.\footnote{%
If $\left\vert f\left( \Theta \right) \right\vert =1$, the implementation
problem can be solved trivially.} Define lower and upper contour sets as
follows.%
\begin{eqnarray*}
\mathcal{L}_{i}\left( y,\theta \right) &=&\left\{ y^{\prime }\in Y:\text{ }%
u_{i}\left( y,\text{ }\theta \right) \geq u_{i}\left( y^{\prime },\text{ }%
\theta \right) \right\} \text{, }\forall y\in Y\text{,} \\
\mathcal{L}_{i}^{\circ }\left( y,\theta \right) &=&\left\{ y^{\prime }\in Y:%
\text{ }u_{i}\left( y,\text{ }\theta \right) >u_{i}\left( y^{\prime },\text{ 
}\theta \right) \right\} \text{, }\forall y\in Y\text{,} \\
\mathcal{U}_{i}^{\circ }\left( y,\theta \right) &=&\left\{ y^{\prime }\in Y:%
\text{ }u_{i}\left( y,\text{ }\theta \right) <u_{i}\left( y^{\prime },\text{ 
}\theta \right) \right\} \text{, }\forall y\in Y\text{.}
\end{eqnarray*}

A mechanism is a tuple $\mathcal{M}=\left\langle M\equiv \times _{i\in 
\mathcal{I}}M_{i}\text{, \ }g:M\longrightarrow Y\right\rangle $, where each $%
M_{i}$ is a countable set, and it denotes the set of strategies for agent $i$
in $\mathcal{M}$. We now define rationalizability and rationalizable
implementation. For every $i\in \mathcal{I}$, define $\mathcal{S}_{i}\equiv
2^{M_{i}}$ and $\mathcal{S}\equiv \times _{i\in \mathcal{I}}\mathcal{S}_{i}$%
. Given any $\left( \mathcal{M},\text{ }\theta \right) $, consider an
operator $b^{\mathcal{M},\text{ }\theta }:\mathcal{S}\longrightarrow 
\mathcal{S}$ with $b^{\mathcal{M},\text{ }\theta }\equiv \left[ b_{i}^{%
\mathcal{M},\text{ }\theta }:\mathcal{S}\longrightarrow \mathcal{S}_{i}%
\right] _{i\in \mathcal{I}}$, where each $b_{i}^{\mathcal{M},\text{ }\theta
} $ is defined as follows. For every $S\in \mathcal{S}$, 
\begin{equation*}
b_{i}^{\mathcal{M},\text{ }\theta }\left( S\right) =\left\{ m_{i}\in M_{i}:%
\begin{tabular}{c}
$\exists \lambda _{-i}\in \triangle \left( M_{-i}\right) $ such that \\ 
(1) $\lambda _{-i}\left( m_{-i}\right) >0$ implies $m_{-i}\in S_{-i}$, and
\\ 
(2) $m_{i}\in \arg \max_{m_{i}^{\prime }\in M_{i}}\Sigma _{m_{-i}\in
M_{-i}}\lambda _{-i}\left( m_{-i}\right) u_{i}\left( g\left( m_{i}^{\prime
},m_{-i}\right) ,\theta \right) $%
\end{tabular}%
\right\} \text{.}
\end{equation*}%
Clearly, $\mathcal{S}$ is a lattice with the order of "set inclusion," and $%
b^{\mathcal{M},\text{ }\theta }$ is increasing, i.e., 
\begin{equation*}
S\subseteq S^{\prime }\text{ }\Longrightarrow \text{ }b^{\mathcal{M},\text{ }%
\theta }\left( S\right) \subseteq b^{\mathcal{M},\text{ }\theta }\left(
S^{\prime }\right) \text{.}
\end{equation*}%
By Tarski's fixed point theorem, a largest fixed point of $b^{\mathcal{M},%
\text{ }\theta }$ exists, and we denote it by $S^{\mathcal{M},\text{ }\theta
}\equiv \left( S_{i}^{\mathcal{M},\text{ }\theta }\right) _{i\in \mathcal{I}%
} $.\footnote{%
That is, $S^{\mathcal{M},\text{ }\theta }=b^{\mathcal{M},\text{ }\theta
}\left( S^{\mathcal{M},\text{ }\theta }\right) $ and for any $S\in \mathcal{S%
}$ with $S=b^{\mathcal{M},\text{ }\theta }\left( S\right) $, we have $%
S\subset S^{\mathcal{M},\text{ }\theta }$.} We say $m_{i}\in M_{i}$ is
rationalizable in $\mathcal{M}$ at $\theta $ if and only if $m_{i}\in S_{i}^{%
\mathcal{M},\text{ }\theta }$, i.e., $S_{i}^{\mathcal{M},\text{ }\theta }$
is the set of rationalizable strategies of agent $i$ in $\mathcal{M}$ at $%
\theta $, and $S^{\mathcal{M},\text{ }\theta }$ is the set of rationalizable
strategy profiles. We say that $S\in \mathcal{S}$ satisfies \emph{the
best-reply property} in $\mathcal{M}$ at $\theta $ if and only if $S\subset
b^{\mathcal{M},\text{ }\theta }\left( S\right) $. Clearly, $S\subset S^{%
\mathcal{M},\text{ }\theta }$, if $S$ satisfies the best-reply property.%
\footnote{%
Suppose $S$ satisfies the best-reply property. Inductively define $\left( b^{%
\mathcal{M},\text{ }\theta }\right) ^{n}\left( S\right) =b^{\mathcal{M},%
\text{ }\theta }\left[ \left( b^{\mathcal{M},\text{ }\theta }\right)
^{n-1}\left( S\right) \right] $. Then, $\cup _{n=1}^{\infty }\left( b^{%
\mathcal{M},\text{ }\theta }\right) ^{n}\left( S\right) $ is a fix point. As
a result, $S\subset \cup _{n=1}^{\infty }\left( b^{\mathcal{M},\text{ }%
\theta }\right) ^{n}\left( S\right) \subset S^{\mathcal{M},\text{ }\theta }$.%
}

\begin{define}
\label{def:implementation}An SCF $f:\Theta \longrightarrow Z$ is
rationalizably implemented by a mechanism $\mathcal{M}$ if%
\begin{equation*}
g\left[ S^{\mathcal{M},\text{ }\theta }\right] =\left\{ f\left( \theta
\right) \right\} \text{, }\forall \theta \in \Theta \text{.}
\end{equation*}%
$f$ is rationalizably implementable if there exists a mechanism that
rationalizably implements $f$.
\end{define}

\section{Motivating examples}

\label{sec:example}

The characterization of rationalizable implementation in BMT hinges
critically on the two conditions defined as follows.

\begin{define}[No worst alternative]
\label{def:NWA}An SCF $f:\Theta \longrightarrow Z$ satisfies "no worst
alternative" (NWA) if, for each $\left( i,\theta \right) \in \mathcal{I}%
\times \Theta $, there exists $z\in Z$ such that%
\begin{equation}
u_{i}\left( f\left( \theta \right) ,\text{ }\theta \right) >u_{i}\left( z,%
\text{ }\theta \right) \text{.}  \label{aaa1}
\end{equation}
\end{define}

\begin{define}[responsiveness]
\label{def:responsive}An SCF $f:\Theta \longrightarrow Z$ is responsive if%
\begin{equation*}
f\left( \theta \right) =f\left( \theta ^{\prime }\right) \text{\ }%
\Longrightarrow \text{ }\theta =\theta ^{\prime }\text{, }\forall \theta
,\theta ^{\prime }\in \Theta \text{.}
\end{equation*}
\end{define}

In this section, we provide examples, showing that we can still characterize
rationalizable implementation, even if any of the two conditions fails.

\subsection{Violation of NWA}

First, we provide an example in which NWA\ fails in a particular way, and we
show that BMT's characterization can still be immediately applied, subject
to some modification. Second, we argue that a similar logic can be applied
to any general violation of NWA.

\begin{equation*}
\begin{tabular}{c}
Example 1.a: $\mathcal{I}=\left\{ i_{1},i_{2},i_{3},i_{4}\right\} $, $\Theta
=\left\{ \theta _{1},\theta _{2},\theta _{3}\right\} $, $Z=\left\{
a,b,c\right\} $, \\ 
\begin{tabular}{c}
state $\theta _{1}$ with $f\left( \theta _{1}\right) =a$, \\ 
$%
\begin{tabular}{|c|c|c|}
\hline
$u_{i}\left( z,\theta _{1}\right) $ & $i=i_{1},i_{2},i_{3}$ & $i=i_{4}$ \\ 
\hline
$z=a$ & $2$ & $0$ \\ \hline
$z=b$ & $1$ & $1$ \\ \hline
$z=c$ & $0$ & $2$ \\ \hline
\end{tabular}%
$%
\end{tabular}%
\begin{tabular}{c}
state $\theta _{2}$ with $f\left( \theta _{2}\right) =b$, \\ 
$%
\begin{tabular}{|c|c|c|}
\hline
$u_{i}\left( z,\theta _{2}\right) $ & $i=i_{1},i_{2},i_{3}$ & $i=i_{4}$ \\ 
\hline
$z=a$ & $0$ & $2$ \\ \hline
$z=b$ & $2$ & $0$ \\ \hline
$z=c$ & $1$ & $1$ \\ \hline
\end{tabular}%
$%
\end{tabular}%
\begin{tabular}{c}
state $\theta _{3}$ with $f\left( \theta _{3}\right) =c$, \\ 
$%
\begin{tabular}{|c|c|c|}
\hline
$u_{i}\left( z,\theta _{3}\right) $ & $i=i_{1},i_{2},i_{3}$ & $i=i_{4}$ \\ 
\hline
$z=a$ & $0$ & $1$ \\ \hline
$z=b$ & $1$ & $2$ \\ \hline
$z=c$ & $2$ & $0$ \\ \hline
\end{tabular}%
$%
\end{tabular}%
\end{tabular}%
\end{equation*}%
The tables record agents' utility of each social outcome at each state. For
example, agent 1 has utility of 2 for the outcome $a$ at state $\theta _{1}$.

In Example 1.a, strict Maskin monotonicity and responsiveness hold, but NWA
is violated (for agent $i_{4}$ at every state). Consider a modified version
of Example 1.a, described as follows---the only difference is that agent $4$
is eliminated in Example 1.b.%
\begin{equation*}
\begin{tabular}{c}
Example 1.b: $\mathcal{I}=\left\{ i_{1},i_{2},i_{3}\right\} $, $\Theta
=\left\{ \theta _{1},\theta _{2},\theta _{3}\right\} $, $Z=\left\{
a,b,c\right\} $, \\ 
\begin{tabular}{c}
state $\theta _{1}$ with $f\left( \theta _{1}\right) =a$, \\ 
$%
\begin{tabular}{|c|c|}
\hline
$u_{i}\left( z,\theta _{1}\right) $ & $i=i_{1},i_{2},i_{3}$ \\ \hline
$z=a$ & $2$ \\ \hline
$z=b$ & $1$ \\ \hline
$z=c$ & $0$ \\ \hline
\end{tabular}%
$%
\end{tabular}%
\begin{tabular}{c}
state $\theta _{2}$ with $f\left( \theta _{2}\right) =b$, \\ 
$%
\begin{tabular}{|c|c|}
\hline
$u_{i}\left( z,\theta _{2}\right) $ & $i=i_{1},i_{2},i_{3}$ \\ \hline
$z=a$ & $0$ \\ \hline
$z=b$ & $2$ \\ \hline
$z=c$ & $1$ \\ \hline
\end{tabular}%
$%
\end{tabular}%
\begin{tabular}{c}
state $\theta _{3}$ with $f\left( \theta _{3}\right) =c$, \\ 
$%
\begin{tabular}{|c|c|}
\hline
$u_{i}\left( z,\theta _{3}\right) $ & $i=i_{1},i_{2},i_{3}$ \\ \hline
$z=a$ & $0$ \\ \hline
$z=b$ & $1$ \\ \hline
$z=c$ & $2$ \\ \hline
\end{tabular}%
$%
\end{tabular}%
\end{tabular}%
\end{equation*}%
In this modified example, strict Maskin monotonicity, responsiveness and NWA
hold, i.e., the sufficient condition in BMT is satisfied. As a result, we
can achieve rationalizable implementation in Example 1.b, which further
implies that we can achieve rationalizable implementation in Example 1.a (by
using the same mechanism for Example 1.b and ignore agent $4$'s
report).---That is, literally, BMT's characterization cannot be applied in
Example 1.a, but it can be applied in Example 1.b., which indirectly
provides characterization in Example 1.a.

Examples 1.a and 1.b shed light on rationalizable implementation when NWA
fails. We say agent $i$ is \emph{inactive} at state $\theta $ if and only if
NWA is violated for agent $i$ at $\theta $.\footnote{%
Rigorously, agent $i$ is inactive at state $\theta $ if and only if $%
u_{i}\left( f\left( \theta \right) ,\text{ }\theta \right) \leq u_{i}\left(
z,\text{ }\theta \right) $ for every $z\in Z$.} The intuition is that, if
agent $i$ is inactive at $\theta $, we can (and should) ignore agent $i$ at
state $\theta $---this intuition is formalized in Lemma \ref{lem:dominated}
in Section \ref{sec:intuition:NWA}. In this simple example, agent $4$ is
inactive at every state, and hence, we can ignore her totally.

However, in more general cases, it may happen that $i$ is inactive at some
state $\theta $, while $j\left( \neq i\right) $ is inactive at some other
state $\theta ^{\prime }$, as described in the following example.%
\begin{equation*}
\begin{tabular}{c}
Example 2: $\mathcal{I}=\left\{ i_{1},i_{2},i_{3},i_{4}\right\} $, $\Theta
=\left\{ \theta _{1},\theta _{2},\theta _{3}\right\} $, $Z=\left\{
a,b,c\right\} $, \\ 
\begin{tabular}{c}
state $\theta _{1}$ with $f\left( \theta _{1}\right) =a$, \\ 
$%
\begin{tabular}{|c|c|c|}
\hline
$u_{i}\left( z,\theta _{1}\right) $ & $i=i_{1},i_{2},i_{3}$ & $i=i_{4}$ \\ 
\hline
$z=a$ & $2$ & $0$ \\ \hline
$z=b$ & $1$ & $1$ \\ \hline
$z=c$ & $0$ & $2$ \\ \hline
\end{tabular}%
$%
\end{tabular}%
\begin{tabular}{c}
state $\theta _{2}$ with $f\left( \theta _{2}\right) =b$, \\ 
$%
\begin{tabular}{|c|c|c|}
\hline
$u_{i}\left( z,\theta _{2}\right) $ & $i=i_{1},i_{2},i_{4}$ & $i=i_{3}$ \\ 
\hline
$z=a$ & $0$ & $2$ \\ \hline
$z=b$ & $2$ & $0$ \\ \hline
$z=c$ & $1$ & $1$ \\ \hline
\end{tabular}%
$%
\end{tabular}%
\begin{tabular}{c}
state $\theta _{3}$ with $f\left( \theta _{3}\right) =c$, \\ 
$%
\begin{tabular}{|c|c|c|}
\hline
$u_{i}\left( z,\theta _{3}\right) $ & $i=i_{1},i_{3},i_{4}$ & $i=i_{2}$ \\ 
\hline
$z=a$ & $0$ & $1$ \\ \hline
$z=b$ & $1$ & $2$ \\ \hline
$z=c$ & $2$ & $0$ \\ \hline
\end{tabular}%
$%
\end{tabular}%
\end{tabular}%
\end{equation*}%
We will show that the same logic as above applies, i.e., we need to ignore
the inactive agent $i_{4}$ at state $\theta _{1}$, the inactive agent $i_{3}$
at state $\theta _{2}$, the inactive agent $i_{2}$ at state $\theta _{3}$.
--- This immediately leads to a technical difficulty: since the mechanism
designer does not know the true state, and hence, \textit{a priori}, cannot
tell when to ignore which agent. Thus, the goal of the mechanism designer is
to build a game in which the reports of all agents collectively determine
the true state, which guides him regarding when to ignore which agent. We
will build a new canonical mechanism which achieves this goal. For instance,
it is intuitive that our canonical mechanism would dictate the following.%
\begin{equation*}
\left[ 
\begin{array}{c}
\text{if agents }i_{1}\text{, }i_{2}\text{ an }i_{3}\text{ report state }%
\theta _{1}\text{, we would ignore agent }i_{4}\text{, and implement }%
f\left( \theta _{1}\right) \text{,} \\ 
\text{if agents }i_{1}\text{, }i_{2}\text{ an }i_{4}\text{ report state }%
\theta _{2}\text{, we would ignore agent }i_{3}\text{, and implement }%
f\left( \theta _{2}\right) \text{,} \\ 
\text{if agents }i_{1}\text{, }i_{3}\text{ an }i_{4}\text{ report state }%
\theta _{3}\text{, we would ignore agent }i_{2}\text{, and implement }%
f\left( \theta _{3}\right) \text{.}%
\end{array}%
\right]
\end{equation*}%
We discuss more intuition of our canonical mechanism in Section \ref%
{sec:illustration:NWA}.

\subsection{Violation of responsiveness}

\label{sec:example:responsive}

Consider the following degenerate example, in which responsiveness fails.%
\begin{equation*}
\begin{tabular}{c}
Example 3.a: $\mathcal{I}=\left\{ i_{1},i_{2},i_{3}\right\} $, $\Theta
=\left\{ \theta _{1},\theta _{2},\theta _{3}\right\} $, $Z=\left\{
a,b,c\right\} $, \\ 
\begin{tabular}{c}
state $\theta _{1}$ with $f\left( \theta _{1}\right) =a$, \\ 
$%
\begin{tabular}{|c|c|}
\hline
$u_{i}\left( z,\theta _{1}\right) $ & $i=i_{1},i_{2},i_{3}$ \\ \hline
$z=a$ & $1$ \\ \hline
$z=b$ & $2$ \\ \hline
$z=c$ & $0$ \\ \hline
\end{tabular}%
$%
\end{tabular}
\ \ 
\begin{tabular}{c}
state $\theta _{2}$ with $f\left( \theta _{2}\right) =a$, \\ 
$%
\begin{tabular}{|c|c|}
\hline
$u_{i}\left( z,\theta _{2}\right) $ & $i=i_{1},i_{2},i_{3}$ \\ \hline
$z=a$ & $1$ \\ \hline
$z=b$ & $2$ \\ \hline
$z=c$ & $0$ \\ \hline
\end{tabular}%
$%
\end{tabular}
\ \ 
\begin{tabular}{c}
state $\theta _{3}$ with $f\left( \theta _{3}\right) =c$, \\ 
$%
\begin{tabular}{|c|c|}
\hline
$u_{i}\left( z,\theta _{3}\right) $ & $i=i_{1},i_{2},i_{3}$ \\ \hline
$z=a$ & $0$ \\ \hline
$z=b$ & $1$ \\ \hline
$z=c$ & $2$ \\ \hline
\end{tabular}%
$%
\end{tabular}%
\end{tabular}%
\end{equation*}%
In this degenerate example, states $\theta _{1}$ and $\theta _{2}$ are the
"same" in the sense that all agents' preferences do not change in the two
states. Consider the following slightly modified version of Example 3.a, in
which state $\theta _{2}$ is eliminated.%
\begin{equation*}
\begin{tabular}{c}
Example 3.b: $\mathcal{I}=\left\{ i_{1},i_{2},i_{3}\right\} $, $\Theta
=\left\{ \theta _{1},\theta _{3}\right\} $, $Z=\left\{ a,b,c\right\} $, \\ 
\begin{tabular}{c}
state $\theta _{1}$ with $f\left( \theta _{1}\right) =a$, \\ 
$%
\begin{tabular}{|c|c|}
\hline
$u_{i}\left( z,\theta _{1}\right) $ & $i=i_{1},i_{2},i_{3}$ \\ \hline
$z=a$ & $1$ \\ \hline
$z=b$ & $2$ \\ \hline
$z=c$ & $0$ \\ \hline
\end{tabular}%
$%
\end{tabular}
\ 
\begin{tabular}{c}
state $\theta _{3}$ with $f\left( \theta _{3}\right) =c$, \\ 
$%
\begin{tabular}{|c|c|}
\hline
$u_{i}\left( z,\theta _{3}\right) $ & $i=i_{1},i_{2},i_{3}$ \\ \hline
$z=a$ & $0$ \\ \hline
$z=b$ & $1$ \\ \hline
$z=c$ & $2$ \\ \hline
\end{tabular}%
$%
\end{tabular}%
\end{tabular}%
\end{equation*}%
Clearly, responsiveness holds in Example 3.b, and BMT's result shows that we
can achieve rationalizable implementation, which further implies that we can
also achieve rationalizable implementation in Example 3.a (by using the same
mechanism for Example 3.b).

Example 3.a is a degenerate case of violation of responsiveness. The
following example shows that we can still achieve rationalizable
implementation in more general cases. Example 3.c differs from Example 3.b
only at state $\theta _{2}^{\prime }$.

\begin{equation*}
\begin{tabular}{c}
Example 3.c: $\mathcal{I}=\left\{ i_{1},i_{2},i_{3}\right\} $, $\Theta
=\left\{ \theta _{1},\theta _{2}^{\prime },\theta _{3}\right\} $, $Z=\left\{
a,b,c\right\} $, \\ 
\begin{tabular}{c}
state $\theta _{1}$ with $f\left( \theta _{1}\right) =a$, \\ 
$%
\begin{tabular}{|c|c|}
\hline
$u_{i}\left( z,\theta _{1}\right) $ & $i=i_{1},i_{2},i_{3}$ \\ \hline
$z=a$ & $1$ \\ \hline
$z=b$ & $2$ \\ \hline
$z=c$ & $0$ \\ \hline
\end{tabular}%
$%
\end{tabular}
\ \ 
\begin{tabular}{c}
state $\theta _{2}$ with $f\left( \theta _{2}^{\prime }\right) =a$, \\ 
$%
\begin{tabular}{|c|c|}
\hline
$u_{i}\left( z,\theta _{2}^{\prime }\right) $ & $i=i_{1},i_{2},i_{3}$ \\ 
\hline
$z=a$ & $2$ \\ \hline
$z=b$ & $1$ \\ \hline
$z=c$ & $0$ \\ \hline
\end{tabular}%
$%
\end{tabular}
\ \ 
\begin{tabular}{c}
state $\theta _{3}$ with $f\left( \theta _{3}\right) =c$, \\ 
$%
\begin{tabular}{|c|c|}
\hline
$u_{i}\left( z,\theta _{3}\right) $ & $i=i_{1},i_{2},i_{3}$ \\ \hline
$z=a$ & $0$ \\ \hline
$z=b$ & $1$ \\ \hline
$z=c$ & $2$ \\ \hline
\end{tabular}%
$%
\end{tabular}%
\end{tabular}%
\end{equation*}%
In Example 3.c, we have 
\begin{equation}
\mathcal{L}_{i}\left( f\left( \theta _{1}\right) ,\theta _{1}\right) \subset 
\mathcal{L}_{i}\left( f\left( \theta _{1}\right) ,\theta _{2}^{\prime
}\right) \text{, }\forall i\in \mathcal{I}\text{,}  \label{tutt1}
\end{equation}%
which immediately implies $S^{\mathcal{M},\text{ }\theta _{1}}=S^{\mathcal{M}%
,\text{ }\theta _{2}^{\prime }}$,\footnote{%
Suppose $F$ is rationalizably implemented by a mechanism $\mathcal{M}$.
Consider any $i\in \mathcal{I}$ and any $m_{i}\in S_{i}^{\mathcal{M},\text{ }%
\theta _{1}}$. Thus, $m_{i}$ is a best reply to some some rationalizable
belief $\lambda _{-i}$, or equivalently, given $\lambda _{-i}$, any of $i$'s
unilateral deviations would induce an outcome in $\mathcal{L}_{i}\left(
f\left( \theta _{1}\right) ,\theta _{1}\right) $. Thus, (\ref{tutt1})
implies that $m_{i}$ remains a best reply to $\lambda _{-i}$ at $\theta
_{2}^{\prime }$. As a result, $S_{i}^{\mathcal{M},\text{ }\theta
_{1}}\subset S_{i}^{\mathcal{M},\text{ }\theta _{2}^{\prime }}$, which,
together with Lemma \ref{lem:yt}, implies $S_{i}^{\mathcal{M},\text{ }\theta
_{1}}=S_{i}^{\mathcal{M},\text{ }\theta _{2}^{\prime }}$.} and hence, we can
also achieve rationalizable implementation in Example 3.c (by using the same 
\emph{canonical} mechanism for Example 3.b).

In fact, given violation of responsiveness, we can achieve rationalizable
implementation when a much weaker condition than (\ref{tutt1}) holds (see
Definition \ref{def:monotonicity**} and Theorem \ref{theorem:non-responsive}%
).

\section{Illustration of the canonical mechanisms}

\label{sec:illustration}

In this section, we describe different canonical mechanisms that are used to
achieve implementation in \cite{em}, BMT, and this paper.

\subsection{The modified revelation principle}

\label{sec:illustration:revelation}

It is well-known that the revelation principle fails in full implementation.
Nevertheless, a modified version holds. To see this, suppose that a
mechanism $g:\times _{i\in \mathcal{I}}M_{i}\longrightarrow Z$ achieves full
implementation in some solution concept (e.g., Nash equilibrium,
rationalizability). Pick any one of the solution ($\left( \phi _{i}:\Theta
\longrightarrow M_{i}\right) _{i\in \mathcal{I}}$) of $g$, and we re-label
each "$\phi _{i}\left( \theta \right) $" to a new message "$\theta $."
Furthermore, denote $\widetilde{M}_{i}\equiv M_{i}\diagdown \phi _{i}\left(
\Theta \right) $, and hence $M_{i}=\left[ \phi _{i}\left( \Theta \right) %
\right] \cup \widetilde{M}_{i}$. Then, the original mechanism $g$ can be
"rephrased" to $\widetilde{g}:\times _{i\in \mathcal{I}}\left( \Theta \cup 
\widetilde{M}_{i}\right) \longrightarrow Z$ by relabeling $\phi _{i}\left(
\theta \right) $ to $\theta $, and $\widetilde{g}$ achieves full
implementation.\footnote{%
Besides the solution $\left[ \phi _{i}\left( \theta \right) \right] _{\theta
\in \Theta ,i\in \mathcal{I}}$ chosen in $g$, there may be other solutions
involving messages in $\left[ \times _{i\in \mathcal{I}}M_{i}\right]
\diagdown \left\{ \theta \in \Theta :\left[ \phi _{i}\left( \theta \right) %
\right] _{i\in \mathcal{I}}\right\} $. Such solutions correspond to Case (2)
and Case (3) in canonical mechanisms discussed below. Thus, our goal is to
choose each $\widetilde{M}_{i}$ carefully so that the solutions involving
Case (2) and Case (3) in canonical mechanisms still achieve full
implementation.} Let us call such $\widetilde{g}$ an \emph{augmented direct
mechanism} (with the augmented messages in $\widetilde{M}_{i}$ for agent $i$%
). Therefore, this establishes a modified revelation principle for full
implementation: it suffers no loss of generality to consider augmented
direct mechanisms.\footnote{%
However, this revelation principle is much weaker than the original
revelation principal for partial implemenation. For the latter, a direct
mechanism is precisely defined, but for the former, an augamented direct
mechanism is vaguely defined (i.e., $\widetilde{M}_{i}$ is not a precisely
defined set).}

This modified revelation principle provides the basis for canonical
mechanisms in full implementation. First, all agents truthfully reporting $%
\theta $ at state $\theta $ is always a solution in the augmented direct
mechanism. Second, the implementation problem is reduced to identifying the
augmented messages in $\widetilde{M}_{i}$ in order to achieve full
implementation. Most papers in the literature of full implementation follow
this idea.

\subsection{The canonical mechanisms}

\label{sec:illustration:mechanism}

There is a generic form for most mechanisms in full implementation, which is
described as follows.

\begin{quote}
Agents are invited to report the true state. There are three cases for
agents' reports: (1) agreement, i.e., all agents report the same state $%
\theta $; (2) unilateral deviation, i.e., all agents except agent $j$ report 
$\theta $; (3) multi-lateral deviation, i.e., this includes all other
scenarios. In Case (1), the canonical mechanism picks $f\left( \theta
\right) $. In Case (2), agent $j$ is allowed to choose any outcome in $%
\mathcal{L}_{j}\left( f\left( \theta \right) ,\theta \right) $, which
ensures that truthful reporting is a Nash equilibrium (and hence,
rationalizable). In Case (3), we first let all agents compete by submitting
a positive integer. The agent who submits the largest integer wins\ (subject
to any tie-breaking rule), and we let the winner choose any outcome in $Z$.
\end{quote}

At the true state $\theta $, "all agents reporting $\theta $" is a "good"
equilibrium (or solution), which induces $f\left( \theta \right) $. In order
to achieve full implementation, we have to further make sure that there is
no "bad" equilibrium in any of Cases (1), (2) and (3).

For different environments and/or solution concepts, we may have to modify
the canonical mechanism above slightly, which is illustrated below.

\subsection{Illustration of \protect\cite{em}}

\label{sec:illustration:Maskin}

\cite{em} adopts the canonical mechanism in Section \ref%
{sec:illustration:mechanism}, and uses Maskin monotonicity to eliminate
"bad" equilibria in Cases (1), and uses no-veto power to eliminate "bad"
equilibria in Cases (2) and (3).

\begin{define}[Maskin monotonicity]
An SCF $f$ satisfies Maskin monotonicity if%
\begin{equation}
f\left( \theta \right) \neq f\left( \theta ^{\prime }\right) \Longrightarrow
\left( 
\begin{array}{c}
\exists j\in \mathcal{I}, \\ 
\mathcal{L}_{j}\left( f\left( \theta \right) ,\theta \right) \cap \mathcal{U}%
_{j}^{\circ }\left( f\left( \theta \right) ,\theta ^{\prime }\right) \neq
\varnothing%
\end{array}%
\right) \text{, }\forall \left( \theta ,\theta ^{\prime }\right) \in \Theta
\times \Theta \text{.}  \label{cdd}
\end{equation}
\end{define}

\begin{define}[no-veto power]
An SCF $f$ satisfies no-veto power if%
\begin{equation*}
\left\vert \left\{ i\in \mathcal{I}:a\in \arg \max_{z\in Z}u_{i}\left(
z,\theta \right) \right\} \right\vert \geq \left\vert \mathcal{I}\right\vert
-1\Longrightarrow a\in f\left( \theta \right) \text{, }\forall \left( \theta
,a\right) \in \Theta \times Z\text{.}
\end{equation*}
\end{define}

Suppose the true state is $\theta ^{\prime }$. A "bad" equilibrium in Case
(1) means that all agents report $\theta $ with $f\left( \theta \right) \neq
f\left( \theta ^{\prime }\right) $. Given Maskin monotonicity, such a
strategy profile cannot be an equilibrium, because (\ref{cdd}) implies agent 
$j$ has a profitable deviation to Case (2), i.e., $j$ can pick $y\in 
\mathcal{L}_{j}\left( f\left( \theta \right) ,\theta \right) \cap \mathcal{U}%
_{j}^{\circ }\left( f\left( \theta \right) ,\theta ^{\prime }\right) $%
.---When this happens, $j$ is called a whistle-blower, and $y\in \mathcal{L}%
_{j}\left( f\left( \theta \right) ,\theta \right) \cap \mathcal{U}%
_{j}^{\circ }\left( f\left( \theta \right) ,\theta ^{\prime }\right) $ is
called $j$'s blocking plan.\footnote{%
That is, $j$ uses $y\in \mathcal{L}_{j}\left( f\left( \theta \right) ,\theta
\right) \cap \mathcal{U}_{j}^{\circ }\left( f\left( \theta \right) ,\theta
^{\prime }\right) $ to inform the mechanism designer that agents $-j$ are
lying by reporting $\theta $. "$y\in \mathcal{L}_{j}\left( f\left( \theta
\right) ,\theta \right) $" ensures that $j$'s information is credible,
because if agents $-j$ were not lying, $y$ would be inferior to $j$ at $%
\theta $. "$y\in \mathcal{U}_{j}^{\circ }\left( f\left( \theta \right)
,\theta ^{\prime }\right) $" ensures that $y$ is indeed a profitable
deviation to $j$ at the true state $\theta ^{\prime }$.}

A "bad" equilibrium in Case (2) or Case (3) is that agent $j$ deviates from
Case (1), and it induces $c\neq f\left( \theta \right) $. Given no-veto
power, such a strategy profile cannot be an equilibrium, because the other $%
\left\vert \mathcal{I}\right\vert -1$ agents (i.e., agents $-j$) can further
deviate to Case (3) , and induce their top outcomes in $Z$ (by submitting a
largest integer). If it were an equilibrium, $c$ would be a top outcome for
the other $\left\vert \mathcal{I}\right\vert -1$ agents, which, together
with no-veto power, implies $c=f\left( \theta \right) $, contradicting $%
c\neq f\left( \theta \right) $.

\subsection{Illustration of BMT}

\label{sec:illustration:bmt}

To achieve rationalizable implementation, BMT uses strict Maskin
monotonicity to eliminate bad solutions in Case (1), when responsiveness
holds.\footnote{%
Suppose the true state in $\theta ^{\ast }$. More precisely, strict Maskin
montonicity implies existence of a whistle-blower whenever agents reach an
agreement on $\theta $ with $f\left( \theta \right) \neq f\left( \theta
^{\ast }\right) $. Furthermore, given responsiveness, "$f\left( \theta
\right) \neq f\left( \theta ^{\ast }\right) $" is equivalent to "$\theta
\neq \theta ^{\ast }$." Therefore, strict Maskin monotonicity eliminates any
bad solutions in Case (1), i.e., agreement on $\theta $ with $\theta \neq
\theta ^{\ast }$.} The intuition is the same as above.

\begin{define}[strict Maskin monotonicity]
\label{def:strict-Maskin}An SCF $f$ satisfies strict Maskin monotonicity if%
\begin{equation}
f\left( \theta \right) \neq f\left( \theta ^{\prime }\right) \Longrightarrow
\left( 
\begin{array}{c}
\exists j\in \mathcal{I}, \\ 
\mathcal{L}_{j}^{\circ }\left( f\left( \theta \right) ,\theta \right) \cap 
\mathcal{U}_{j}^{\circ }\left( f\left( \theta \right) ,\theta ^{\prime
}\right) \neq \varnothing%
\end{array}%
\right) \text{, }\forall \left( \theta ,\theta ^{\prime }\right) \in \Theta
\times \Theta \text{.}  \label{kf23}
\end{equation}
\end{define}

BMT uses NWA to eliminate bad solutions in Cases (2) and (3). NWA\ implies
existence of $\underline{y}\in Y$ such that 
\begin{equation}
\underline{y}\notin \dbigcup\limits_{\theta \in \Theta
}\dbigcup\limits_{i\in \mathcal{I}}\arg \max_{y\in Y}u_{i}\left( y,\theta
\right) \text{,}  \label{ktt1}
\end{equation}%
i.e., $\underline{y}$ is never a top outcome for any agent at any state.
Furthermore, BMT prove that NWA and strict Maskin monotonicity implies
existence of $z_{i}\left( \theta ,\theta \right) \in \mathcal{L}_{i}\left(
f\left( \theta \right) ,\theta \right) $ for each $\left( \theta ,i\right)
\in \Theta \times \mathcal{I}$ such that%
\begin{equation}
\max_{y\in \mathcal{L}_{i}\left( f\left( \theta \right) ,\theta \right)
}u_{i}\left( y,\theta ^{\ast }\right) >u_{i}\left( z_{i}\left( \theta
,\theta \right) ,\theta ^{\ast }\right) \text{, }\forall \theta ^{\ast }\in
\Theta \text{.}  \label{ktt2}
\end{equation}%
Furthermore, BMT modify Cases (2) and (3) in the canonical mechanism as
follows. In Case (2), agent $j$ is allowed to choose any $y\in \mathcal{L}%
_{j}\left( f\left( \theta \right) ,\theta \right) $ and any positive integer 
$n$, and the mechanism picks%
\begin{equation}
\frac{n-1}{n}\times y+\frac{1}{n}\times z_{i}\left( \theta ,\theta \right)
\in \mathcal{L}_{j}\left( f\left( \theta \right) ,\theta \right)
\label{uut1}
\end{equation}%
In Case (3), the agent who submits the largest integer is allowed to choose
any $z\in Z$ and any positive integer $n$, and the mechanism picks%
\begin{equation}
\frac{n-1}{n}\times z+\frac{1}{n}\times \underline{y}\text{.}  \label{uut2}
\end{equation}%
(\ref{ktt1}), (\ref{ktt2}), (\ref{uut1}) and (\ref{uut2}) imply that all
agents can never have a best reply in Cases (2) and (3), and hence, no "bad"
solution.

\subsection{Illustration of our canonical mechanism: NWA}

\label{sec:illustration:NWA}

When neither NWA\ nor responsiveness is imposed, the full characterization
of rationalizable implementation is complicated. For expositional ease, we
treat the two technical problems separately, and develop tools to fully
characterize rationalizable implementation when only one of the two
assumptions holds.

Suppose responsiveness holds, while NWA\ does not. The innovation of our
canonical mechanism is introducing the notion of "active agents."

\begin{gather*}
\mathcal{I}^{\theta }=\left\{ i\in \mathcal{I}:\exists z\in Z\text{ such
that }u_{i}\left( f\left( \theta \right) ,\text{ }\theta \right)
>u_{i}\left( z,\text{ }\theta \right) \right\} \text{, }\forall \theta \in
\Theta \text{,} \\
\text{and }\mathcal{I}^{E}=\dbigcap\limits_{\theta \in E}\mathcal{I}^{\theta
}\text{, }\forall E\in \left[ 2^{\Theta }\backslash \left\{ \varnothing
\right\} \right] \text{.}
\end{gather*}%
That is, $\mathcal{I}^{\theta }$ is the set of agents who can make condition
(\ref{aaa1}) in NWA hold at state $\theta $. We call agents in $\mathcal{I}%
^{\theta }$ \emph{active agents }at state $\theta $. Clearly, NWA is
equivalent to requiring $\mathcal{I}^{\Theta }=\mathcal{I}$.

Without NWA, what goes wrong in BMT's canonical mechanism? Precisely,
conditions (\ref{ktt1}) and (\ref{ktt2}) do not hold, and BMT's proof breaks
down. However, by incorporating the notion of active agents, modified
versions of (\ref{ktt1}) and (\ref{ktt2}) hold.%
\begin{equation}
\underline{y}\notin \dbigcup\limits_{\theta \in \Theta
}\dbigcup\limits_{i\in \mathcal{I}^{\theta }}\arg \max_{y\in Y}u_{i}\left(
y,\theta \right) \text{. }  \label{ktt3}
\end{equation}%
\begin{equation}
\max_{y\in \mathcal{L}_{i}\left( f\left( \theta \right) ,\theta \right)
}u_{i}\left( y,\theta ^{\ast }\right) >u_{i}\left( z_{i}\left( \theta
,\theta \right) ,\theta ^{\ast }\right) \text{, }\forall \left( \theta
,\theta ^{\ast }\right) \in \Theta \times \Theta \text{, }\forall i\in 
\mathcal{I}^{\theta ^{\ast }}\text{.}  \label{ktt4}
\end{equation}%
I.e., (\ref{ktt1}) and (\ref{ktt2}) hold only for active agents at each
state.

With this being sorted out, there is only one major difference between BMT's
canonical mechanism in Section \ref{sec:illustration:bmt} and ours: how is
an agreement defined in case (1)?

Specifically, we modify the canonical mechanism as follows. In case (1), all
agents in $\mathcal{I}^{\theta }$\ report the same state $\theta $,\footnote{%
To see why an agreement is determined by agents in $\mathcal{I}^{\theta }$ 
\emph{only}, suppose that a canonical mechanism achieves rationalizable
implementation. At state $\theta $, consider any inactive agent $j\notin $ $%
\mathcal{I}^{\theta }$. Pick any rationalizable strategy of agent $j$, and
it is a best reply to a rationalizable conjecture, which induces the worst
outcome $f\left( \theta \right) $ for $j$. This immediately implies any
other strategy must also be a best reply to the same rationalizable
conjecture, i.e., all the other strategies are rationalizable for $j$. Or
equivalently, any of $j$'s report of the true state is not informative.
Therefore, an agreement is determined by agents in $\mathcal{I}^{\theta }$
only.} and the mechanism picks $f\left( \theta \right) $\textbf{.} The rest
of the canonical mechanism remains the same as above.

We will use strict event monotonicity and dictator monotonicity (see
Definitions \ref{def:strict-group-monotonicity} and \ref{def:dictator}) to
eliminate "bad" solutions in Case (1). And as above, all agents never have a
best reply in Cases (2) and (3), i.e., no "bad" solution.

\subsection{Illustration of our canonical mechanism: responsiveness}

\label{sec:illustration:responsive}

Suppose NWA\ holds, while responsiveness does not. What goes wrong in BMT's
canonical mechanism? To see this, consider a concrete example.%
\begin{equation*}
\Theta =\left\{ \theta ^{1},\theta ^{2},\theta ^{3}\right\} \text{ such that 
}f\left( \theta ^{1}\right) \neq f\left( \theta ^{2}\right) =f\left( \theta
^{3}\right) \text{.}
\end{equation*}%
Suppose that the true state is $\theta ^{2}$. Strict Maskin monotonicity
ensures that agreement on $\theta ^{1}$ in the canonical mechanism will not
be a Nash equilibrium (or more precisely, will not be rationalizable).
However, strict Maskin monotonicity does not preclude the possibility that
agreement on $\theta ^{3}$ in the canonical mechanism is a Nash equilibrium.%
\footnote{%
Strict Maskin monotonicity kicks in only when $f\left( \theta ^{2}\right)
\neq f\left( \theta ^{3}\right) $, but, here, we have $f\left( \theta
^{2}\right) =f\left( \theta ^{3}\right) $.} Such a possibility does not
destroy Nash implementation, because $f\left( \theta ^{2}\right) =f\left(
\theta ^{3}\right) $. However, it destroys rationalizable implementation,
which is due to a distinct feature of rationalizability. When both "all
agents reporting $\theta ^{2}$" and "all agents reporting $\theta ^{3}$" are
Nash equilibria, one rationalizable strategy profile could be: "odd-indexed
agents reporting $\theta ^{2}$, and even-indexed agents reporting $\theta
^{3}$," which would trigger either case (2) or case (3), and induce
undesired outcomes $z_{i}\left( \theta ,\theta \right) $ or $\underline{y}$
with positive probability, i.e., rationalizable implementation is not
achieved.

\subsubsection{BMT's attempt}

BMT\ provide a first attempt to characterize rationalizable implementation,
when responsiveness is violated. Suppose the true state is $\theta ^{2}$.
Strict Maskin monotonicity provides tools for a whistle-blower to block 
\emph{only} "reporting $\theta $" with $f\left( \theta \right) \neq f\left(
\theta ^{2}\right) $. The example above shows that the problem comes from $%
\theta ^{\prime }$ with $f\left( \theta ^{\prime }\right) =f\left( \theta
^{2}\right) $. Thus, if there is no whistle-blower who is able to block
"reporting $\theta ^{\prime }$" with $f\left( \theta ^{\prime }\right)
=f\left( \theta ^{2}\right) $, we have to identify $\theta ^{\prime }$ and $%
\theta ^{2}$, in order to avoid (the undesired outcomes in) Cases (2) and
(3).\footnote{%
That is, when all agents report $\theta ^{\prime }$ or $\theta ^{2}$, we
regard it as Case (1), and the canonical mechanism picks $f\left( \theta
^{\prime }\right) =f\left( \theta ^{2}\right) $.} Or equivalently, we must
form a partition $\mathcal{P}^{\ast }$ on $\Theta $ such that%
\begin{equation*}
\mathcal{P}^{\ast }\left( \theta \right) =\mathcal{P}^{\ast }\left( \theta
^{\prime }\right) \Longrightarrow f\left( \theta \right) =f\left( \theta
^{\prime }\right) \text{, }\forall \left( \theta ,\theta ^{\prime }\right)
\in \Theta \times \Theta \text{,}
\end{equation*}%
and at any true state $\theta $, reporting any state in $\mathcal{P}^{\ast
}\left( \theta \right) $ must be rationalizable in the canonical mechanism
for all agents at $\theta $. This immediately leads to an additional
requirement: 
\begin{equation}
\mathcal{P}^{\ast }\left( \theta \right) \neq \mathcal{P}^{\ast }\left(
\theta ^{\prime }\right) \Longrightarrow \left( 
\begin{array}{c}
\text{at the true state }\theta ^{\prime }\text{,} \\ 
\text{there exists a whistle-blower }j\in \mathcal{I}\text{ such that} \\ 
j\text{ can block "agents }-j\text{ reporting }\widehat{\theta }\text{"} \\ 
\text{\emph{simultaneously} for any }\widehat{\theta }\in \mathcal{P}^{\ast
}\left( \theta \right)%
\end{array}%
\right) \text{, }\forall \left( \theta ,\theta ^{\prime }\right) \in \Theta
\times \Theta \text{.}  \label{tutu1}
\end{equation}%
By reporting $\widehat{\theta }\in \mathcal{P}^{\ast }\left( \theta \right) $%
, agents $-j$ disclose that any state in $\mathcal{P}^{\ast }\left( \theta
\right) $ might be the true state, and hence, the whistle-blower must block
all of the false states in $\mathcal{P}^{\ast }\left( \theta \right) $
simultaneously.

One critical issue is how we should formalize "simultaneously" in (\ref%
{tutu1}). BMT adopt the following formalization, i.e., strict Maskin
monotonicity$^{\ast }$ (Definition \ref{def:monotonicity*}).%
\begin{equation*}
\mathcal{P}^{\ast }\left( \theta \right) \neq \mathcal{P}^{\ast }\left(
\theta ^{\prime }\right) \Longrightarrow \left( 
\begin{array}{c}
\exists j\in \mathcal{I}, \\ 
\left[ \dbigcap\limits_{\widehat{\theta }\in \mathcal{P}^{\ast }\left(
\theta \right) }\mathcal{L}_{j}^{\circ }\left( f\left( \widehat{\theta }%
\right) ,\widehat{\theta }\right) \right] \cap \mathcal{U}_{j}^{\circ
}\left( f\left( \theta \right) ,\theta ^{\prime }\right) \neq \varnothing%
\end{array}%
\right) \text{, }\forall \left( \theta ,\theta ^{\prime }\right) \in \Theta
\times \Theta \text{.}
\end{equation*}%
That is, at the true state $\theta ^{\prime }$, when all agents report $%
\widehat{\theta }\in \mathcal{P}^{\ast }\left( \theta \right) $, there must
exist a whistle-blower $j$ with a blocking plan $y\in \left[
\dbigcap\limits_{\widehat{\theta }\in \mathcal{P}^{\ast }\left( \theta
\right) }\mathcal{L}_{j}^{\circ }\left( f\left( \widehat{\theta }\right) ,%
\widehat{\theta }\right) \right] \cap \mathcal{U}_{j}^{\circ }\left( f\left(
\theta \right) ,\theta ^{\prime }\right) $, i.e., $y$ is credible at \emph{%
all states in} $\mathcal{P}^{\ast }\left( \theta \right) $, and is strictly
profitable at $\theta ^{\prime }$.

Specifically, BMT modify the canonical mechanism as follows. In case (1),
all agents\ report some states in $\mathcal{P}^{\ast }\left( \theta \right) $%
, the mechanism picks $f\left( \theta \right) $\textbf{.} In case (2), all
agents except agent $j$ report some states in $\mathcal{P}^{\ast }\left(
\theta \right) $, then we let agent $j$ choose any outcome $y\in \left[
\dbigcap\limits_{\widehat{\theta }\in \mathcal{P}^{\ast }\left( \theta
\right) }\mathcal{L}_{j}^{\circ }\left( f\left( \widehat{\theta }\right) ,%
\widehat{\theta }\right) \right] $ and any positive integer $n$, and the
mechanism picks $\frac{n-1}{n}\times y+\frac{1}{n}\times z_{i}\left( \theta
,\theta \right) \in \mathcal{L}_{i}\left( f\left( \theta \right) ,\theta
\right) $. The rest of the canonical mechanism remains the same.

BMT uses strict Maskin monotonicity$^{\ast }$ to eliminate "bad" solutions
in Case (1). And as above, all agents never have a best reply in Cases (2)
and (3), i.e., no "bad" solution.

\subsubsection{Our canonical mechanism}

We take a different formalization of "simultaneously" in (\ref{tutu1}) as
follows (i.e., strict Maskin monotonicity$^{\ast \ast }$ in Definition \ref%
{def:monotonicity**}).%
\begin{equation}
\mathcal{P}^{\ast }\left( \theta \right) \neq \mathcal{P}^{\ast }\left(
\theta ^{\prime }\right) \Longrightarrow \left( 
\begin{array}{c}
\exists j\in \mathcal{I}\text{, }\exists \phi :\Theta \rightarrow Y\text{,}
\\ 
\phi \left( \widehat{\theta }\right) \in \mathcal{L}_{j}^{\circ }\left(
f\left( \widehat{\theta }\right) ,\widehat{\theta }\right) \cap \mathcal{U}%
_{j}^{\circ }\left( f\left( \theta \right) ,\theta ^{\prime }\right) \neq
\varnothing \text{,} \\ 
\forall \widehat{\theta }\in \mathcal{P}^{\ast }\left( \theta \right) \text{,%
}%
\end{array}%
\right) \text{, }\forall \left( \theta ,\theta ^{\prime }\right) \in \Theta
\times \Theta \text{.}  \label{tutu2}
\end{equation}%
Our innovation is that we allow a whistle-blower to adopt a \emph{%
state-contingent blocking plan}, i.e., $\phi :\Theta \rightarrow Y$ in (\ref%
{tutu2}).

Specifically, we modify the canonical mechanism as follows. In case (1), all
agents\ report some states in $\mathcal{P}^{\ast }\left( \theta \right) $,
the mechanism picks $f\left( \theta \right) $\textbf{.} In case (2), all
agents except agent $j$ report some states in $\mathcal{P}^{\ast }\left(
\theta \right) $, then we let agent $j$ choose any $\phi :\Theta \rightarrow
Y$ such that $\phi \left( \widehat{\theta }\right) \in \mathcal{L}%
_{j}^{\circ }\left( f\left( \widehat{\theta }\right) ,\widehat{\theta }%
\right) $ for every $\widehat{\theta }\in \mathcal{P}^{\ast }\left( \theta
\right) $ and any positive integer $n$, and the mechanism picks $\frac{n-1}{n%
}\times \phi \left( \theta ^{j+1}\right) +\frac{1}{n}\times z_{i}\left(
\theta ^{j+1},\theta ^{j+1}\right) \in \mathcal{L}_{i}\left( f\left( \theta
^{j+1}\right) ,\theta ^{j+1}\right) $, where $\theta ^{j+1}$ denotes the
report of agent $\left( j+1\right) $ module $I$. The rest of the mechanism
remains the same.

We use strict Maskin monotonicity$^{\ast \ast }$ to eliminate "bad"
solutions in Case (1). And as above, all agents never have a best reply in
Cases (2) and (3), i.e., no "bad" solution.

\section{How to deal with violation of responsiveness?}

\label{sec:responsive}

In this section, we drop responsiveness, and fully characterize
rationalizable implementation when only NWA\ is assumed.

\subsection{A summary of the full characterization}

\label{sec:summary:responsive}

Let $\mathcal{P}_{f}$ denote the partition on $\Theta $ induced by $f$,
which is defined as follows.%
\begin{equation*}
\mathcal{P}_{f}\left( \theta \right) =\left\{ \theta ^{\prime }\in \Theta
:f\left( \theta ^{\prime }\right) =f\left( \theta \right) \right\} \text{, }%
\forall \theta \in \Theta \text{.}
\end{equation*}

Given NWA, BMT show that strict Maskin monotonicity$^{\ast }$ defined below
is sufficient for rationalizable implementation.

\begin{define}[strict Maskin monotonicity*]
\label{def:monotonicity*}An SCF $f:\Theta \longrightarrow Z$ satisfies
strict Maskin monotonicity$^{\ast }$ if there exists a partition $\mathcal{P}
$ on $\Theta $ finer than $\mathcal{P}_{f}$ such that for any $\left( \theta
,\theta ^{\prime }\right) \in \Theta \times \Theta $,%
\begin{equation*}
\theta ^{\prime }\in \mathcal{P}\left( \theta \right) \text{ }\Longleftarrow 
\text{\ }\left[ 
\begin{array}{c}
\forall \left( y,i\right) \in Y\times \mathcal{I}\text{,} \\ 
\left( 
\begin{array}{c}
u_{i}\left( f\left( \theta \right) ,\text{ }\widehat{\theta }\right)
>u_{i}\left( y,\text{ }\widehat{\theta }\right) \text{,} \\ 
\forall \widehat{\theta }\in \mathcal{P}\left( \theta \right)%
\end{array}%
\right) \Longrightarrow \text{ }u_{i}\left( f\left( \theta \right) ,\text{ }%
\theta ^{\prime }\right) \geq u_{i}\left( y,\text{ }\theta ^{\prime }\right)%
\end{array}%
\right] \text{,}
\end{equation*}

or, equivalently,%
\begin{equation}
\theta ^{\prime }\notin \mathcal{P}\left( \theta \right) \text{ }%
\Longrightarrow \text{ }\left[ 
\begin{array}{c}
\exists \left( y,i\right) \in Y\times \mathcal{I}\text{,} \\ 
\left( 
\begin{array}{c}
u_{i}\left( f\left( \theta \right) ,\text{ }\widehat{\theta }\right)
>u_{i}\left( y,\text{ }\widehat{\theta }\right) \text{,} \\ 
\forall \widehat{\theta }\in \mathcal{P}\left( \theta \right)%
\end{array}%
\right) \text{ and }u_{i}\left( y,\text{ }\theta ^{\prime }\right)
>u_{i}\left( f\left( \theta \right) ,\text{ }\theta ^{\prime }\right)%
\end{array}%
\right] \text{.}  \label{wbp1}
\end{equation}
\end{define}

We propose a new axiom which is weaker than strict Maskin monotonicity$%
^{\ast }$.

\begin{define}[strict Maskin monotonicity$^{\ast \ast }$]
\label{def:monotonicity**}An SCF $f:\Theta \longrightarrow Z$ satisfies
strict Maskin monotonicity$^{\ast \ast }$ if there exists a partition $%
\mathcal{P}$ on $\Theta $ finer than $\mathcal{P}_{f}$ such that for any $%
\left( \theta ,\theta ^{\prime }\right) \in \Theta \times \Theta $,%
\begin{equation}
\theta ^{\prime }\in \mathcal{P}\left( \theta \right) \text{ }\Longleftarrow 
\text{\ }\left[ 
\begin{array}{c}
\forall i\in \mathcal{I}\text{, \ }\exists \widehat{\theta }\in \mathcal{P}%
\left( \theta \right) \text{,}\forall y\in Y\text{,} \\ 
u_{i}\left( f\left( \theta \right) ,\text{ }\widehat{\theta }\right)
>u_{i}\left( y,\text{ }\widehat{\theta }\right) \Longrightarrow \text{ }%
u_{i}\left( f\left( \theta \right) ,\text{ }\theta ^{\prime }\right) \geq
u_{i}\left( y,\text{ }\theta ^{\prime }\right)%
\end{array}%
\right] \text{,}  \label{wbp3}
\end{equation}

or, equivalently,%
\begin{equation}
\theta ^{\prime }\notin \mathcal{P}\left( \theta \right) \text{ }%
\Longrightarrow \text{ }\left[ 
\begin{array}{c}
\exists i\in \mathcal{I}\text{ such that\ }\forall \widehat{\theta }\in 
\mathcal{P}\left( \theta \right) \text{, }\exists y^{\widehat{\theta }}\in Y%
\text{,} \\ 
u_{i}\left( f\left( \theta \right) ,\text{ }\widehat{\theta }\right)
>u_{i}\left( y^{\widehat{\theta }},\text{ }\widehat{\theta }\right) \text{
and }u_{i}\left( y^{\widehat{\theta }},\text{ }\theta ^{\prime }\right)
>u_{i}\left( f\left( \theta \right) ,\text{ }\theta ^{\prime }\right) \text{.%
}%
\end{array}%
\right] \text{.}  \label{wbp2}
\end{equation}
\end{define}

Conditions (\ref{wbp1}) and (\ref{wbp2}) speak out the difference between
the two axioms: when $\theta ^{\prime }\notin \mathcal{P}\left( \theta
\right) $ (i.e., the true state is $\theta ^{\prime }$ and all agents
falsely report states in $\mathcal{P}\left( \theta \right) $), strict Maskin
monotonicity$^{\ast }$ requires existence of a whistle-blower $i$ and a 
\emph{common} blocking plan $y$ which works for every states $\widehat{%
\theta }\in \mathcal{P}\left( \theta \right) $, while strict Maskin
monotonicity$^{\ast \ast }$ requires existence of a whistle-blower $i$ and a 
\emph{state-contingent} blocking plan $y^{\widehat{\theta }}$ which works
for each state $\widehat{\theta }\in \mathcal{P}\left( \theta \right) $.
Clearly, strict Maskin monotonicity$^{\ast }$ implies strict Maskin
monotonicity$^{\ast \ast }$, and the latter also suffices for rationalizable
implementation.

\begin{theo}
\label{theorem:non-responsive}Suppose that an SCF $f:\Theta \longrightarrow
Z $ satisfies NWA. Then, $f$\ is rationalizably implementable if and only if 
$f $ satisfies strict Maskin monotonicity$^{\ast \ast }$.
\end{theo}

The "only if" and "if" parts of Theorem \ref{theorem:non-responsive} are
proved in Sections \ref{sec:proof:responsive:necessity} and \ref%
{sec:proof:theorem:non-responsive}.

Given a condition called "the best-response property," BMT show that strict
Maskin monotonicity$^{\ast }$ is necessary for rationalizable
implementation. However, the best-response property is not defined on
primitives, and hence it remains an open question regarding whether it
suffers loss of generality to assume the best-response property? Example 4
in Appendix \ref{sec:the-best-reply} provides a negative answer for this
question. Specifically, NWA and strict Maskin monotonicity$^{\ast \ast }$
hold in this exampe, but strict Maskin monotonicity$^{\ast }$ fails. By
Theorem \ref{theorem:non-responsive}, we can achieve rationalizable
implementation in this example, even though Maskin monotonicity$^{\ast }$
doses not hold.

\subsection{The proof of the "only if" part of Theorem \protect\ref%
{theorem:non-responsive}}

\label{sec:proof:responsive:necessity}

Suppose that $f$ is rationalizably implemented by a mechanism $\mathcal{M}%
=\left\langle M\text{, \ }g:M\longrightarrow Y\right\rangle $. Consider the
partition defined as follows.%
\begin{equation*}
\mathcal{P}\left( \theta \right) =\left\{ \widetilde{\theta }\in \Theta :S^{%
\mathcal{M},\text{ }\widetilde{\theta }}=S^{\mathcal{M},\text{ }\theta
}\right\} \text{, }\forall \theta \in \Theta \text{.}
\end{equation*}%
Since $f$ is rationalizably implemented by $\mathcal{M}$, the partition $%
\mathcal{P}$ defined above is finer than $\mathcal{P}_{f}$. Furthermore, for
any $\left( \theta ,\theta ^{\prime }\right) \in \Theta \times \Theta $,
suppose that the right-hand side of (\ref{wbp3})\ holds and we aim to prove $%
\theta ^{\prime }\in \mathcal{P}\left( \theta \right) $ (i.e., strict Maskin
monotonicity$^{\ast \ast }$ holds). We need the following result, and its
proof is relegated to Appendix \ref{sec:lem:yt}.

\begin{lemma}
\label{lem:yt}If an SCF $f$ is rationalizably implemented by a mechanism $%
\mathcal{M}$, we have%
\begin{equation*}
S^{\mathcal{M},\text{ }\theta }\subset S^{\mathcal{M},\text{ }\theta
^{\prime }}\text{\ }\Longrightarrow \text{ }S^{\mathcal{M},\text{ }\theta
}=S^{\mathcal{M},\text{ }\theta ^{\prime }}\text{, }\forall \left( \theta
,\theta ^{\prime }\right) \in \Theta \times \Theta \text{.}
\end{equation*}
\end{lemma}

We will show that $S^{\mathcal{M},\text{ }\theta }$ satisfies the best-reply
property in $\mathcal{M}$ at state $\theta ^{\prime }$, i.e., $S^{\mathcal{M}%
,\text{ }\theta }\subset S^{\mathcal{M},\text{ }\theta ^{\prime }}$. By
Lemma \ref{lem:yt}, we have $S^{\mathcal{M},\text{ }\theta }=S^{\mathcal{M},%
\text{ }\theta ^{\prime }}$, and hence, $\theta ^{\prime }\in \mathcal{P}%
\left( \theta \right) $.

Consider any $i\in \mathcal{I}$, and pick any $m_{i}\in S_{i}^{\mathcal{M},%
\text{ }\theta }$. By the right-hand side of (\ref{wbp3}), there exists $%
\exists \widehat{\theta }\in \mathcal{P}\left( \theta \right) $ such that%
\begin{equation}
u_{i}\left( f\left( \theta \right) ,\text{ }\widehat{\theta }\right)
>u_{i}\left( y,\text{ }\widehat{\theta }\right) \Longrightarrow \text{ }%
u_{i}\left( f\left( \theta \right) ,\text{ }\theta ^{\prime }\right) \geq
u_{i}\left( y,\text{ }\theta ^{\prime }\right) \text{, }\forall y\in Y\text{.%
}  \label{cdda2}
\end{equation}%
Since $\widehat{\theta }\in \mathcal{P}\left( \theta \right) $, we have $%
m_{i}\in S_{i}^{\mathcal{M},\text{ }\theta }=S_{i}^{\mathcal{M},\text{ }%
\widehat{\theta }}$, and hence, there exists $\lambda _{-i}\in \triangle
\left( S_{-i}^{\mathcal{M},\text{ }\widehat{\theta }}\right) $ such that $%
m_{i}$ is a best reply to $\lambda _{-i}$ for agent $i$ at State $\widehat{%
\theta }$, i.e.,%
\begin{equation*}
u_{i}\left( g\left( m_{i},\lambda _{-i}\right) ,\text{ }\widehat{\theta }%
\right) =u_{i}\left( f\left( \widehat{\theta }\right) ,\text{ }\widehat{%
\theta }\right) =u_{i}\left( f\left( \theta \right) ,\text{ }\widehat{\theta 
}\right) \geq u_{i}\left( g\left( \widetilde{m}_{i},\lambda _{-i}\right) ,%
\text{ }\widehat{\theta }\right) \text{, }\forall \widetilde{m}_{i}\in M_{i}%
\text{,}
\end{equation*}%
which further implies%
\begin{equation}
u_{i}\left( g\left( m_{i},\lambda _{-i}\right) ,\text{ }\widehat{\theta }%
\right) =u_{i}\left( f\left( \theta \right) ,\text{ }\widehat{\theta }%
\right) >u_{i}\left( g\left( \widetilde{m}_{i},\lambda _{-i}\right) ,\text{ }%
\widehat{\theta }\right) \text{, }\forall \widetilde{m}_{i}\in
M_{i}\diagdown S_{i}^{\mathcal{M},\text{ }\widehat{\theta }}\text{,}
\label{cdda4}
\end{equation}%
\begin{equation}
\text{and }g\left( m_{i},\lambda _{-i}\right) =g\left( \widetilde{m}%
_{i},\lambda _{-i}\right) =f\left( \widehat{\theta }\right) \text{, }\forall 
\widetilde{m}_{i}\in S_{i}^{\mathcal{M},\text{ }\widehat{\theta }}\text{.}
\label{cdda5}
\end{equation}%
Then, (\ref{cdda2}) and (\ref{cdda4}) imply%
\begin{equation}
u_{i}\left( g\left( m_{i},\lambda _{-i}\right) ,\text{ }\theta ^{\prime
}\right) =u_{i}\left( f\left( \theta \right) ,\text{ }\theta ^{\prime
}\right) =u_{i}\left( f\left( \widehat{\theta }\right) ,\text{ }\theta
^{\prime }\right) \geq u_{i}\left( g\left( \widetilde{m}_{i},\lambda
_{-i}\right) ,\text{ }\theta ^{\prime }\right) \text{, }\forall \widetilde{m}%
_{i}\in M_{i}\diagdown S_{i}^{\mathcal{M},\text{ }\widehat{\theta }}\text{.}
\label{cdda6}
\end{equation}%
Finally, (\ref{cdda5}) and (\ref{cdda6}) imply%
\begin{equation*}
u_{i}\left( g\left( m_{i},\lambda _{-i}\right) ,\text{ }\theta ^{\prime
}\right) =u_{i}\left( f\left( \widehat{\theta }\right) ,\text{ }\theta
^{\prime }\right) \geq u_{i}\left( g\left( \widetilde{m}_{i},\lambda
_{-i}\right) ,\text{ }\theta ^{\prime }\right) \text{, }\forall \widetilde{m}%
_{i}\in M_{i}\text{,}
\end{equation*}%
i.e., $m_{i}$ is a best reply to $\lambda _{-i}$ for $i$ at $\theta ^{\prime
}$. Thus, $S^{\mathcal{M},\text{ }\theta }$ satisfies the best reply
property at $\theta ^{\prime }$.$\blacksquare $

\section{How to deal with violation of NWA?}

\label{sec:NWA}

In this section, we drop NWA, and fully characterize rationalizable
implementation when only responsiveness\ is assumed.

\subsection{A summary of the full characterization}

\label{sec:summary:NWA}

We fully characterize rationalizable implementation by two new axioms.
First, we propose an axiom called "strict event monotonicity," which
strengthens strict Maskin monotonicity (Definition \ref{def:strict-Maskin}),
and when NWA holds, the two notions coincide.

\begin{define}[strict event monotonicity]
\label{def:strict-group-monotonicity}An SCF $f:\Theta \longrightarrow Z$
satisfies strict event monotonicity if for every $\left( \theta ^{\prime
},E\right) \in \Theta \times \left[ 2^{\Theta }\backslash \left\{
\varnothing \right\} \right] :$

$\left\{ f\left( \theta ^{\prime }\right) \right\} =f\left( E\right) $
whenever%
\begin{equation}
u_{i}\left( f\left( \theta \right) ,\text{ }\theta \right) >u_{i}\left( y,%
\text{ }\theta \right) \text{ }\Longrightarrow \text{ }u_{i}\left( f\left(
\theta \right) ,\text{ }\theta ^{\prime }\right) \geq u_{i}\left( y,\text{ }%
\theta ^{\prime }\right) \text{, }\forall \left( \theta ,y,i,\right) \in
E\times Y\times \mathcal{I}^{E}\text{,}  \label{kf11}
\end{equation}%
or, equivalently,

$\left\{ f\left( \theta ^{\prime }\right) \right\} \neq f\left( E\right) $
implies%
\begin{equation}
u_{i}\left( f\left( \theta \right) ,\text{ }\theta \right) >u_{i}\left( y,%
\text{ }\theta \right) \text{ and }u_{i}\left( y,\text{ }\theta ^{\prime
}\right) >u_{i}\left( f\left( \theta \right) ,\text{ }\theta ^{\prime
}\right) \text{, for some }\left( \theta ,y,i,\right) \in E\times Y\times 
\mathcal{I}^{E}\text{.}  \label{tat1}
\end{equation}
\end{define}

There are two subtle differences between strict event monotonicity and
strict Maskin monotonicity. First, pairwise comparison between states (i.e.,
"$\theta ^{\prime }$ $Vs$ $\theta $") is conducted in strict Maskin
monotonicity, while a state is compared to a group of states (i.e., "$\theta
^{\prime }$ $Vs$ $E$") in strict event monotonicity. Second, as shown in
conditions (\ref{kf23}) and (\ref{tat1}), the whistle-blower is required to
be an active agent in $\mathcal{I}^{E}$ in strict event monotonicity, while
he or she could be anyone in $\mathcal{I}$ in strict Maskin monotonicity. It
is straightforward to show that, given NWA, strict event monotonicity is
equivalent to strict Maskin monotonicity.

\begin{define}[dictator monotonicity]
\label{def:dictator}Agent $i\in \mathcal{I}$ is a dictator if $\left\{
i\right\} =\mathcal{I}^{\theta }$. An SCF $f:\Theta \longrightarrow Z$
satisfies dictator monotonicity if for every $\left( i,\theta ,\theta
^{\prime },\theta ^{\prime \prime }\right) \in \mathcal{I}\times \Theta
\times \Theta \times \Theta $,%
\begin{equation}
\left[ 
\begin{array}{c}
\left\{ i\right\} =\mathcal{I}^{\theta }\text{ } \\ 
\text{and }f\left( \theta \right) \neq f\left( \theta ^{\prime }\right)%
\end{array}%
\right] \Longrightarrow \left[ 
\begin{array}{c}
\exists y\in Y\text{ such that} \\ 
u_{i}\left( f\left( \theta ^{\prime \prime }\right) ,\text{ }\theta ^{\prime
\prime }\right) \geq u_{i}\left( y,\text{ }\theta ^{\prime \prime }\right) 
\text{ and }u_{i}\left( y,\text{ }\theta ^{\prime }\right) >u_{i}\left(
f\left( \theta \right) ,\text{ }\theta ^{\prime }\right)%
\end{array}%
\right] \text{.}  \label{ttaa1}
\end{equation}
\end{define}

\begin{theo}
\label{theorem:NWA}A responsive SCF $f:\Theta \longrightarrow Z$\ is
rationalizably implementable if and only if $f$ satisfies strict event
monotonicity and dictator monotonicity.
\end{theo}

We provide an intuition of Theorem \ref{theorem:NWA} in Sections \ref%
{sec:intuition:NWA:dictator} and \ref{sec:intuition:NWA:group}, and the
proofs are presented in Sections \ref{sec:proof:NWA:dictator}, \ref%
{sec:proof:NWA:strict-group} and \ref{sec:proof:theorem:NWA}.

\subsection{A crucial observation}

\label{sec:intuition:NWA}

We first offer a crucial observation, which provides a powerful tool in
establishing both the necessity and the sufficiency parts of Theorem \ref%
{theorem:NWA}.

\begin{lemma}
\label{lem:dominated}Suppose that a social choice function $f:\Theta
\longrightarrow Z$ is rationalizably implemented by a mechanism $\mathcal{M}%
=\left\langle M\text{, \ }g:M\longrightarrow Y\right\rangle $. Then, for
every $\left( i,\theta \right) \in \mathcal{I}\times \Theta $,%
\begin{equation}
i\notin \mathcal{I}^{\theta }\text{ }\Longrightarrow \left[ 
\begin{array}{c}
\text{ }S_{i}^{\mathcal{M},\text{ }\theta }=M_{i}\text{ and} \\ 
g\left( m_{i},m_{-i}\right) =f\left( \theta \right) \text{, }\forall \left(
m_{i},m_{-i}\right) \in M_{i}\times S_{-i}^{\mathcal{M},\text{ }\theta }%
\text{.}%
\end{array}%
\right] \text{.}  \label{aaa2}
\end{equation}
\end{lemma}

Given $f$ being rationalizably implemented by $\mathcal{M}$, Lemma \ref%
{lem:dominated} says that at any state, all strategies are rationalizable
for an inactive agent. The proof is straightforward: given $i\notin \mathcal{%
I}^{\theta }$, pick any $m_{i}\in $ $S_{i}^{\mathcal{M},\text{ }\theta }$,
and there exists $\lambda _{-i}\in \triangle \left( S_{-i}^{\mathcal{M},%
\text{ }\theta }\right) $ such that $m_{i}$ is a best reply to $\lambda
_{-i} $ and $g\left( m_{i},\lambda _{-i}\right) =f\left( \theta \right) $.
Given $i\notin \mathcal{I}^{\theta }$, $f\left( \theta \right) $ is a worst
outcome for $i$ at $\theta $, and hence, any other strategy is a best reply
to $\lambda _{-i}$. Therefore, $S_{i}^{\mathcal{M},\text{ }\theta }=M_{i}$.

Lemma \ref{lem:dominated} sheds light on the canonical mechanism that
rationalizably implements $f$: at true state $\theta $, we should let active
agents in $\mathcal{I}^{\theta }$ \emph{only} to determine the outcome of
the mechanism, and ignore agents in $\mathcal{I}\backslash \mathcal{I}%
^{\theta }$, because they are not informative.

\subsection{The meaning of dictator monotonicity}

\label{sec:intuition:NWA:dictator}

\subsubsection{The sufficiency part of dictator monotonicity: intuition}

We first show why we need dictator monotonicity when we prove the
sufficiency part of Theorem \ref{theorem:NWA}. If agent $i$ is a dictator at 
$\theta $ and $i$ reports $\theta $ in the canonical mechanism, by Lemma \ref%
{lem:dominated}, we should trust $i$ and ignore other agents' reports, and
pick $f\left( \theta \right) $. In particular, consider the following
scenario: at true state $\theta ^{\prime }$, we aim to implements $f\left(
\theta ^{\prime }\right) $, but agent $i$ with $\left\{ i\right\} =\mathcal{I%
}^{\theta }$ reports $\theta $ with $f\left( \theta \right) \neq f\left(
\theta ^{\prime }\right) $, while all the other agents report $\theta
^{\prime \prime }$---in this scenario, $f\left( \theta \right) $ "should" be
chosen by Lemma \ref{lem:dominated}. Since $f\left( \theta ^{\prime }\right)
\neq f\left( \theta \right) $, a whistle-blower must exist to block this
false reporting, but who should this whistle-blower be? Recall Lemma \ref%
{lem:dominated}, which says that we should ignore all the other agents'
reports, when $i$ is a dictator at $\theta $ and reports $\theta $. As a
result, the only possible whistle-blower must be agent $i$. To ensure that
the whistle-blower $i$ has a blocking plan, we need%
\begin{equation}
\left[ 
\begin{array}{c}
\exists y\in Y\text{ such that} \\ 
u_{i}\left( f\left( \theta ^{\prime \prime }\right) ,\text{ }\theta ^{\prime
\prime }\right) \geq u_{i}\left( y,\text{ }\theta ^{\prime \prime }\right) 
\text{ and }u_{i}\left( y,\text{ }\theta ^{\prime }\right) >u_{i}\left(
f\left( \theta \right) ,\text{ }\theta ^{\prime }\right)%
\end{array}%
\right] \text{.}  \label{ttef2}
\end{equation}%
That is, this legitimate blocking plan for $i$ must satisfy two conditions.
First, given agents $-i$ reporting $\theta ^{\prime \prime }$, the blocking
plan $y$ must be credible,\footnote{%
Agent $i$ uses "$y$" to report that agents $-i$ are lying by reporting $%
\theta ^{\prime \prime }$. It is credible because if the true state were $%
\theta ^{\prime \prime }$, $y$ would be inferior to $f\left( \theta ^{\prime
\prime }\right) $ for agent $i$.}%
\begin{equation}
u_{i}\left( f\left( \theta ^{\prime \prime }\right) ,\text{ }\theta ^{\prime
\prime }\right) \geq u_{i}\left( y,\text{ }\theta ^{\prime \prime }\right) 
\text{.}  \label{ttyu1}
\end{equation}%
Second, the blocking plan must be strictly profitable at $\theta ^{\prime }$%
, i.e., 
\begin{equation}
u_{i}\left( y,\text{ }\theta ^{\prime }\right) >u_{i}\left( f\left( \theta
\right) ,\text{ }\theta ^{\prime }\right) \text{.}  \label{ttyu2}
\end{equation}%
(\ref{ttyu1}) and (\ref{ttyu2}) imply (\ref{ttef2}), i.e., the dictator
monotonicity in Definition \ref{def:dictator} (precisely, (\ref{ttaa1})).

\subsubsection{The necessity part of dictator monotonicity: proof}

\label{sec:proof:NWA:dictator}

To prove the necessity of dictator monotonicity, we show a contrapositive
statement of (\ref{ttaa1}): for every for every $\left( i,\theta ,\theta
^{\prime },\theta ^{\prime \prime }\right) \in \mathcal{I}\times \Theta
\times \Theta \times \Theta $,

\begin{equation}
\left[ 
\begin{array}{c}
\left\{ i\right\} =\mathcal{I}^{\theta }\text{ and} \\ 
\left( 
\begin{array}{c}
\forall y\in Y\text{,} \\ 
u_{i}\left( f\left( \theta ^{\prime \prime }\right) ,\text{ }\theta ^{\prime
\prime }\right) \geq u_{i}\left( y,\text{ }\theta ^{\prime \prime }\right) 
\text{ $\Longrightarrow $ }u_{i}\left( f\left( \theta \right) ,\text{ }%
\theta ^{\prime }\right) \geq u_{i}\left( y,\text{ }\theta ^{\prime }\right)%
\end{array}%
\right)%
\end{array}%
\right] \Longrightarrow f\left( \theta \right) =f\left( \theta ^{\prime
}\right) \text{.}  \label{aaa2b}
\end{equation}

Suppose that $f$ is rationalizably implemented by a mechanism $\mathcal{M}%
=\left\langle M\text{, \ }g:M\longrightarrow Y\right\rangle $, and that the
left-hand-side of (\ref{aaa2b}) holds. We aim to show $f\left( \theta
\right) =f\left( \theta ^{\prime }\right) $.

Pick any $\left( m_{i},m_{i}^{\prime \prime }\right) \in S_{i}^{\mathcal{M},%
\text{ }\theta }\times S_{i}^{\mathcal{M},\text{ }\theta ^{\prime \prime }}$%
, and there exists $\lambda _{-i}^{\prime \prime }\in \triangle \left(
S_{-i}^{\mathcal{M},\text{ }\theta ^{\prime \prime }}\right) $ such that $%
m_{i}^{\prime \prime }$ is a best reply to $\lambda _{-i}^{\prime \prime }$
for $i$ at $\theta ^{\prime \prime }$, i.e.,%
\begin{equation}
u_{i}\left( g\left( m_{i}^{\prime \prime },\lambda _{-i}^{\prime \prime
}\right) ,\text{ }\theta ^{\prime \prime }\right) =u_{i}\left( f\left(
\theta ^{\prime \prime }\right) ,\text{ }\theta ^{\prime \prime }\right)
\geq u_{i}\left( g\left( m_{i},\lambda _{-i}^{\prime \prime }\right) ,\text{ 
}\theta ^{\prime \prime }\right) \text{, }\forall m_{i}\in M_{i}\text{.}
\label{aaa2e}
\end{equation}%
By $\left\{ i\right\} =\mathcal{I}^{\theta }$ and Lemma \ref{lem:dominated},
we have%
\begin{equation}
g\left( m_{i},m_{-i}\right) =f\left( \theta \right) =g\left( m_{i},\lambda
_{-i}^{\prime \prime }\right) \text{, }\forall m_{-i}\in M_{-i}\text{,}
\label{aab1}
\end{equation}%
which, together with (\ref{aaa2e}) and the left-hand side (\ref{aaa2b}),
implies%
\begin{equation}
u_{i}\left( g\left( m_{i},\lambda _{-i}^{\prime \prime }\right) ,\text{ }%
\theta ^{\prime }\right) =u_{i}\left( f\left( \theta \right) ,\text{ }\theta
^{\prime }\right) \geq u_{i}\left( g\left( m_{i},\lambda _{-i}^{\prime
\prime }\right) ,\text{ }\theta ^{\prime }\right) \text{, }\forall m_{i}\in
M_{i}\text{.}  \label{aab2}
\end{equation}%
For $\left( m_{i},\lambda _{-i}^{\prime \prime }\right) $, (\ref{aab2})
shows that $i$ does not have a profitable deviation at $\theta ^{\prime }$,
and (\ref{aab1}) shows that agents $-i$ do not have a profitable deviation
at $\theta ^{\prime }$. Therefore, $\left( m_{i},\lambda _{-i}^{\prime
\prime }\right) $ is a Nash equilibrium at $\theta ^{\prime }$, which
induces $f\left( \theta \right) $. Therefore, $f\left( \theta \right)
=f\left( \theta ^{\prime }\right) $.$\blacksquare $

\subsection{The meaning of strict event monotonicity}

\label{sec:intuition:NWA:group}

\subsubsection{The sufficiency part of strict event monotonicity: intuition}

To illustrate strict event monotonicity, we consider an alternative and
equivalent notion.

\begin{define}[strict iterated-elimination monotonicity]
An SCF $f:\Theta \longrightarrow Z$ satisfies strict iterated-elimination
monotonicity if for every $\theta ^{\prime }\in \Theta $, there exists $%
\left( \theta ^{1},\theta ^{2},...,\theta ^{n}\right) $ such that%
\begin{gather*}
\left\{ \theta ^{1},\theta ^{2},...,\theta ^{n}\right\} =\Theta \text{,} \\
\theta ^{n}=\theta ^{\prime }\text{,}
\end{gather*}%
and for every $k\in \left\{ 1,2,...,n-1\right\} $, 
\begin{equation}
u_{i}\left( f\left( \theta ^{k}\right) ,\text{ }\theta ^{k}\right)
>u_{i}\left( y,\text{ }\theta ^{k}\right) \text{ and }u_{i}\left( y,\text{ }%
\theta ^{\prime }\right) >u_{i}\left( f\left( \theta ^{k}\right) ,\text{ }%
\theta ^{\prime }\right) \text{, for some }\left( y,i\right) \in Y\times 
\mathcal{I}^{\left\{ \theta ^{k},\theta ^{k+1},...,\theta ^{n}\right\} }%
\text{.}  \label{kf33}
\end{equation}
\end{define}

\begin{prop}
\label{prop:NWA}A responsive SCF $f:\Theta \longrightarrow Z$ satisfies
strict event monotonicity if and only if $f$ satisfies strict
iterated-elimination monotonicity.
\end{prop}

The proof of Proposition \ref{prop:NWA} is relegated to Appendix \ref%
{sec:proof:prop:NWA}. To see the intuition of the "if" part of Theorem \ref%
{theorem:NWA}, we first recall the canonical mechanism in Sections \ref%
{sec:illustration:NWA}. In this mechanism, we invite agents to report the
true state, and there are three cases: (1) agreement, (2) unilateral
deviation, and (3) multi-lateral deviation. A distinct feature of this
mechanism is that agents do not have a best reply when Cases (2) and (3) are
triggered. As a result, a strategy can be rationalized \emph{only in Case (1)%
}. We now show that "truthful report" is the only rationalizable strategy in
this mechanism. Suppose the true state is $\theta ^{\prime }$. We start a
iterative process of deletion with $\Theta =\left\{ \theta ^{1},\theta
^{2},...,\theta ^{n}\right\} $. First, suppose all agents report $\theta
^{1} $. By (\ref{kf33}) in Proposition \ref{prop:NWA}, a whistle-blower $%
i\in \mathcal{I}^{\Theta }$ finds it strictly profitable to deviate to Case
(2), and pick $y\in \mathcal{L}_{i}^{\circ }\left( f\left( \theta
^{1}\right) ,\theta \right) \cap \mathcal{U}_{i}^{\circ }\left( f\left(
\theta ^{1}\right) ,\theta ^{\prime }\right) $. Thus, reporting $\theta ^{1}$
is not rationalizable for $i$, and as a result, $\theta ^{1}$ can be deleted
from the rationalizable set of every agent in $\mathcal{I}^{\Theta }$.%
\footnote{%
Given $\theta ^{1}$ being not rationalizable for agent $i$, agents in $%
\mathcal{I}^{\Theta }\diagdown \left\{ i\right\} $ can rationalize
"reporting $\theta ^{1}$" only in Cases (2) and (3), in which a best reply
does not exist.} Second, with $\Theta ^{\prime }=\left\{ \theta ^{2},\theta
^{3},...,\theta ^{n}\right\} $, suppose all agents report $\theta ^{2}$.
Similarly, by Proposition \ref{prop:NWA}, reporting $\theta ^{2}$ is not
rationalizable for some whistle-blower $i^{\prime }\in \mathcal{I}^{\Theta
^{\prime }}$, and hence not rationalizable for all agents in $\mathcal{I}%
^{\Theta ^{\prime }}$....we continue this iterative process of deletion
until we delete $\theta ^{n-1}$. As a result, only reporting $\theta
^{n}=\theta ^{\prime }$ is rationalizable for agents in $\mathcal{I}%
^{\left\{ \theta ^{\prime }\right\} }$, which induces $f\left( \theta
^{\prime }\right) $ at $\theta ^{\prime }$, i.e., we achieve rationalizable
implementation.

\subsubsection{The necessity part of strict event monotonicity: proof}

\label{sec:proof:NWA:strict-group}

Suppose that $f$ is rationalizably implemented by a mechanism $\mathcal{M}%
=\left\langle M\text{, \ }g:M\longrightarrow Y\right\rangle $, and that (\ref%
{kf11}) holds for some $\left( \theta ^{\prime },E\right) \in \Theta \times %
\left[ 2^{\Theta }\backslash \left\{ \varnothing \right\} \right] $, i.e.,%
\begin{equation}
u_{i}\left( f\left( \theta \right) ,\text{ }\theta \right) >u_{i}\left( y,%
\text{ }\theta \right) \text{ }\Longrightarrow \text{ }u_{i}\left( f\left(
\theta \right) ,\text{ }\theta ^{\prime }\right) \geq u_{i}\left( y,\text{ }%
\theta ^{\prime }\right) \text{, }\forall \left( \theta ,y,i,\right) \in
E\times Y\times \mathcal{I}^{E}\text{,}  \label{bbb1}
\end{equation}%
We aim to show $\left\{ f\left( \theta ^{\prime }\right) \right\} =f\left(
E\right) $, which establishes strict event monotonicity. Consider%
\begin{equation*}
S^{\mathcal{M},\text{ }E}\equiv \left( S_{i}^{\mathcal{M},\text{ }E}\equiv
\left( \dbigcup_{\theta \in E}S_{i}^{\mathcal{M},\text{ }\theta }\right)
\right) _{i\in \mathcal{I}}\text{.}
\end{equation*}%
We will show that $S^{\mathcal{M},\text{ }E}$ satisfies the best-reply
property in $\mathcal{M}$ at state $\theta ^{\prime }$, which further
implies $S^{\mathcal{M},\text{ }E}\subset S^{\mathcal{M},\text{ }\theta
^{\prime }\text{ }}$, and hence, $\left\{ f\left( \theta ^{\prime }\right)
\right\} =f\left( E\right) $.

First, consider any $i\notin \mathcal{I}^{E}$, i.e., $i\notin \mathcal{I}%
^{\theta }$ for some $\theta \in E$. By Lemma \ref{lem:dominated}, we have $%
S_{i}^{\mathcal{M},\text{ }\theta }=M_{i}$, and hence, 
\begin{equation*}
M_{i}=S_{i}^{\mathcal{M},\text{ }\theta }\subset S_{i}^{\mathcal{M},\text{ }%
E}\subset M_{i}\text{,}
\end{equation*}%
i.e., $S_{i}^{\mathcal{M},\text{ }E}=M_{i}$. Pick any $\widehat{m}_{-i}\in
S_{-i}^{\mathcal{M},\text{ }\theta }$, Lemma \ref{lem:dominated} implies%
\begin{equation*}
g\left( m_{i},\widehat{m}_{-i}\right) =f\left( \theta \right) \text{, }%
\forall m_{i}\in M_{i}\text{,}
\end{equation*}%
i.e., every $m_{i}\in M_{i}=S_{i}^{\mathcal{M},\text{ }E}$ is a best reply
to $\widehat{m}_{-i}\in S_{-i}^{\mathcal{M},\text{ }\theta }\subset S_{-i}^{%
\mathcal{M},\text{ }E}$ for agent $i$ at state $\theta ^{\prime }$.

Second, consider any $i\in \mathcal{I}^{E}$. Pick any $\theta \in E$ and any 
$m_{i}\in S_{i}^{\mathcal{M},\text{ }\theta }$, and we will show that $m_{i}$
is a best reply for agent $i$ at state $\theta ^{\prime }$ to some $\lambda
_{-i}\in \triangle \left( S_{-i}^{\mathcal{M},\text{ }E}\right) $, which
would establish the best-reply property of $S^{\mathcal{M},\text{ }E}$ at
state $\theta ^{\prime }$.

Since $m_{i}\in S_{i}^{\mathcal{M},\text{ }\theta }$, there exists $%
\widetilde{\lambda }_{-i}\in \triangle \left( S_{-i}^{\mathcal{M},\text{ }%
\theta }\right) $ such that $m_{i}$ is a best reply to $\widetilde{\lambda }%
_{-i}$ for $i$ at $\theta $, i.e.,%
\begin{equation*}
u_{i}\left( g\left( m_{i},\widetilde{\lambda }_{-i}\right) ,\text{ }\theta
\right) =u_{i}\left( f\left( \theta \right) ,\text{ }\theta \right) \geq
u_{i}\left( g\left( \overline{m}_{i},\widetilde{\lambda }_{-i}\right) ,\text{
}\theta \right) \text{, }\forall \overline{m}_{i}\in M_{i}\text{,}
\end{equation*}%
and more precisely,%
\begin{equation}
g\left( m_{i},\widetilde{\lambda }_{-i}\right) =f\left( \theta \right)
=g\left( \overline{m}_{i},\widetilde{\lambda }_{-i}\right) \text{, }\forall 
\overline{m}_{i}\in S_{i}^{\mathcal{M},\text{ }\theta }\text{,}  \label{thj1}
\end{equation}%
\begin{equation}
u_{i}\left( g\left( m_{i},\widetilde{\lambda }_{-i}\right) ,\text{ }\theta
\right) =u_{i}\left( f\left( \theta \right) ,\text{ }\theta \right)
>u_{i}\left( g\left( \overline{m}_{i},\widetilde{\lambda }_{-i}\right) ,%
\text{ }\theta \right) \text{, }\forall \overline{m}_{i}\in M_{i}\diagdown
S_{i}^{\mathcal{M},\text{ }\theta }\text{.}  \label{thj2}
\end{equation}%
Thus, (\ref{thj1}) implies%
\begin{equation}
u_{i}\left( g\left( m_{i},\widetilde{\lambda }_{-i}\right) ,\text{ }\theta
^{\prime }\right) =u_{i}\left( f\left( \theta \right) ,\text{ }\theta
^{\prime }\right) =u_{i}\left( g\left( \overline{m}_{i},\widetilde{\lambda }%
_{-i}\right) ,\text{ }\theta ^{\prime }\right) \text{, }\forall \overline{m}%
_{i}\in S_{i}^{\mathcal{M},\text{ }\theta }\text{,}  \label{thj3}
\end{equation}%
and (\ref{bbb1}) and (\ref{thj2}) imply%
\begin{equation}
u_{i}\left( g\left( m_{i},\widetilde{\lambda }_{-i}\right) ,\text{ }\theta
^{\prime }\right) =u_{i}\left( f\left( \theta \right) ,\text{ }\theta
^{\prime }\right) \geq u_{i}\left( g\left( \overline{m}_{i},\widetilde{%
\lambda }_{-i}\right) ,\text{ }\theta ^{\prime }\right) \text{, }\forall 
\overline{m}_{i}\in M_{i}\diagdown S_{i}^{\mathcal{M},\text{ }\theta }\text{.%
}  \label{thj4}
\end{equation}%
(\ref{thj3}) and (\ref{thj4}) imply $m_{i}$ is a best reply to $\widetilde{%
\lambda }_{-i}\in \triangle \left( S_{-i}^{\mathcal{M},\text{ }\theta
}\right) \subset \triangle \left( S_{-i}^{\mathcal{M},\text{ }E}\right) $
for $i$ at $\theta ^{\prime }$.$\blacksquare $

\section{A full characterization of rationalizable implementation}

\label{sec:no-condition}

In this section, we combine the techniques developed in Sections \ref%
{sec:responsive} and \ref{sec:NWA}, and fully characterize rationalizable
implementation when neither NWA nor responsiveness is imposed.

From the analysis in\ Section \ref{sec:responsive}, we learn that a
necessary and sufficient condition requires existence of a partition $%
\mathcal{P}$ finer than $\mathcal{P}_{f}$ such that for any two states, $%
\theta $ and $\theta ^{\prime }$, with $\mathcal{P}\left( \theta \right)
\neq \mathcal{P}\left( \theta ^{\prime }\right) $, there must exist a
whistle-blower $i$, who, at true state $\theta ^{\prime }$, can always block
any false state $\widehat{\theta }\in \mathcal{P}\left( \theta \right) $
reported by agent $-i$. Furthermore, from the analysis in\ Section \ref%
{sec:NWA}, we learn that three additional modification should be imposed:
(1) we should compare $E\left( \subset \Theta \right) $ with $\theta
^{\prime }$ (rather than "$\theta $ Vs $\theta ^{\prime }$" ), and (2) the
whistle-blower must be active at any state in $E$, and (3)\ dictator
monotonicity holds.

\begin{define}[strict event monotonicity$^{\ast \ast }$]
\label{def:strict-group-monotonicity**}A SCF $f:\Theta \longrightarrow Y$
satisfies strict event monotonicity$^{\ast \ast }$ if there exists a
partition $\mathcal{P}$ of $\Theta $ finer than $\mathcal{P}_{f}$ such that
the following two conditions hold.

\begin{enumerate}
\item (strict event monotonicity) for every $\left( \theta ^{\prime
},E\right) \in \Theta \times \left[ 2^{\Theta }\backslash \left\{
\varnothing \right\} \right] $,%
\begin{equation*}
\mathcal{P}\left( \theta ^{\prime }\right) =\dbigcup\limits_{\theta \in E}%
\mathcal{P}\left( \theta \right) \Longleftarrow \left[ 
\begin{array}{c}
\forall \left( \theta ,i\right) \in E\times \mathcal{I}^{\left[ \cup
_{\theta \in E}\mathcal{P}\left( \theta \right) \right] }\text{, \ }\exists 
\widehat{\theta }\in \mathcal{P}\left( \theta \right) \text{,}\forall y\in Y%
\text{,} \\ 
u_{i}\left( f\left( \theta \right) ,\text{ }\widehat{\theta }\right)
>u_{i}\left( y,\text{ }\widehat{\theta }\right) \Longrightarrow \text{ }%
u_{i}\left( f\left( \theta \right) ,\text{ }\theta ^{\prime }\right) \geq
u_{i}\left( y,\text{ }\theta ^{\prime }\right)%
\end{array}%
\right] \text{,}
\end{equation*}

\item (dictator monotonicity) for every $\left( i,\theta ,\theta ^{\prime
},\theta ^{\prime \prime }\right) \in \mathcal{I}\times \Theta \times \Theta
\times \Theta $, 
\begin{equation*}
\left[ 
\begin{array}{c}
\left\{ i\right\} =\mathcal{I}^{\mathcal{P}\left( \theta \right) }\text{ }
\\ 
\text{and }\mathcal{P}\left( \theta \right) \neq \mathcal{P}\left( \theta
^{\prime }\right)%
\end{array}%
\right] \Longrightarrow \left[ 
\begin{array}{c}
\exists y\in Y\text{ such that} \\ 
u_{i}\left( f\left( \theta ^{\prime \prime }\right) ,\text{ }\theta ^{\prime
\prime }\right) \geq u_{i}\left( y,\text{ }\theta ^{\prime \prime }\right) 
\text{ and }u_{i}\left( y,\text{ }\theta ^{\prime }\right) >u_{i}\left(
f\left( \theta \right) ,\text{ }\theta ^{\prime }\right)%
\end{array}%
\right] \text{.}
\end{equation*}
\end{enumerate}
\end{define}

\bigskip Two points are worth noting. First, strict event monotonicity$%
^{\ast \ast }$ combines strict event monotonicity and dictator monotonicity.
Without responsiveness, all axioms must be based on a \emph{common}
partition on $\Theta $. Because of this, we have to write both strict event
monotonicity and dictator monotonicity into one single axiom, which is based
on a common partition on $\Theta $. With abuse of notation, we call this new
axiom, strict event monotonicity$^{\ast \ast }$.

Second, $\mathcal{P}\left( \theta \right) $ represents the set of states
which are indistinguishable from $\theta $ (regarding players'
rationalizable strategies in canonical mechanisms). Parts 1 and 2 of
Definition \ref{def:strict-group-monotonicity**} are simply the
corresponding versions of Definitions \ref{def:strict-group-monotonicity}
and \ref{def:dictator}, respectively, incorporating this idea of equivalent
class of states (induced by the partition $\mathcal{P}$).

\begin{theo}
\label{theorem:SCF:full}An SCF $f$\ is rationalizably implementable if and
only if $f$ satisfies strict event monotonicity$^{\ast \ast }$.
\end{theo}

The proof of Theorem \ref{theorem:SCF:full} is similar to those of Theorems %
\ref{theorem:non-responsive} and \ref{theorem:NWA}, and we omit it.

\appendix

\section{Proofs}

\subsection{Proof of Lemma \protect\ref{lem:yt}}

\label{sec:lem:yt}

Suppose that $f$ is rationalizably implemented by a mechanism $\mathcal{M}%
=\left\langle M\text{, \ }g:M\longrightarrow Y\right\rangle $ and that $S^{%
\mathcal{M},\text{ }\theta }\subset S^{\mathcal{M},\text{ }\theta ^{\prime
}} $. We aim to show $S^{\mathcal{M},\text{ }\theta }=S^{\mathcal{M},\text{ }%
\theta ^{\prime }}$. Clearly, $S^{\mathcal{M},\text{ }\theta }\subset S^{%
\mathcal{M},\text{ }\theta ^{\prime }}$ implies%
\begin{equation}
f\left( \theta \right) =f\left( \theta ^{\prime }\right) \text{.}
\label{vgg}
\end{equation}%
We will show $S^{\mathcal{M},\text{ }\theta ^{\prime }}$ satisfies the
best-reply property at state $\theta $, which would imply $S^{\mathcal{M},%
\text{ }\theta ^{\prime }}\subset S^{\mathcal{M},\text{ }\theta }$, and
hence, also $S^{\mathcal{M},\text{ }\theta }=S^{\mathcal{M},\text{ }\theta
^{\prime }}$. For any $i\in \mathcal{I}$, pick any $m_{i}\in S_{i}^{\mathcal{%
M},\text{ }\theta }$, and there exists $\lambda _{-i}\in \triangle \left(
S_{-i}^{\mathcal{M},\text{ }\theta }\right) \subset \triangle \left( S_{-i}^{%
\mathcal{M},\text{ }\theta ^{\prime }}\right) $ such that%
\begin{equation}
u_{i}\left( g\left( m_{i},\lambda _{-i}\right) ,\text{ }\theta \right)
=u_{i}\left( f\left( \theta \right) ,\text{ }\theta \right) \geq u_{i}\left(
g\left( \widetilde{m}_{i},\lambda _{-i}\right) ,\text{ }\theta \right) \text{%
, }\forall \widetilde{m}_{i}\in M_{i}\text{.}  \label{vgg1}
\end{equation}%
Pick any $m_{i}^{\prime }\in S_{i}^{\mathcal{M},\text{ }\theta ^{\prime }}$.
Then, $S^{\mathcal{M},\text{ }\theta }\subset S^{\mathcal{M},\text{ }\theta
^{\prime }}$ and (\ref{vgg}) imply 
\begin{equation}
u_{i}\left( g\left( m_{i}^{\prime },\lambda _{-i}\right) ,\text{ }\theta
\right) =u_{i}\left( f\left( \theta ^{\prime }\right) ,\text{ }\theta
\right) =u_{i}\left( f\left( \theta \right) ,\text{ }\theta \right) \text{.}
\label{vgg2}
\end{equation}%
Thus, (\ref{vgg1}) and (\ref{vgg2}) imply%
\begin{equation*}
u_{i}\left( g\left( m_{i}^{\prime },\lambda _{-i}\right) ,\text{ }\theta
\right) =u_{i}\left( f\left( \theta \right) ,\text{ }\theta \right) \geq
u_{i}\left( g\left( \widetilde{m}_{i},\lambda _{-i}\right) ,\text{ }\theta
\right) \text{, }\forall \widetilde{m}_{i}\in M_{i}\text{,}
\end{equation*}%
i.e., $m_{i}^{\prime }$ is a best reply to $\lambda _{-i}$ for $i$ at $%
\theta $. And, $S^{\mathcal{M},\text{ }\theta ^{\prime }}$ satisfies the
best-reply property at $\theta $.$\blacksquare $

\subsection{A useful lemma}

\label{sec:proof:lem:y}

Following a similar construction as in BMT, we get the following lemma.

\begin{lemma}
\label{lem:y}There exist lotteries%
\begin{gather*}
\underline{y}\in Y\text{,} \\
\left\{ y_{i}^{\ast }\left( \theta \right) \in Y:\left( \theta ,i\right) \in
\Theta \times \mathcal{I}\right\} \text{,} \\
\left\{ z_{i}\left( \theta ,\theta ^{\prime }\right) \in Y:\left( \theta
,\theta ^{\prime }\right) \in \Theta \times \Theta \text{ and }i\in \mathcal{%
I}\right\} \text{,}
\end{gather*}%
such that%
\begin{equation}
u_{i}\left( y_{i}^{\ast }\left( \theta \right) ,\theta \right) >u_{i}\left( 
\underline{y},\theta \right) \text{, }\forall \theta \in \Theta \text{, }%
\forall i\in \mathcal{I}^{\theta }\text{,}  \label{cck1}
\end{equation}%
\begin{equation}
u_{i}\left( f\left( \theta ^{\prime }\right) ,\theta ^{\prime }\right)
>u_{i}\left( z_{i}\left( \theta ,\theta ^{\prime }\right) ,\theta ^{\prime
}\right) \text{, }\forall \left( \theta ,\theta ^{\prime }\right) \in \Theta
\times \Theta \text{, }\forall i\in \mathcal{I}^{\theta ^{\prime }}\text{,}
\label{cck2}
\end{equation}%
\begin{equation}
u_{i}\left( z_{i}\left( \theta ,\theta ^{\prime }\right) ,\theta \right)
>u_{i}\left( z_{i}\left( \theta ^{\prime },\theta ^{\prime }\right) ,\theta
\right) \text{, }\forall \left( \theta ,\theta ^{\prime }\right) \in \Theta
\times \Theta \text{ with }\theta \neq \theta ^{\prime }\text{, }\forall
i\in \mathcal{I}^{\theta }\text{.}  \label{cck3}
\end{equation}
\end{lemma}

If NWA is imposed, i.e., $\mathcal{I}^{\Theta }=\mathcal{I}$, Lemma \ref%
{lem:y} reduces to Lemma 2 in BMT.

\noindent \textbf{Proof of Lemma \ref{lem:y}:} We can find a set $\left\{ 
\underline{y}_{i}\left( \theta \right) :\left( \theta ,i\right) \in \Theta
\times \mathcal{I}\right\} $ such that%
\begin{equation}
u_{i}\left( f\left( \theta \right) ,\text{ }\theta \right) >u_{i}\left( 
\underline{y}_{i}\left( \theta \right) ,\text{ }\theta \right) \text{, }%
\forall \theta \in \Theta \text{, }\forall i\in \mathcal{I}^{\theta }\text{.}
\label{hkkk1}
\end{equation}%
Define%
\begin{gather*}
\underline{y}_{i}\overset{\triangle }{=}\frac{1}{\left\vert \Theta
\right\vert }\dsum\limits_{\theta \in \Theta }\underline{y}_{i}\left( \theta
\right) \text{ and} \\
y_{i}\left( \theta \right) \overset{\triangle }{=}\frac{1}{\left\vert \Theta
\right\vert }\dsum\limits_{\widehat{\theta }\in \Theta \diagdown \left\{
\theta \right\} }\underline{y}_{i}\left( \widehat{\theta }\right) +\frac{1}{%
\left\vert \Theta \right\vert }f\left( \theta \right) \text{, }\forall
\left( \theta ,i\right) \in \Theta \times \mathcal{I}\text{,}
\end{gather*}%
which, together with (\ref{hkkk1}), imply%
\begin{equation}
u_{i}\left( y_{i}\left( \theta \right) ,\text{ }\theta \right) >u_{i}\left( 
\underline{y}_{i},\text{ }\theta \right) \text{, }\forall \theta \in \Theta 
\text{, }\forall i\in \mathcal{I}^{\theta }\text{.}  \label{hkkk2}
\end{equation}%
Furthermore, define%
\begin{gather*}
\underline{y}\overset{\triangle }{=}\frac{1}{\left\vert \mathcal{I}%
\right\vert }\dsum\limits_{i\in \mathcal{I}}\underline{y}_{i}\text{ and} \\
y_{i}^{\ast }\left( \theta \right) \overset{\triangle }{=}\frac{1}{%
\left\vert \mathcal{I}\right\vert }\dsum\limits_{j\in \mathcal{I}\diagdown
\left\{ i\right\} }\underline{y}_{j}+\frac{1}{\left\vert \mathcal{I}%
\right\vert }y_{i}\left( \theta \right) \text{, }\forall \left( \theta
,i\right) \in \Theta \times \mathcal{I}\text{,}
\end{gather*}%
which, together with (\ref{hkkk2}), imply%
\begin{equation*}
u_{i}\left( y_{i}^{\ast }\left( \theta \right) ,\text{ }\theta \right)
>u_{i}\left( \underline{y},\text{ }\theta \right) \text{, }\forall \theta
\in \Theta \text{, }\forall i\in \mathcal{I}^{\theta }\text{,}
\end{equation*}%
which shows (\ref{cck1})\ in Lemma \ref{lem:y}. With $\varepsilon >0$, define%
\begin{equation}
z_{i}\left( \theta ^{\prime },\theta ^{\prime }\right) \overset{\triangle }{=%
}\left( 1-\varepsilon \right) \underline{y}_{i}\left( \theta ^{\prime
}\right) +\varepsilon \underline{y}_{i}\text{, }\forall \left( \theta
^{\prime },i\right) \in \Theta \times \mathcal{I}\text{ and}  \label{hkkk4}
\end{equation}%
\begin{equation}
z_{i}\left( \theta ,\theta ^{\prime }\right) \overset{\triangle }{=}\left(
1-\varepsilon \right) \underline{y}_{i}\left( \theta ^{\prime }\right) +%
\frac{\varepsilon }{\left\vert \Theta \right\vert }\left( \dsum\limits_{%
\widehat{\theta }\in \Theta \diagdown \left\{ \theta \right\} }\underline{y}%
_{i}\left( \widehat{\theta }\right) +f\left( \theta \right) \right) \text{, }%
\forall \left( \theta ,\theta ^{\prime },i\right) \in \Theta \times \Theta
\times \mathcal{I}\text{ with }\theta \neq \theta ^{\prime }\text{.}
\label{hkkk5}
\end{equation}%
By (\ref{hkkk1}), we have 
\begin{eqnarray}
u_{i}\left( f\left( \theta ^{\prime }\right) ,\text{ }\theta ^{\prime
}\right) &>&u_{i}\left( \underline{y}_{i}\left( \theta ^{\prime }\right) ,%
\text{ }\theta ^{\prime }\right) \text{, }\forall \theta ^{\prime }\in
\Theta \text{, }\forall i\in \mathcal{I}^{\theta ^{\prime }}\text{,}
\label{hkkk6} \\
u_{i}\left( f\left( \theta \right) ,\text{ }\theta \right) &>&u_{i}\left( 
\underline{y}_{i}\left( \theta \right) ,\text{ }\theta \right) \text{, }%
\forall \theta \in \Theta \text{, }\forall i\in \mathcal{I}^{\theta }\text{.}
\label{hkkk8}
\end{eqnarray}%
Then, (\ref{hkkk4}), (\ref{hkkk5}) and (\ref{hkkk8}) imply%
\begin{equation*}
u_{i}\left( z_{i}\left( \theta ,\theta ^{\prime }\right) ,\theta \right)
>u_{i}\left( z_{i}\left( \theta ^{\prime },\theta ^{\prime }\right) ,\theta
\right) \text{, }\forall \left( \theta ,\theta ^{\prime }\right) \in \Theta
\times \Theta \text{ with }\theta \neq \theta ^{\prime }\text{, }\forall
i\in \mathcal{I}^{\theta }\text{,}
\end{equation*}%
which establish (\ref{cck3})\ in Lemma \ref{lem:y}.

Furthermore, by choosing sufficient small $\varepsilon >0$, (\ref{hkkk4}), (%
\ref{hkkk5}) and (\ref{hkkk6}) imply%
\begin{equation*}
u_{i}\left( f\left( \theta ^{\prime }\right) ,\theta ^{\prime }\right)
>u_{i}\left( z_{i}\left( \theta ,\theta ^{\prime }\right) ,\theta ^{\prime
}\right) \text{, }\forall \left( \theta ,\theta ^{\prime }\right) \in \Theta
\times \Theta \text{, }\forall i\in \mathcal{I}^{\theta ^{\prime }}\text{,}
\end{equation*}%
which establish (\ref{cck2})\ in Lemma \ref{lem:y}.$\blacksquare $

\subsection{Proof: the "if" part of Theorem \protect\ref%
{theorem:non-responsive}}

\label{sec:proof:theorem:non-responsive}

Suppose that NWA and strict Maskin monotonicity$^{\ast \ast }$ hold. Thus,
there exists a partition $\mathcal{P}$ of $\Theta $ finer than $\mathcal{P}%
_{f}$ such that for any $\left( \theta ,\theta ^{\prime }\right) \in \Theta
\times \Theta $,%
\begin{equation}
\theta ^{\prime }\notin \mathcal{P}\left( \theta \right) \text{ implies }%
\left[ 
\begin{array}{c}
\exists i\in \mathcal{I}\text{ such that\ }\forall \widehat{\theta }\in 
\mathcal{P}\left( \theta \right) \text{, }\exists y^{\widehat{\theta }}\in Y%
\text{,} \\ 
u_{i}\left( f\left( \widehat{\theta }\right) ,\text{ }\widehat{\theta }%
\right) >u_{i}\left( y^{\widehat{\theta }},\text{ }\widehat{\theta }\right) 
\text{ and }u_{i}\left( y^{\widehat{\theta }},\text{ }\theta ^{\prime
}\right) >u_{i}\left( f\left( \widehat{\theta }\right) ,\text{ }\theta
^{\prime }\right) \text{.}%
\end{array}%
\right]  \label{tfe1}
\end{equation}%
That is, $y^{\widehat{\theta }}$ in (\ref{tfe1})\ is the blocking plan for
agent $i$, when the true state is $\theta ^{\prime }$ and all other agents
report $\widehat{\theta }$. Let $\mathcal{B}$ denote the finite set of all
such $y^{\widehat{\theta }}$. Since $\mathcal{I}\times \Theta $ is finite,
there exists a finite set $\Sigma \subset Y$ such that%
\begin{gather}
\mathcal{B}\cup f\left( \Theta \right) \cup \left\{ z_{i}\left( \theta
,\theta ^{\prime }\right) \in Y:\left( \theta ,\theta ^{\prime }\right) \in
\Theta \times \Theta \right\} \subset \Sigma \text{ and}  \notag \\
\theta ^{\prime }\notin \mathcal{P}\left( \theta \right) \text{ implies }%
\left[ 
\begin{array}{c}
\exists i\in \mathcal{I}\text{ such that\ }\forall \widehat{\theta }\in 
\mathcal{P}\left( \theta \right) \text{, }\exists y^{\widehat{\theta }}\in
\Sigma \text{,} \\ 
u_{i}\left( f\left( \widehat{\theta }\right) ,\text{ }\widehat{\theta }%
\right) >u_{i}\left( y^{\widehat{\theta }},\text{ }\widehat{\theta }\right) 
\text{ and }u_{i}\left( y^{\widehat{\theta }},\text{ }\theta ^{\prime
}\right) >u_{i}\left( f\left( \widehat{\theta }\right) ,\text{ }\theta
^{\prime }\right) \text{.}%
\end{array}%
\right]  \label{tfe2}
\end{gather}%
where $z_{i}\left( \theta ,\theta ^{\prime }\right) $ are defined in Lemma %
\ref{lem:y}. That is, $\Sigma $ is a finite set which contains all of $%
f\left( \theta \right) $, $z_{i}\left( \theta ,\theta ^{\prime }\right) $
and potential blocking plans for all possible profiles of $\left( \theta
^{\prime },\text{ }\widehat{\theta }\right) $.---Our canonical mechanism is
required to be a countable-action game, and hence, we should focus on $%
\Sigma $.

We use $\mathcal{M}=\left\langle M=\left( M_{i}\right) _{i\in \mathcal{I}}%
\text{, \ }g:M\longrightarrow Y\right\rangle $ defined as follows to
implement $f$. In this mechanism, each agent $i$ sends a message $m_{i}=%
\left[ m_{i}^{1},m_{i}^{2},m_{i}^{3},m_{i}^{4}\right] \in M_{i}$, where 
\begin{eqnarray*}
m_{i}^{1} &\in &\Theta , \\
m_{i}^{2} &\in &\mathbb{N}, \\
m_{i}^{3} &\in &\Sigma ^{\Theta }, \\
m_{i}^{4} &\in &Z\text{.}
\end{eqnarray*}%
The innovation is that $m_{i}^{3}\in \Sigma ^{\Theta }$ is a
state-contingent blocking plan. As usual, we partition $M$ into three sets:
agreement, unilateral deviation and multi-lateral deviation.%
\begin{equation*}
M^{\prime }=\left\{ \left( m_{i}=\left[
m_{i}^{1},m_{i}^{2},m_{i}^{3},m_{i}^{4}\right] \right) _{i\in \mathcal{I}%
}\in M:\exists \theta \in \Theta \text{, }m_{i}^{1}\in \mathcal{P}\left(
\theta \right) \text{ and }m_{i}^{2}=1\text{, }\forall i\in \mathcal{I}%
\right\} \text{,}
\end{equation*}%
\begin{equation*}
M^{\prime \prime }=\left\{ m_{i}\in M\diagdown M^{\prime }:\exists \left(
\theta ,i\right) \in \Theta \times \mathcal{I}\text{, }m_{j}^{1}\in \mathcal{%
P}\left( \theta \right) \text{ and }m_{j}^{2}=1\text{, }\forall j\in 
\mathcal{I\diagdown }\left\{ i\right\} \right\} \text{,}
\end{equation*}%
\begin{equation*}
M^{\prime \prime \prime }=M\diagdown \left( M^{\prime }\cup M^{\prime \prime
}\right) \text{.}
\end{equation*}

Then, $g$ is defined by the following rules.

\begin{quotation}
\textbf{Rule 1 (agreement): }when $m\in M^{\prime }$: there exists $\theta
\in \Theta $, $m_{i}^{1}\in \mathcal{P}\left( \theta \right) $ and $%
m_{i}^{2}=1$ for every $i\in \mathcal{I}$. In particular, $f\left( \theta
\right) $ is unique, and $g$ picks $f\left( \theta \right) $;

\textbf{Rule 2 (unilateral deviation): }when $m\in M^{\prime \prime }$:
there exists $\left( \theta ,i\right) \in \Theta \times \mathcal{I}$ such
that $m_{j}^{1}\in \mathcal{P}\left( \theta \right) $and $m_{j}^{2}=1$ for
every $j\in \mathcal{I\diagdown }\left\{ i\right\} $. In particular, such $%
f\left( \theta \right) $ is unique. For notational ease, consider agent $i+1$
(module $I$)\footnote{%
That is, agent $\left( I+1\right) $ is agent $1$.} and set $\widehat{\theta }%
\equiv m_{i+1}^{1}$ --- the interpretation is: even though agents $-i$ may
report different states in $\mathcal{P}\left( \theta \right) $, we
"hypothetically" regard they all reporting $\widehat{\theta }$, when agent $%
i $ is the whistle-blower. We further distinguish two sub-cases:

\ \ \ \ \textbf{Rule (2.a):} if $u_{i}\left( f\left( \widehat{\theta }%
\right) ,\widehat{\theta }\right) \geq u_{i}\left( m_{i}^{3}\left( \widehat{%
\theta }\right) ,\widehat{\theta }\right) $, then $g$ picks $m_{i}^{3}\left( 
\widehat{\theta }\right) $ with probability $1-\frac{1}{m_{i}^{2}+1}$ and $g$
picks $z_{i}\left( \widehat{\theta },\widehat{\theta }\right) $ with
probability $\frac{1}{m_{i}^{2}+1}$;

\ \ \ \ \ \ \textbf{Rule (2.b):} otherwise, $g$ picks $z_{i}\left( \widehat{%
\theta },\widehat{\theta }\right) $;

\textbf{Rule 3 (multi-lateral deviation): }when $m\in M^{\prime \prime
\prime }$: consider agent $j=max\left[ \arg \max_{h\in \mathcal{I}}m_{h}^{2}%
\right] $, i.e., $j$ is the largest-numbered agent who reports the largest
integer in the second dimension. Then, $g$ picks $m_{j}^{4}$ with
probability $1-\frac{1}{m_{j}^{2}+1}$ and $g$ picks $\underline{y}$ with
probability $\frac{1}{m_{j}^{2}+1}$.
\end{quotation}

From now on, let us assume that the true state is $\theta ^{\ast }$, and we
will show that reporting states in $\mathcal{P}\left( \theta ^{\ast }\right) 
$ are the only rationalizable strategies for all agents. Before starting our
proof, we first show that there exist best challenging schemes in Rules 2
and 3. Fix any%
\begin{gather}
\widehat{m}_{i}^{3}\left( \theta \right) \in \arg \max_{y\in \left\{ 
\widetilde{y}\in \Sigma \text{ }:\text{ }u_{i}\left( f\left( \theta \right)
,\theta \right) \geq u_{i}\left( \widetilde{y},\theta \right) \right\}
}u_{i}\left( y,\theta ^{\ast }\right) \text{, }\forall \theta \in \Theta 
\text{,}  \label{xtx1} \\
\widehat{m}_{i}^{3}\equiv \left[ \widehat{m}_{i}^{3}\left( \theta \right) %
\right] _{\theta \in \Theta }\text{,}  \notag \\
\widehat{m}_{i}^{4}\in \arg \max_{z\in Z}u_{i}\left( z,\theta ^{\ast
}\right) \text{.}  \label{xtx3}
\end{gather}%
That is, $\widehat{m}_{i}^{3}\equiv \left[ \widehat{m}_{i}^{3}\left( \theta
\right) \right] _{\theta \in \Theta }$ and $\widehat{m}_{i}^{4}$ are the
best options for agent $i$ if Rules 2 and 3 are triggered, respectively.
Specifically, we have 
\begin{eqnarray}
u_{i}\left( \widehat{m}_{i}^{4},\theta ^{\ast }\right) &\geq &u_{i}\left(
y_{i}\left( \theta ^{\ast }\right) ,\theta ^{\ast }\right) >u_{i}\left( 
\underline{y},\theta ^{\ast }\right) \text{, }\forall \theta \in \Theta 
\text{,}  \label{ckk2} \\
u_{i}\left( \widehat{m}_{i}^{4},\theta ^{\ast }\right) &\geq &u_{i}\left( 
\widehat{m}_{i}^{3}\left( \theta \right) ,\theta ^{\ast }\right)
>u_{i}\left( z_{i}\left( \theta ,\theta \right) ,\theta ^{\ast }\right) 
\text{, }\forall \theta \in \Theta \text{,}  \label{ckk1}
\end{eqnarray}%
where $y_{i}\left( \theta ^{\ast }\right) $, $\underline{y}$ and $%
z_{i}\left( \theta ,\theta \right) $ are defined in Lemma \ref{lem:y}. In
particular, the weak inequalities in (\ref{ckk2}) and (\ref{ckk1}) follow
from (\ref{xtx3}), and the strict inequality in (\ref{ckk2}) from (\ref{cck1}%
) in\ Lemma \ref{lem:y}. To see strict inequality in (\ref{ckk1}), first
suppose $\theta =\theta ^{\ast }$, and we have%
\begin{equation}
u_{i}\left( \widehat{m}_{i}^{3}\left( \theta ^{\ast }\right) ,\theta ^{\ast
}\right) \geq u_{i}\left( f\left( \theta ^{\ast }\right) ,\theta ^{\ast
}\right) >u_{i}\left( z_{i}\left( \theta ^{\ast },\theta ^{\ast }\right)
,\theta ^{\ast }\right) \text{,}  \label{dde1}
\end{equation}%
where the weak inequality follows from $f\left( \theta ^{\ast }\right) \in
\left\{ \widetilde{y}\in \Sigma :u_{i}\left( f\left( \theta ^{\ast }\right)
,\theta ^{\ast }\right) \geq u_{i}\left( \widetilde{y},\theta ^{\ast
}\right) \right\} $ and the strict inequality follows from (\ref{cck2}) in\
Lemma \ref{lem:y}. Second, suppose $\theta \neq \theta ^{\ast }$, and we have%
\begin{equation}
u_{i}\left( \widehat{m}_{i}^{3}\left( \theta \right) ,\theta ^{\ast }\right)
\geq u_{i}\left( z_{i}\left( \theta ^{\ast },\theta \right) ,\theta ^{\ast
}\right) >u_{i}\left( z_{i}\left( \theta ,\theta \right) ,\theta ^{\ast
}\right) \text{,}  \label{dde2}
\end{equation}%
where the weak inequality follows from $z_{i}\left( \theta ^{\ast },\theta
\right) \in \left\{ \widetilde{y}\in \Sigma :u_{i}\left( f\left( \theta
\right) ,\theta \right) \geq u_{i}\left( \widetilde{y},\theta \right)
\right\} $ due to (\ref{cck2})\ in Lemma \ref{lem:y} and the strict
inequality follows from (\ref{cck3}) in\ Lemma \ref{lem:y}. Hence, (\ref%
{dde1})\ and (\ref{dde2})\ imply the strict inequality in (\ref{ckk1}).

When either Rule 2 or Rule 3 is triggered, the induced payoffs are listed as
follows.

\begin{eqnarray*}
\text{Rule 2}\text{: } &&\left[ 
\begin{array}{c}
\frac{\left( n-1\right) \times u_{i}\left( y,\theta ^{\ast }\right)
+u_{i}\left( z_{i}\left( \theta ,\theta \right) ,\theta ^{\ast }\right) }{n}%
\text{for some }\left( n,\theta ,y\right) \in \mathbb{N}\times \Theta \times
\Sigma \\ 
\text{such that }u_{i}\left( f\left( \theta \right) ,\theta \right) \geq
u_{i}\left( y,\theta \right)%
\end{array}%
\right] \text{,} \\
\text{Rule 3} &\text{:}&\text{ }\left[ \frac{n\times u_{i}\left( z,\theta
^{\ast }\right) +u_{i}\left( \underline{y},\theta ^{\ast }\right) }{n+1}%
\text{ for some }\left( n,z\right) \in \mathbb{N}\times Z\right] \text{,}
\end{eqnarray*}

Then, (\ref{ckk1}) implies%
\begin{gather}
u_{i}\left( \widehat{m}_{i}^{4},\theta ^{\ast }\right) \geq u_{i}\left( 
\widehat{m}_{i}^{3}\left( \theta \right) ,\theta ^{\ast }\right) >\frac{%
\left( n-1\right) \times u_{i}\left( y,\theta ^{\ast }\right) +u_{i}\left(
z_{i}\left( \theta ,\theta \right) ,\theta ^{\ast }\right) }{n}\text{,}
\label{tef1} \\
\forall \left( n,\theta ,y\right) \in \mathbb{N}\times \Theta \times \Sigma 
\text{ such that }u_{i}\left( f\left( \theta \right) ,\theta \right) \geq
u_{i}\left( y,\theta \right) \text{,}  \notag
\end{gather}%
and (\ref{ckk2})\ implies%
\begin{gather}
u_{i}\left( \widehat{m}_{i}^{4},\theta ^{\ast }\right) >\frac{n\times
u_{i}\left( z,\theta ^{\ast }\right) +u_{i}\left( \underline{y},\theta
^{\ast }\right) }{n+1}\text{,}  \label{tef2} \\
\forall \left( n,z\right) \in \mathbb{N}\times Z\text{,}  \notag
\end{gather}%
i.e., $\widehat{m}_{i}^{3}$ and $\widehat{m}_{i}^{4}$ are strictly better
than any induced payoffs in Rule 2 and 3, respectively.

\medskip

In five steps, we now prove that $\mathcal{M}$ rationalizably implements $f$.

\textbf{Step 1: \ \ \ \ \ \ \ \ \ at true state }$\theta ^{\ast }\in \Theta $%
\textbf{, any }$\left[ m_{i}^{1},m_{i}^{2},m_{i}^{3},m_{i}^{4}\right] _{i\in 
\mathcal{I}}$\textbf{\ with }$\left( m_{i}^{1},m_{i}^{2}\right) =\left(
\theta ^{\ast },1\right) $\textbf{\ for every }$i\in I$\textbf{\ is a Nash
equilibrium, which induces }$f\left( \theta ^{\ast }\right) $\textbf{\ as
dictated by Rule 1.}

For every $i\in \mathcal{I}$, any deviation of $i$ would either stay in Rule
1 and induce the same outcome $f\left( \theta ^{\ast }\right) $, or trigger
Rule 2, which induces either $z_{i}\left( \theta ^{\ast },\theta ^{\ast
}\right) $ or a mixture of $z_{i}\left( \theta ^{\ast },\theta ^{\ast
}\right) $ and $m_{i}^{3}\left( \theta ^{\ast }\right) $ with $u_{i}\left(
f\left( \theta ^{\ast }\right) ,\theta ^{\ast }\right) \geq u_{i}\left(
m_{i}^{3}\left( \theta ^{\ast }\right) ,\theta ^{\ast }\right) $. Clearly, $%
z_{i}\left( \theta ^{\ast },\theta ^{\ast }\right) $ is worse than $f\left(
\theta ^{\ast }\right) $ by Lemma \ref{lem:y} (precisely, (\ref{cck2})).
Therefore, any deviation of $i$ is not a profitable.

\medskip

\textbf{Step 2: \ \ \ \ \ \ \ \ \ at true state }$\theta ^{\ast }\in \Theta $%
\textbf{, for every }$i\in \mathcal{I}$,\textbf{\ if any }$m_{i}\in M_{i}$%
\textbf{\ is a best reply to } $\lambda _{-i}\in \triangle \left(
M_{-i}\right) $\textbf{, then }$\left( m_{i},\lambda _{-i}\right) $\textbf{\
induces Rules 2 and 3 with probability 0.}

We prove this by contradiction. Suppose $\left( m_{i},\lambda _{-i}\right) $
induces Rules 2 or 3 with a positive probability. We thus partition $M_{-i}$
as follows.%
\begin{equation*}
M_{-i}=\left( \dbigcup\limits_{\theta \in \Theta }M_{-i}^{\theta }\right)
\cup \left( \dbigcup\limits_{\left( n,\theta ,y\right) \in \mathbb{N}\times
\Theta \times \Sigma }M_{-i}^{\left( n,\theta ,y\right) }\right) \cup \left(
\dbigcup\limits_{\left( n,z\right) \in \mathbb{N}\times Z}M_{-i}^{\left(
n,z\right) }\right) \text{,}
\end{equation*}

\begin{eqnarray*}
\text{where }M_{-i}^{\theta } &\equiv &\left\{ m_{-i}\in M_{-i}:%
\begin{tabular}{l}
$\left( m_{i},m_{-i}\right) \text{ triggers Rule 1 and induces payoff}$ \\ 
$u_{i}\left( f\left( \theta \right) ,\theta ^{\ast }\right) $ for agent $i$%
\end{tabular}%
\right\} \text{,} \\
\text{where }M_{-i}^{\left( n,\theta ,y\right) } &\equiv &\left\{ m_{-i}\in
M_{-i}:%
\begin{tabular}{l}
$u_{i}\left( f\left( \theta \right) ,\theta \right) \geq u_{i}\left(
y,\theta \right) $ and \\ 
$\left( m_{i},m_{-i}\right) \text{ triggers Rule 2 and induces payoff}$ \\ 
$\frac{\left( n-1\right) \times u_{i}\left( y,\theta ^{\ast }\right)
+u_{i}\left( z_{i}\left( \theta ,\theta \right) ,\theta ^{\ast }\right) }{n}$
for agent $i$%
\end{tabular}%
\right\} \text{,} \\
\text{and }M_{-i}^{\left( n,z\right) } &\equiv &\left\{ m_{-i}\in M_{-i}:%
\begin{tabular}{l}
$\left( m_{i},m_{-i}\right) \text{ triggers Rule 3 and induces payoff}$ \\ 
$\frac{n\times u_{i}\left( z,\theta ^{\ast }\right) +u_{i}\left( \underline{y%
},\theta ^{\ast }\right) }{n+1}$ for agent $i$%
\end{tabular}%
\right\} \text{.}
\end{eqnarray*}%
Suppose agent $i$ deviates from $m_{i}$ to $\left[ m_{i}^{1},\widetilde{m}%
_{i}^{2},\widehat{m}_{i}^{3},\widehat{m}_{i}^{4}\right] $ with $\widetilde{m}%
_{i}^{2}\rightarrow \infty $, where $\widehat{m}_{i}^{3}$ and $\widehat{m}%
_{i}^{4}$ are defined in (\ref{xtx1}) and (\ref{xtx3}). The payoff changes
are listed in the following table.%
\begin{equation*}
\begin{tabular}{|c|c|c|c|}
\hline
as $\widetilde{m}_{i}^{2}\rightarrow \infty $ & payoff under $m_{i}$ &  & 
\begin{tabular}{c}
the supremum of payoffs \\ 
under $\left[ m_{i}^{1},\widetilde{m}_{i}^{2},\widehat{m}_{i}^{3},\widehat{m}%
_{i}^{4}\right] $%
\end{tabular}
\\ \hline
$m_{-i}\in M_{-i}^{\theta }$ & $u_{i}\left( f\left( \theta \right) ,\theta
^{\ast }\right) $ & $\leq $ & $u_{i}\left( \widehat{m}_{i}^{3}\left( \theta
\right) ,\theta ^{\ast }\right) $ \\ \hline
$m_{-i}\in M_{-i}^{\left( n,\theta ,y\right) }\neq \varnothing $ & $\frac{%
\left( n-1\right) \times u_{i}\left( y,\theta ^{\ast }\right) +u_{i}\left(
z_{i}\left( \theta ,\theta \right) ,\theta ^{\ast }\right) }{n}$ & $<$ & $%
u_{i}\left( \widehat{m}_{i}^{3}\left( \theta \right) ,\theta ^{\ast }\right) 
$ \\ \hline
$m_{-i}\in M_{-i}^{\left( n,z\right) }$ & $\frac{n\times u_{i}\left(
y,\theta ^{\ast }\right) +u_{i}\left( \underline{y},\theta ^{\ast }\right) }{%
n+1}$ & $<$ & $u_{i}\left( \widehat{m}_{i}^{4},\theta ^{\ast }\right) $ \\ 
\hline
\end{tabular}%
\end{equation*}%
where the strict inequality follows from (\ref{tef1}) and (\ref{tef2}).
Hence, for any mixed strategy $\lambda _{-i}\in \triangle \left(
M_{-i}\right) $ which induces Rules 2 and 3 with a positive probability, we
have%
\begin{equation*}
u_{i}\left( g\left( \left[ m_{i}^{1},m_{i}^{2},m_{i}^{3},m_{i}^{4}\right]
,\lambda _{-i}\right) ,\theta ^{\ast }\right) <\lim_{\widetilde{m}%
_{i}^{2}\rightarrow \infty }u_{i}\left( g\left( \left[ m_{i}^{1},\widetilde{m%
}_{i}^{2},\widehat{m}_{i}^{3},\widehat{m}_{i}^{4}\right] ,\lambda
_{-i}\right) ,\theta ^{\ast }\right)
\end{equation*}%
and as a result, there exists $\widehat{m}_{i}^{2}\in \mathbb{N}$ such that%
\begin{equation*}
u_{i}\left( g\left( \left[ m_{i}^{1},m_{i}^{2},m_{i}^{3},m_{i}^{4}\right]
,\lambda _{-i}\right) ,\theta ^{\ast }\right) <u_{i}\left( g\left( \left[
m_{i}^{1},\widehat{m}_{i}^{2},\widehat{m}_{i}^{3},\widehat{m}_{i}^{4}\right]
,\lambda _{-i}\right) ,\theta ^{\ast }\right) \text{,}
\end{equation*}%
contradicting $m_{i}$ being a best reply to $\lambda _{-i}$.

\medskip

\textbf{Step 3: \ \ \ \ \ \ \ \ \ at true state }$\theta ^{\ast }\in \Theta $%
\textbf{, for every }$i\in \mathcal{I}$,\textbf{\ any strategy }$m_{i}=\left[
m_{i}^{1},m_{i}^{2},m_{i}^{3},m_{i}^{4}\right] $\textbf{\ with }$m_{i}^{2}>1$%
\textbf{\ is not rationalizable for agent }$i$\textbf{.}

This follows from Step 2, because $m_{i}^{2}>1$ induces rules 2 or 3 with
probability 1.

\medskip

\textbf{Step 4: \ \ \ \ \ \ \ \ \ at true state }$\theta ^{\ast }\in \Theta $%
\textbf{, for any }$\overline{\theta }\in \Theta \diagdown \mathcal{P}\left(
\theta ^{\ast }\right) $\textbf{\ (or equivalently, }$\theta ^{\ast }\notin 
\mathcal{P}\left( \overline{\theta }\right) $\textbf{), there exists }$j\in 
\mathcal{I}$\textbf{\ such that any strategy }$m_{j}=\left[
m_{j}^{1},m_{j}^{2},m_{j}^{3},m_{j}^{4}\right] $\textbf{\ with }$%
m_{j}^{1}\in \mathcal{P}\left( \overline{\theta }\right) $\textbf{\ is not
rationalizable for agent }$j$\textbf{.}

By our construction of $\Sigma $ above and (\ref{tfe2}), we have%
\begin{gather*}
\forall \widetilde{\theta }\in \mathcal{P}\left( \overline{\theta }\right) 
\text{, }\exists y^{\widetilde{\theta }}\in \Sigma \text{,} \\
u_{j}\left( f\left( \widetilde{\theta }\right) ,\text{ }\widetilde{\theta }%
\right) >u_{j}\left( y^{\widetilde{\theta }},\text{ }\widetilde{\theta }%
\right) \text{ and }u_{j}\left( y^{\widetilde{\theta }},\text{ }\theta
^{\ast }\right) >u_{j}\left( f\left( \widetilde{\theta }\right) ,\text{ }%
\theta ^{\ast }\right) \text{.}
\end{gather*}%
and furthermore, by our definition of $\widehat{m}_{j}^{3}\left( \theta
\right) $ above, we have%
\begin{gather}
\forall \widetilde{\theta }\in \mathcal{P}\left( \overline{\theta }\right) 
\text{,}  \notag \\
u_{j}\left( f\left( \widetilde{\theta }\right) ,\text{ }\widetilde{\theta }%
\right) \geq u_{j}\left( \widehat{m}_{j}^{3}\left( \widetilde{\theta }%
\right) ,\text{ }\widetilde{\theta }\right) \text{ and }u_{j}\left( \widehat{%
m}_{j}^{3}\left( \widetilde{\theta }\right) ,\text{ }\theta ^{\ast }\right)
>u_{j}\left( f\left( \widetilde{\theta }\right) ,\text{ }\theta ^{\ast
}\right) \text{.}  \label{gre1}
\end{gather}

For any $m_{j}=\left[ m_{j}^{1},m_{j}^{2},m_{j}^{3},m_{j}^{4}\right] $ with $%
m_{j}^{1}\in \mathcal{P}\left( \overline{\theta }\right) $, we prove by
contradiction that it is not rationalizable for $j$. Suppose otherwise,
i.e., $m_{j}$ is a best reply to some $\lambda _{-j}\in \triangle \left(
S_{-j}^{\mathcal{M},\text{ }\theta ^{\ast }}\right) $. By Step 2, $\left(
m_{j},\lambda _{-j}\right) $ must induce Rule 1 with probability 1. Thus,
every agent $i$ report $m_{i}^{1}\in \mathcal{P}\left( \overline{\theta }%
\right) $ and $m_{i}^{2}=1$, which induces $f\left( \overline{\theta }%
\right) $. As a result, agent $j$'s payoff is $u_{j}\left( f\left( \overline{%
\theta }\right) ,\text{ }\theta ^{\ast }\right) $. Then, agent $j$ would
like to deviate from $m_{j}$ to $\left[ m_{j}^{1},\widetilde{m}_{j}^{2},%
\widehat{m}_{j}^{3},\widehat{m}_{j}^{4}\right] $ with $\widetilde{m}%
_{j}^{2}\rightarrow \infty $, which would induce Rule 2 and%
\begin{equation*}
u_{j}\left( g\left( m_{j},\lambda _{-j}\right) ,\theta ^{\ast }\right)
=u_{j}\left( f\left( \overline{\theta }\right) ,\text{ }\theta ^{\ast
}\right) <\lim_{\widetilde{m}_{j}^{2}\rightarrow \infty }u_{i}\left( g\left( %
\left[ m_{j}^{1},\widetilde{m}_{j}^{2},\widehat{m}_{j}^{3},\widehat{m}%
_{j}^{4}\right] ,\lambda _{-j}\right) ,\theta ^{\ast }\right)
\end{equation*}%
where the inequality follows from (\ref{gre1}). Hence, there exists $%
\widehat{m}_{j}^{2}\in \mathbb{N}$ such that%
\begin{equation*}
u_{j}\left( g\left( m_{j},\lambda _{-j}\right) ,\theta ^{\ast }\right)
=u_{j}\left( f\left( \overline{\theta }\right) ,\text{ }\theta ^{\ast
}\right) <u_{i}\left( g\left( \left[ m_{j}^{1},\widehat{m}_{j}^{2},\widehat{m%
}_{j}^{3},\widehat{m}_{j}^{4}\right] ,\lambda _{-j}\right) ,\theta ^{\ast
}\right)
\end{equation*}%
contradicting $m_{j}$ being a best reply to $\lambda _{-j}$.

\medskip

\textbf{Step 5: \ \ \ \ \ \ \ \ \ at true state }$\theta ^{\ast }\in \Theta $%
\textbf{, for any }$\left( i,\overline{\theta }\right) \in \mathcal{I\times }%
\left[ \Theta \diagdown \mathcal{P}\left( \theta ^{\ast }\right) \right] $%
\textbf{, any }$m_{i}=\left[ m_{i}^{1},m_{i}^{2},m_{i}^{3},m_{i}^{4}\right] $%
\textbf{\ with }$m_{i}^{1}\in \mathcal{P}\left( \overline{\theta }\right) $%
\textbf{\ is not rationalizable for agent }$i$\textbf{.}

First, this is true for the agent $j$ identified in Step 4. Second, consider
any $i\neq j$. We prove it by contradiction. Suppose $m_{i}=\left[
m_{i}^{1},m_{i}^{2},m_{i}^{3},m_{i}^{4}\right] $\ with $m_{i}^{1}\in P\left( 
\overline{\theta }\right) $\ is rationalizable for agent $i$. Then, $m_{i}$
is a best reply to some rationalizable conjecture $\lambda _{-i}\in
\triangle \left( S_{-i}^{\mathcal{M},\text{ }\theta ^{\ast }}\right) $. By
Step 2, $\left( m_{i},\lambda _{-i}\right) $ must induce Rule 1 with
probability 1, or equivalently, with probability 1, every agent $h\in 
\mathcal{I}$ reports $m_{h}^{1}\in \mathcal{P}\left( \overline{\theta }%
\right) $, including agent $j$, contradicting Step 4.

\medskip

To sum, Step 1 shows%
\begin{equation*}
S^{\mathcal{M},\text{ }\theta ^{\ast }}\supset \dprod\limits_{i\in \mathcal{I%
}}\left\{ \left( m_{i}^{1},m_{i}^{2},m_{i}^{3},m_{i}^{4}\right) \in
M_{i}:\left( m_{i}^{1},m_{i}^{2}\right) =\left( \theta ^{\ast },1\right)
\right\} \text{,}
\end{equation*}%
and Steps 2-5 show%
\begin{equation*}
S^{\mathcal{M},\text{ }\theta ^{\ast }}\subset \dprod\limits_{i\in \mathcal{I%
}}\left\{ \left( m_{i}^{1},m_{i}^{2},m_{i}^{3},m_{i}^{4}\right) \in
M_{i}:m_{i}^{1}\in \mathcal{P}\left( \theta ^{\ast }\right) \text{ and }%
m_{i}^{2}=1\right\} \text{.}
\end{equation*}%
Thus, every $m\in S^{\mathcal{M},\text{ }\theta ^{\ast }}$ triggers Rule 1
and induces $f\left( \theta ^{\ast }\right) $, i.e., $g\left( S^{\mathcal{M},%
\text{ }\theta ^{\ast }}\right) =\left\{ f\left( \theta ^{\ast }\right)
\right\} $.$\blacksquare $

\subsection{Proofs of Proposition \protect\ref{prop:NWA}}

\label{sec:proof:prop:NWA}

Consider any responsive SCF $f:\Theta \longrightarrow Z$. First, suppose
strict event monotonicity, and we show strict iterated-elimination
monotonicity. Fix any $\theta ^{\prime }\in \Theta $, and we will define a
sequence $\left( \theta ^{1},\theta ^{2},...,\theta ^{n}\right) $
inductively. Define $\theta ^{n}=\theta ^{\prime }$, and apply strict event
monotonicity on $E=\Theta $. Given responsiveness, we have $\left\{ f\left(
\theta ^{\prime }\right) \right\} \neq f\left( E\right) $ and hence, strict
event monotonicity implies 
\begin{gather*}
u_{i}\left( f\left( \theta ^{1}\right) ,\text{ }\theta ^{1}\right)
>u_{i}\left( y,\text{ }\theta ^{1}\right) \text{ and }u_{i}\left( y,\text{ }%
\theta ^{\prime }\right) >u_{i}\left( f\left( \theta ^{1}\right) ,\text{ }%
\theta ^{\prime }\right) \text{, } \\
\text{for some }\left( \theta ^{1},y,i\right) \in \Theta \times Y\times 
\mathcal{I}^{\Theta }\text{.}
\end{gather*}%
Inductively, for each $k\in \left\{ 2,...,n-1\right\} $, apply strict group
monotonicity on 
\begin{equation*}
E=\Theta \diagdown \left\{ \theta ^{1},...,\theta ^{k-1}\right\} ,
\end{equation*}%
and we get%
\begin{gather*}
u_{i}\left( f\left( \theta ^{k}\right) ,\text{ }\theta ^{k}\right)
>u_{i}\left( y,\text{ }\theta ^{k}\right) \text{ and }u_{i}\left( y,\text{ }%
\theta ^{\prime }\right) >u_{i}\left( f\left( \theta ^{k}\right) ,\text{ }%
\theta ^{\prime }\right) \text{, } \\
\text{for some }\left( \theta ^{k},y,i\right) \in \left[ \Theta \diagdown
\left\{ \theta ^{1},...,\theta ^{k-1}\right\} \right] \times Y\times 
\mathcal{I}^{\Theta \diagdown \left\{ \theta ^{1},...,\theta ^{k-1}\right\} }%
\text{,}
\end{gather*}%
i.e., strict iterated-elimination monotonicity holds.

Second, suppose strict iterated-elimination monotonicity, and we show strict
event monotonicity. For any $\left( \theta ^{\prime },E\right) $ with $%
\left\{ f\left( \theta ^{\prime }\right) \right\} \neq f\left( E\right) $,
and we aim to show%
\begin{equation}
u_{i}\left( f\left( \theta \right) ,\text{ }\theta \right) >u_{i}\left( y,%
\text{ }\theta \right) \text{ and }u_{i}\left( y,\text{ }\theta \right)
>u_{i}\left( f\left( \theta \right) ,\text{ }\theta ^{\prime }\right) \text{%
, for some }\left( \theta ,y,i,\right) \in E\times Y\times \mathcal{I}^{E}%
\text{.}  \label{ccc4}
\end{equation}%
Given strict iterated-elimination monotonicity, there exists $\left( \theta
^{1},\theta ^{2},...,\theta ^{n}\right) $ such that%
\begin{gather*}
\left\{ \theta ^{1},\theta ^{2},...,\theta ^{n}\right\} =\Theta \text{,} \\
\theta ^{n}=\theta ^{\prime }\text{,}
\end{gather*}%
and for every $k\in \left\{ 1,2,...,n-1\right\} $,%
\begin{equation}
u_{i}\left( f\left( \theta ^{k}\right) ,\text{ }\theta ^{k}\right)
>u_{i}\left( y,\text{ }\theta ^{k}\right) \text{ and }u_{i}\left( y,\text{ }%
\theta ^{\prime }\right) >u_{i}\left( f\left( \theta ^{k}\right) ,\text{ }%
\theta ^{\prime }\right) \text{, for some }\left( y,i\right) \in Y\times 
\mathcal{I}^{\left\{ \theta ^{k},\theta ^{k+1},...,\theta ^{n}\right\} }%
\text{.}  \label{ccc3}
\end{equation}%
We write $E=\left\{ \theta ^{k_{1}},...\theta ^{k_{n}}\right\} \subset
\Theta $ with $k_{1}<...<k_{n}$ and $k_{1}<n$ due to $\left\{ f\left( \theta
^{\prime }\right) \right\} \neq f\left( E\right) $. Then, (\ref{ccc3})\
implies%
\begin{gather}
u_{i}\left( f\left( \theta ^{k_{1}}\right) ,\text{ }\theta ^{k_{1}}\right)
>u_{i}\left( y,\text{ }\theta ^{k_{1}}\right) \text{ and }u_{i}\left( y,%
\text{ }\theta ^{\prime }\right) >u_{i}\left( f\left( \theta ^{k_{1}}\right)
,\text{ }\theta ^{\prime }\right) \text{, }  \label{ccc5} \\
\text{for some }\left( y,i\right) \in Y\times \mathcal{I}^{\left\{ \theta
^{k_{1}},\theta ^{k_{1}+1},...,\theta ^{n}\right\} }\text{.}  \notag
\end{gather}%
Note that $\theta ^{k_{1}}\in E$ and $E=\left\{ \theta ^{k_{1}},...\theta
^{k_{n}}\right\} \subset \left\{ \theta ^{k_{1}},\theta
^{k_{1}+1},...,\theta ^{n}\right\} $ and hence, $\mathcal{I}^{\left\{ \theta
^{k_{1}},\theta ^{k_{1}+1},...,\theta ^{n}\right\} }\subset \mathcal{I}^{E}$%
. As a result, (\ref{ccc5})\ implies (\ref{ccc4}).$\blacksquare $

\subsection{Proof: the "if" part of Theorem \protect\ref{theorem:NWA}}

\label{sec:proof:theorem:NWA}

Suppose that $f$\ satisfies responsiveness, strict event monotonicity and
dictator monotonicity. As argued above, since $\mathcal{I}\times \Theta $ is
finite, there exists a finite set $\Sigma \subset Y$ such that%
\begin{equation*}
f\left( \Theta \right) \cup \left\{ z_{i}\left( \theta ,\theta ^{\prime
}\right) \in Y:\left( \theta ,\theta ^{\prime }\right) \in \Theta \times
\Theta \right\} \subset \Sigma \text{, }
\end{equation*}%
\begin{gather}
\text{and }\forall \left( \theta ^{\prime },E\right) \in \Theta \times \left[
2^{\Theta }\backslash \left\{ \varnothing \right\} \right] \text{,}
\label{tsn1} \\
\left\{ f\left( \theta ^{\prime }\right) \right\} \neq f\left( E\right) 
\text{ implies }\left[ 
\begin{array}{c}
u_{i}\left( f\left( \theta \right) ,\text{ }\theta \right) >u_{i}\left( y,%
\text{ }\theta \right) \text{ and }u_{i}\left( y,\text{ }\theta ^{\prime
}\right) >u_{i}\left( f\left( \theta \right) ,\text{ }\theta ^{\prime
}\right) \text{,} \\ 
\text{ for some }\left( \theta ,y,i,\right) \in E\times \Sigma \times 
\mathcal{I}^{E}\text{,}%
\end{array}%
\right] \text{,}  \notag
\end{gather}%
\begin{gather}
\text{and }\forall \left( i,\theta ,\theta ^{\prime },\theta ^{\prime \prime
}\right) \in \mathcal{I}\times \Theta \times \Theta \times \Theta ,  \notag
\\
\left[ 
\begin{array}{c}
\left\{ i\right\} =\mathcal{I}^{\theta }\text{ } \\ 
\text{and }f\left( \theta \right) \neq f\left( \theta ^{\prime }\right)%
\end{array}%
\right] \Longrightarrow \left[ 
\begin{array}{c}
\exists y\in \Sigma \text{ such that} \\ 
u_{i}\left( f\left( \theta ^{\prime \prime }\right) ,\text{ }\theta ^{\prime
\prime }\right) \geq u_{i}\left( y,\text{ }\theta ^{\prime \prime }\right) 
\text{ and }u_{i}\left( y,\text{ }\theta ^{\prime }\right) >u_{i}\left(
f\left( \theta \right) ,\text{ }\theta ^{\prime }\right)%
\end{array}%
\right] \text{,}  \label{tsn2} \\
\left[ 
\begin{array}{c}
\left\{ i\right\} =\mathcal{I}^{\theta }\text{ } \\ 
\text{and }f\left( \theta \right) \neq f\left( \theta ^{\prime }\right)%
\end{array}%
\right] \Longrightarrow u_{i}\left( f\left( \theta ^{\prime }\right) ,\text{ 
}\theta ^{\prime }\right) >u_{i}\left( f\left( \theta \right) ,\text{ }%
\theta ^{\prime }\right) \text{,}  \label{tsn3}
\end{gather}%
where $z_{i}\left( \theta ,\theta ^{\prime }\right) $ are defined in Lemma %
\ref{lem:y}, and (\ref{tsn1})\ follows from strict event monotonicity, and (%
\ref{tsn2}) from dictator monotonicity, and (\ref{tsn3}) from (\ref{tsn2})
by considering $\theta ^{\prime }=\theta ^{\prime \prime }$. That is, $%
\Sigma $ is a finite set which contains all of $f\left( \theta \right) $, $%
z_{i}\left( \theta ,\theta ^{\prime }\right) $ and all potential blocking
plans. In particular, when we take $E=\Theta $, we get $\left\{ f\left(
\theta ^{\prime }\right) \right\} \neq f\left( E\right) $, and (\ref{tsn1})
implies%
\begin{equation}
\mathcal{I}^{\Theta }\neq \varnothing \text{.}  \label{hjr1}
\end{equation}

We use $\mathcal{M}=\left\langle M=\left( M_{i}\right) _{i\in \mathcal{I}}%
\text{, \ }g:M\longrightarrow Y\right\rangle $ defined as follows to
implement $f$. In this mechanism, each agent $i$ sends a message $m_{i}=%
\left[ m_{i}^{1},m_{i}^{2},m_{i}^{3},m_{i}^{4}\right] \in M_{i}$, where 
\begin{eqnarray*}
m_{i}^{1} &\in &\Theta , \\
m_{i}^{2} &\in &\mathbb{N}, \\
m_{i}^{3} &\in &\Sigma ^{\Theta }, \\
m_{i}^{4} &\in &Z\text{.}
\end{eqnarray*}%
We partition $M$ as follows: agreement, unilateral deviation and
multi-lateral deviation.%
\begin{equation*}
M^{\prime }=\left\{ \left( m_{i}=\left[
m_{i}^{1},m_{i}^{2},m_{i}^{3},m_{i}^{4}\right] \right) _{i\in \mathcal{I}%
}\in M:\exists \theta \in \Theta \text{, }\left( m_{i}^{1},m_{i}^{2}\right)
=\left( \theta ,1\right) \text{, }\forall i\in \mathcal{I}^{\theta }\right\} 
\text{,}
\end{equation*}%
\begin{equation*}
M^{\prime \prime }=\left\{ m_{i}\in M\diagdown M^{\prime }:\exists \left(
\theta ,i\right) \in \Theta \times \mathcal{I}\text{, }\left(
m_{j}^{1},m_{j}^{2}\right) =\left( \theta ,1\right) \text{, }\forall j\in 
\mathcal{I\diagdown }\left\{ i\right\} \right\} \text{,}
\end{equation*}%
\begin{equation*}
M^{\prime \prime \prime }=M\diagdown \left( M^{\prime }\cup M^{\prime \prime
}\right) \text{.}
\end{equation*}%
It is worth noting that $\mathcal{I}^{\theta }$ is used in the definition of 
$M^{\prime }$, and $\mathcal{I}$ is used in the definition of $M^{\prime
\prime }$. Specifically, agreement is defined as all agents in $\mathcal{I}%
^{\theta }$ reporting $\left( \theta ,1\right) $ in the first two
dimensions. Furthermore, unilateral deviation refers to the unilateral
deviation from all agents in $\mathcal{I}$ reporting the same $\left( \theta
,1\right) $ for some $\theta $. Thus, a unilateral deviation from a message
profile in $M^{\prime }$ may induce a message profile in $M^{\prime \prime
\prime }$.

Then, $g$ is defined by the following rules.

\begin{quotation}
\textbf{Rule 1 (agreement): }when $m\in M^{\prime }$: there exists $\theta
\in \Theta $ such that $\left( m_{i}^{1},m_{i}^{2}\right) =\left( \theta
,1\right) $ for every $i\in \mathcal{I}^{\theta }$. By (\ref{hjr1}), such $%
\theta $ is unique. Then, $g$ picks $f\left( \theta \right) $;

\textbf{Rule 2 (unilateral deviation): }when $m\in M^{\prime \prime }$:
there exists $\left( \theta ,i\right) \in \Theta \times \mathcal{I}$ such
that $\left( m_{j}^{1},m_{j}^{2}\right) =\left( \theta ,1\right) $ for every 
$j\in \mathcal{I\diagdown }\left\{ i\right\} $, and such $\left( \theta
,i\right) $ is unique due to $\left\vert \mathcal{I}\right\vert \geq 3$. We
further distinguish two sub-cases:

\ \ \ \ \textbf{Rule (2.a):} if $u_{i}\left( f\left( \theta \right) ,\theta
\right) \geq u_{i}\left( m_{i}^{3}\left( \theta \right) ,\theta \right) $,
then $g$ picks $m_{i}^{3}\left( \theta \right) $ with probability $1-\frac{1%
}{m_{i}^{2}+1}$ and $g$ picks $z_{i}\left( \theta ,\theta \right) $ with
probability $\frac{1}{m_{i}^{2}+1}$;

\ \ \ \ \textbf{Rule (2.b):} if $u_{i}\left( f\left( \theta \right) ,\theta
\right) <u_{i}\left( m_{i}^{3}\left( \theta \right) ,\theta \right) $, then $%
g$ picks $z_{i}\left( \theta ,\theta \right) $;

\textbf{Rule 3 (multi-lateral deviation): }when $m\in M^{\prime \prime
\prime }$: consider agent $j=max\left[ \arg \max_{h\in \mathcal{I}}m_{h}^{2}%
\right] $, i.e., $j$ is the largest-numbered agent who report the largest
integer in the second dimension. Then, $g$ picks $m_{j}^{4}$ with
probability $1-\frac{1}{m_{j}^{2}+1}$ and $g$ picks $\underline{y}$ with
probability $\frac{1}{m_{j}^{2}+1}$.
\end{quotation}

From now on, fix any true state is $\theta ^{\ast }\in \Theta $. As above,
fix any%
\begin{gather*}
\widehat{m}_{i}^{3}\left( \theta \right) \in \arg \max_{y\in \left\{ 
\widetilde{y}\in \Sigma \text{ }:\text{ }u_{i}\left( f\left( \theta \right)
,\theta \right) \geq u_{i}\left( \widetilde{y},\theta \right) \right\}
}u_{i}\left( y,\theta ^{\ast }\right) \text{, }\forall \theta \in \Theta 
\text{,} \\
\widehat{m}_{i}^{3}\equiv \left[ \widehat{m}_{i}^{3}\left( \theta \right) %
\right] _{\theta \in \Theta }\text{,} \\
\widehat{m}_{i}^{4}\in \arg \max_{z\in Z}u_{i}\left( z,\theta ^{\ast
}\right) \text{.}
\end{gather*}%
Using the same argument as in Appendix \ref{sec:proof:theorem:non-responsive}%
, we can show%
\begin{gather}
u_{i}\left( \widehat{m}_{i}^{4},\theta ^{\ast }\right) \geq u_{i}\left( 
\widehat{m}_{i}^{3}\left( \theta \right) ,\theta ^{\ast }\right) >\frac{%
\left( n-1\right) \times u_{i}\left( y,\theta ^{\ast }\right) +u_{i}\left(
z_{i}\left( \theta ,\theta \right) ,\theta ^{\ast }\right) }{n}\text{,}
\label{teff1} \\
\forall i\in \mathcal{I}^{\theta ^{\ast }}\text{, }\forall \left( n,\theta
,y\right) \in \mathbb{N}\times \Theta \times \Sigma \text{ such that }%
u_{i}\left( f\left( \theta \right) ,\theta \right) \geq u_{i}\left( y,\theta
\right) \text{,}  \notag
\end{gather}%
\begin{gather}
\text{and }u_{i}\left( \widehat{m}_{i}^{4},\theta ^{\ast }\right) >\frac{%
n\times u_{i}\left( z,\theta ^{\ast }\right) +u_{i}\left( \underline{y}%
,\theta ^{\ast }\right) }{n+1}\text{,}  \label{teff2} \\
\forall i\in \mathcal{I}^{\theta ^{\ast }}\text{, }\forall \left( n,z\right)
\in \mathbb{N}\times Z\text{.}  \notag
\end{gather}

\medskip

In five steps, we now prove that $\mathcal{M}$ rationalizably implements $f$.

\textbf{Step 1: \ \ \ \ \ \ \ \ \ at true state }$\theta ^{\ast }\in \Theta $%
\textbf{, any }$\left[ m_{i}^{1},m_{i}^{2},m_{i}^{3},m_{i}^{4}\right] _{i\in 
\mathcal{I}}$\textbf{\ with }$\left( m_{i}^{1},m_{i}^{2}\right) =\left(
\theta ^{\ast },1\right) $\textbf{\ for every }$i\in I$\textbf{\ is a Nash
equilibrium, which induces }$f\left( \theta ^{\ast }\right) $\textbf{\ as
dictated by Rule 1.}

By following this strategy profile, agent $i$ gets payoff $u_{i}\left(
f\left( \theta ^{\ast }\right) ,\theta ^{\ast }\right) $. For any $i\notin 
\mathcal{I}^{\theta ^{\ast }}$, any deviation of $i$ would still induce $%
f\left( \theta ^{\ast }\right) $, i.e., not a profitable deviation. For any $%
i\in \mathcal{I}^{\theta ^{\ast }}$, consider any of $i$'s deviation $\left( 
\widehat{m}_{i}^{1},\widehat{m}_{i}^{2},\widehat{m}_{i}^{3},\widehat{m}%
_{i}^{4}\right) $ that changes the outcome chosen by $g$. This deviation
would either trigger Rule 1, when $\left( \widehat{m}_{i}^{1},\widehat{m}%
_{i}^{2}\right) =\left( \widehat{\theta },1\right) $ with $\widehat{\theta }%
\neq \theta ^{\ast }$ and $\left\{ i\right\} =\mathcal{I}^{\widehat{\theta }%
} $, or trigger Rule 2 otherwise. For the former case, this is not
profitable because of dictator monotonicity (precisely, (\ref{tsn3})). In
the latter case, $g$ picks either $z_{i}\left( \theta ^{\ast },\theta ^{\ast
}\right) $, or a mixture of $z_{i}\left( \theta ^{\ast },\theta ^{\ast
}\right) $ and $\widehat{m}_{i}^{3}\left( \theta ^{\ast }\right) $ with $%
u_{i}\left( f\left( \theta ^{\ast }\right) ,\theta ^{\ast }\right) \geq
u_{i}\left( m_{i}^{3}\left( \theta ^{\ast }\right) ,\theta ^{\ast }\right) $%
, all of which are worse than $f\left( \theta ^{\ast }\right) $ for $i$ at $%
\theta ^{\ast }$ by (\ref{cck2}) in Lemma \ref{lem:y}, i.e., not a
profitable deviation.

\medskip

\textbf{Step 2: \ \ \ \ \ \ \ \ \ at true state }$\theta ^{\ast }\in \Theta $%
\textbf{, for every }$i\in \mathcal{I}^{\theta ^{\ast }}$,\textbf{\ if any }$%
m_{i}\in M_{i}$\textbf{\ is a best reply to } $\lambda _{-i}\in \triangle
\left( M_{-i}\right) $\textbf{, then }$\left( m_{i},\lambda _{-i}\right) $%
\textbf{\ induces Rules 2 and 3 with probability 0.}

The proof is the same as Step 2 in Appendix \ref%
{sec:proof:theorem:non-responsive}, and we omit it.

\medskip

\textbf{Step 3: \ \ \ \ \ \ \ \ \ at true state }$\theta ^{\ast }\in \Theta $%
\textbf{, for any }$\overline{\theta }\in \Theta \diagdown \left\{ \theta
^{\ast }\right\} $\textbf{, there exists }$j\in \mathcal{I}^{\overline{%
\theta }}\cap \mathcal{I}^{\theta ^{\ast }}$\textbf{\ such that any }$m_{j}=%
\left[ m_{j}^{1},m_{j}^{2},m_{j}^{3},m_{j}^{4}\right] $\textbf{\ with }$%
\left( m_{j}^{1},m_{j}^{2}\right) =\left( \overline{\theta },1\right) $%
\textbf{\ is not rationalizable for agent }$j$\textbf{.}

By Proposition \ref{prop:NWA}, strict event monotonicity is equivalent to
strict iterated-elimination monotonicity, which means that there exists $%
\left( \theta ^{1},\theta ^{2},...,\theta ^{n}\right) $ such that%
\begin{gather*}
\left\{ \theta ^{1},\theta ^{2},...,\theta ^{n}\right\} =\Theta \text{,} \\
\theta ^{n}=\theta ^{\ast }\text{,}
\end{gather*}%
and for every $k\in \left\{ 1,2,...,n-1\right\} $,%
\begin{equation}
u_{j}\left( f\left( \theta ^{k}\right) ,\text{ }\theta ^{k}\right)
>u_{j}\left( y,\text{ }\theta ^{k}\right) \text{ and }u_{j}\left( y,\text{ }%
\theta ^{\ast }\right) >u_{j}\left( f\left( \theta ^{k}\right) ,\text{ }%
\theta ^{\ast }\right) \text{, for some }\left( y,j\right) \in \Sigma \times 
\mathcal{I}^{\left\{ \theta ^{k},\theta ^{k+1},...,\theta ^{n}\right\} }%
\text{.}  \label{ycy1b}
\end{equation}%
Then, inductively, for each $k\in \left\{ 1,2,...,n-1\right\} $, we will
show that it is not rationalizable for agent $j$ (identified in (\ref{ycy1b}%
)) to report $\left( \theta ^{k},1\right) $ in the first two dimensions.

Clearly, $j\in \mathcal{I}^{\left\{ \theta ^{k},\theta ^{k+1},...,\theta
^{n}\right\} }\subset \mathcal{I}^{\theta ^{n}}=\mathcal{I}^{\theta ^{\ast
}} $. Furthermore, (\ref{ycy1b}) immediately implies

\begin{equation}
u_{j}\left( \widehat{m}_{j}^{4},\text{ }\theta ^{\ast }\right) \geq
u_{j}\left( \widehat{m}_{j}^{3}\left( \theta ^{k}\right) ,\text{ }\theta
^{\ast }\right) \geq u_{j}\left( y,\text{ }\theta ^{\ast }\right)
>u_{j}\left( f\left( \theta ^{k}\right) ,\text{ }\theta ^{\ast }\right) 
\text{,}  \label{tttf}
\end{equation}%
where the strict inequality follows from (\ref{ycy1b}), the first weak
inequality from (\ref{teff1}), the second weak inequality from the
definition of $\widehat{m}_{j}^{3}\left( \theta ^{k}\right) $ and $y\in
\left\{ \widetilde{y}\in \Sigma :u_{i}\left( f\left( \theta ^{k}\right) ,%
\text{ }\theta ^{k}\right) \geq u_{i}\left( \widetilde{y},\theta ^{k}\right)
\right\} $.

We now consider two cases: (a) $\left\{ j\right\} =\mathcal{I}^{\theta ^{k}}$%
, (b) $\left\{ j\right\} \neq \mathcal{I}^{\theta ^{k}}$.

\textbf{Step 3.a: \ \ \ \ \ \ \ \ \ when }$\left\{ j\right\} =\mathcal{I}%
^{\theta ^{k}}$\textbf{, any }$m_{j}=\left(
m_{j}^{1},m_{j}^{2},m_{j}^{3},m_{j}^{4}\right) $\textbf{\ with }$\left(
m_{j}^{1},m_{j}^{2}\right) =\left( \theta ^{k},1\right) $\textbf{\ is not
rationalizable for }$j$\textbf{.}

We prove it by contradiction. Given $\left\{ j\right\} =\mathcal{I}^{\theta
^{k}}$, suppose $m_{j}=\left( m_{j}^{1},m_{j}^{2},m_{j}^{3},m_{j}^{4}\right) 
$\ with $\left( m_{j}^{1},m_{j}^{2}\right) =\left( \theta ^{k},1\right) $\
is rationalizable for agent $j$. Then, $m_{j}$ is a best reply to some $%
\lambda _{-j}\in \triangle \left( S_{-j}^{\mathcal{M},\text{ }\theta ^{\ast
}}\right) $. Since $\left\{ j\right\} =\mathcal{I}^{\theta ^{k}}$, we reach
agreement (i.e., Rule 1) under the strategy profile $\left( m_{j},\lambda
_{-j}\right) $, which induces the outcome $f\left( \theta ^{k}\right) $.

Given $\theta ^{k}\neq \theta ^{\ast }$ and responsiveness, dictator
monotonicity (i.e., (\ref{tsn2})) implies%
\begin{gather*}
\forall \theta ^{\prime \prime }\in \Theta ,\text{ }\exists y\in \Sigma 
\text{ such that} \\
u_{i}\left( f\left( \theta ^{\prime \prime }\right) ,\text{ }\theta ^{\prime
\prime }\right) \geq u_{i}\left( y,\text{ }\theta ^{\prime \prime }\right) 
\text{ and }u_{i}\left( y,\text{ }\theta ^{\ast }\right) >u_{i}\left(
f\left( \theta ^{k}\right) ,\text{ }\theta ^{\ast }\right) \text{,}
\end{gather*}%
which further implies%
\begin{equation}
u_{j}\left( \widehat{m}_{j}^{3}\left( \theta ^{\prime \prime }\right) ,\text{
}\theta ^{\ast }\right) >u_{j}\left( f\left( \theta ^{k}\right) ,\text{ }%
\theta ^{\ast }\right) \text{, }\forall \theta ^{\prime \prime }\in \Theta 
\text{,}  \label{gtew1}
\end{equation}%
i.e., agent $j$ always finds it profitable to use $\widehat{m}_{j}^{3}$ to
deviate to Rule 2. Also, (\ref{tttf})\ implies%
\begin{equation}
u_{j}\left( \widehat{m}_{j}^{4},\text{ }\theta ^{\ast }\right) >u_{j}\left(
f\left( \theta ^{k}\right) ,\text{ }\theta ^{\ast }\right) \text{,}
\label{gtew2}
\end{equation}%
i.e., whenever possible, agent $j$ always finds it profitable to use the
blocking plan $\widehat{m}_{j}^{4}$ to deviate to Rule 3. Then, agent $j$
would like to deviate from $m_{j}$ to $\left[ m_{j}^{1},\widetilde{m}%
_{j}^{2},\widehat{m}_{j}^{3},\widehat{m}_{j}^{4}\right] $ with $\widetilde{m}%
_{j}^{2}\rightarrow \infty $, which would induce either Rule 2 or Rule 3 and%
\begin{equation*}
u_{j}\left( g\left( m_{j},\lambda _{-j}\right) ,\theta ^{\ast }\right)
=u_{j}\left( f\left( \theta ^{k}\right) ,\text{ }\theta ^{\ast }\right)
<\lim_{\widetilde{m}_{j}^{2}\rightarrow \infty }u_{i}\left( g\left( \left[
m_{j}^{1},\widetilde{m}_{j}^{2},\widehat{m}_{j}^{3},\widehat{m}_{j}^{4}%
\right] ,\lambda _{-j}\right) ,\theta ^{\ast }\right)
\end{equation*}%
where the inequality follows from (\ref{gtew1}) and (\ref{gtew2}). Thus,
there exists $\widehat{m}_{j}^{2}\in \mathbb{N}$ such that%
\begin{equation*}
u_{j}\left( g\left( m_{j},\lambda _{-j}\right) ,\theta ^{\ast }\right)
=u_{j}\left( f\left( \theta ^{k}\right) ,\text{ }\theta ^{\ast }\right)
<u_{i}\left( g\left( \left[ m_{j}^{1},\widehat{m}_{j}^{2},\widehat{m}%
_{j}^{3},\widehat{m}_{j}^{4}\right] ,\lambda _{-j}\right) ,\theta ^{\ast
}\right)
\end{equation*}%
contradicting $m_{j}$ being a best reply to $\lambda _{-j}$.

\textbf{Step 3.b: \ \ \ \ \ \ \ \ \ when }$\left\{ j\right\} \neq \mathcal{I}%
^{\theta ^{k}}$\textbf{, any strategy }$m_{j}=\left(
m_{j}^{1},m_{j}^{2},m_{j}^{3},m_{j}^{4}\right) $\textbf{\ with }$\left(
m_{j}^{1},m_{j}^{2}\right) =\left( \theta ^{k},1\right) $\textbf{\ is not
rationalizable for agent }$j$\textbf{.}

We prove it by contradiction. Suppose $m_{j}=\left(
m_{j}^{1},m_{j}^{2},m_{j}^{3},m_{j}^{4}\right) $\ with $\left(
m_{j}^{1},m_{j}^{2}\right) =\left( \theta ^{k},1\right) $\ is rationalizable
for agent $j$. Then, $m_{j}$ is a best reply to some $\lambda _{-j}\in
\triangle \left( S_{-j}^{\mathcal{M},\text{ }\theta ^{\ast }}\right) $.

Recall $j\in \mathcal{I}^{\left\{ \theta ^{k},\theta ^{k+1},...,\theta
^{I}\right\} }\subset \mathcal{I}^{\theta ^{k}}\cap \mathcal{I}^{\theta
^{\ast }}$. By Step 2, the strategy profile $\left( m_{j},\lambda
_{-j}\right) $ induces Rule 1 with probability 1. Given $\left(
m_{j}^{1},m_{j}^{2}\right) =\left( \theta ^{k},1\right) $, all agents in $%
\mathcal{I}^{\theta ^{k}}$ must report $\left( \theta ^{k},1\right) $ under $%
\left( m_{j},\lambda _{-j}\right) $.\footnote{%
By the induction hypothesis, we cannot reach agreement on any state in $%
\left\{ \theta ^{1},\theta ^{2},...,\theta ^{k-1}\right\} $.} With $\mathcal{%
I}^{\theta ^{k}}\diagdown \left\{ j\right\} \neq \varnothing $, we consider
two sub-cases (1) agents $-j$ all report $\left( \theta ^{k},1\right) $ and
(2) otherwise. In Case (1) agent $j$ can deviate to Rule 2 and in Case (2)
agent $j$ can deviate to Rule 3. Note that (\ref{tttf}) implies 
\begin{eqnarray}
u_{j}\left( \widehat{m}_{j}^{3}\left( \theta ^{k}\right) ,\text{ }\theta
^{\ast }\right) &>&u_{j}\left( f\left( \theta ^{k}\right) ,\text{ }\theta
^{\ast }\right) \text{,}  \label{txw1} \\
u_{j}\left( \widehat{m}_{j}^{4},\text{ }\theta ^{\ast }\right)
&>&u_{j}\left( f\left( \theta ^{k}\right) ,\text{ }\theta ^{\ast }\right) 
\text{,}  \label{txw2}
\end{eqnarray}%
i.e., (\ref{txw1}) says that, in Case (1), agent $j$ always finds it
profitable to use the blocking plan $\widehat{m}_{j}^{3}$ to deviate to Rule
2; (\ref{txw2}) says that, in Case (2), agent $j$ always finds it profitable
to use the blocking plan $\widehat{m}_{j}^{4}$ to deviate to Rule 3. Thus,
agent $j$ would like to deviate from $m_{j}$ to $\left[ m_{j}^{1},\widetilde{%
m}_{j}^{2},\widehat{m}_{j}^{3},\widehat{m}_{j}^{4}\right] $ with $\widetilde{%
m}_{j}^{2}\rightarrow \infty $, which would induce either Rule 2 or Rule 3
and%
\begin{equation*}
u_{j}\left( g\left( m_{j},\lambda _{-j}\right) ,\theta ^{\ast }\right)
=u_{j}\left( f\left( \theta ^{k}\right) ,\text{ }\theta ^{\ast }\right)
<\lim_{\widetilde{m}_{j}^{2}\rightarrow \infty }u_{i}\left( g\left( \left[
m_{j}^{1},\widetilde{m}_{j}^{2},\widehat{m}_{j}^{3},\widehat{m}_{j}^{4}%
\right] ,\lambda _{-j}\right) ,\theta ^{\ast }\right)
\end{equation*}%
where the inequality follows from (\ref{txw1}) and (\ref{txw2}). Thus, there
exists $\widehat{m}_{j}^{2}\in \mathbb{N}$ such that%
\begin{equation*}
u_{j}\left( g\left( m_{j},\lambda _{-j}\right) ,\theta ^{\ast }\right)
=u_{j}\left( f\left( \theta ^{k}\right) ,\text{ }\theta ^{\ast }\right)
<u_{i}\left( g\left( \left[ m_{j}^{1},\widehat{m}_{j}^{2},\widehat{m}%
_{j}^{3},\widehat{m}_{j}^{4}\right] ,\lambda _{-j}\right) ,\theta ^{\ast
}\right)
\end{equation*}%
contradicting $m_{j}$ being a best reply to $\lambda _{-j}$.

\medskip

\textbf{Step 4: \ \ \ \ \ \ \ \ \ at true state }$\theta ^{\ast }\in \Theta $%
\textbf{, for every }$i\in \mathcal{I}^{\Theta }$\textbf{, any strategy }$%
m_{i}=\left( m_{i}^{1},m_{i}^{2},m_{i}^{3},m_{i}^{4}\right) $\textbf{\ with }%
$\left( m_{i}^{1},m_{i}^{2}\right) \neq \left( \theta ^{\ast },1\right) $%
\textbf{\ is not rationalizable for agent }$i$\textbf{.}

For any $i\in \mathcal{I}^{\Theta }$, pick any $m_{i}=\left(
m_{i}^{1},m_{i}^{2},m_{i}^{3},m_{i}^{4}\right) \in S_{i}^{\mathcal{M},\text{ 
}\theta ^{\ast }}$. Then, $m_{i}$ is a best reply to some $\lambda _{-i}\in
\triangle \left( S_{-i}^{\mathcal{M},\text{ }\theta ^{\ast }}\right) $%
\textbf{. }By Step 2, $\left( m_{i},\lambda _{-i}\right) $ induces Rule 1
with probability 1. However, by Step 3, any agreement on $\overline{\theta }%
\in \Theta \diagdown \left\{ \theta ^{\ast }\right\} $ cannot be reached.
Thus, the only possible is agreement on $\theta ^{\ast }$, i.e. all agents
in $\mathcal{I}^{\theta ^{\ast }}$ report $\left( \theta ^{\ast },1\right) $%
. As a result, only $m_{i}=\left(
m_{i}^{1},m_{i}^{2},m_{i}^{3},m_{i}^{4}\right) $ with $\left(
m_{i}^{1},m_{i}^{2}\right) =\left( \theta ^{\ast },1\right) $ is
rationalizable for $i\in \mathcal{I}^{\Theta }$ at $\theta ^{\ast }$.

\medskip

\textbf{Step 5: \ \ \ \ \ \ \ \ \ at true state }$\theta ^{\ast }\in \Theta $%
\textbf{, for every }$i\in \mathcal{I}^{\theta ^{\ast }}$\textbf{, any
strategy }$m_{i}=\left( m_{i}^{1},m_{i}^{2},m_{i}^{3},m_{i}^{4}\right) $%
\textbf{\ with }$\left( m_{i}^{1},m_{i}^{2}\right) \neq \left( \theta ^{\ast
},1\right) $\textbf{\ is not rationalizable for agent }$i$\textbf{.}

For any $i\in \mathcal{I}^{\theta ^{\ast }}$, pick any $m_{i}=\left(
m_{i}^{1},m_{i}^{2},m_{i}^{3},m_{i}^{4}\right) \in S_{i}^{\mathcal{M},\text{ 
}\theta ^{\ast }}$. Then, $m_{i}$ is a best reply to some $\lambda _{-i}\in
\triangle \left( S_{-i}^{\mathcal{M},\text{ }\theta ^{\ast }}\right) $%
\textbf{. }By Step 2, $\left( m_{i},\lambda _{-i}\right) $ induces Rule 1
with probability 1. Since $\mathcal{I}^{\Theta }\neq \varnothing $, by Step
4, it must be the case that all agents in $\mathcal{I}^{\Theta }$ report $%
\left( \theta ^{\ast },1\right) $. Thus, the only possibility is agreement
on $\theta ^{\ast }$, i.e. all agents in $\mathcal{I}^{\theta ^{\ast }}$
report $\left( \theta ^{\ast },1\right) $. Therefore, only $m_{i}=\left(
m_{i}^{1},m_{i}^{2},m_{i}^{3},m_{i}^{4}\right) $ with $\left(
m_{i}^{1},m_{i}^{2}\right) =\left( \theta ^{\ast },1\right) $ is
rationalizable for agent $i\in \mathcal{I}^{\theta ^{\ast }}$ at $\theta
^{\ast }$.

\medskip

To sum, Step 1 shows%
\begin{equation*}
S^{\mathcal{M},\text{ }\theta ^{\ast }}\supset \dprod\limits_{i\in \mathcal{I%
}}\left\{ \left( m_{i}^{1},m_{i}^{2},m_{i}^{3},m_{i}^{4}\right) \in
M_{i}:\left( m_{i}^{1},m_{i}^{2}\right) =\left( \theta ^{\ast },1\right)
\right\} \text{,}
\end{equation*}%
and Steps 2-5 show%
\begin{equation*}
S^{\mathcal{M},\text{ }\theta ^{\ast }}\subset \left( \dprod\limits_{i\in 
\mathcal{I}^{\theta ^{\ast }}}\left\{ \left(
m_{i}^{1},m_{i}^{2},m_{i}^{3},m_{i}^{4}\right) \in M_{i}:\left(
m_{i}^{1},m_{i}^{2}\right) =\left( \theta ^{\ast },1\right) \right\} \right)
\times \left( \dprod\limits_{i\in \mathcal{I}\diagdown \mathcal{I}^{\theta
^{\ast }}}M_{j}\right) \text{.}
\end{equation*}%
As a result, we have $g\left( S^{\mathcal{M},\text{ }\theta ^{\ast }}\right)
=\left\{ f\left( \theta ^{\ast }\right) \right\} $, i.e., rationalizable
implementation is achieved.$\blacksquare $

\subsection{The best-response property}

\label{sec:the-best-reply}

Given NWA, \cite{bmt} prove that strict Maskin monotonicity$^{\ast }$
suffices for rationalizable implementation, and given an additional
assumption called "the best-response property," strict Maskin monotonicity$%
^{\ast }$ is also necessary. However, the following example shows that it
suffers loss of generality to impose the best-response property.%
\begin{gather*}
\begin{tabular}{c}
Example 4: $\mathcal{I}=\left\{ i_{1},i_{2},i_{3},i_{4}\right\} $, $\Theta
=\left\{ \theta _{1},\theta _{2},\theta _{3},\theta _{4}\right\} $, $%
Z=\left\{ a,b,c\right\} $, \\ 
\multicolumn{1}{l}{$%
\begin{tabular}{ccc}
$f\left( \theta _{1}\right) =a$, & $f\left( \theta _{2}\right) =a$, & $%
f\left( \theta _{3}\right) =a$. \\ 
$%
\begin{tabular}{|c|c|c|c|}
\hline
$u_{i}\left( z,\theta _{1}\right) $ & $i=1$ & $=2$ & $=3$ \\ \hline
$z=a$ & $0$ & $0$ & $1$ \\ \hline
$z=b$ & $-1$ & $1$ & $0$ \\ \hline
$z=c$ & $1$ & $-1$ & $-1$ \\ \hline
\end{tabular}%
$ & $%
\begin{tabular}{|c|c|c|c|}
\hline
$u_{i}\left( z,\theta _{2}\right) $ & $i=1$ & $=2$ & $=3$ \\ \hline
$z=a$ & $0$ & $0$ & $1$ \\ \hline
$z=b$ & $1$ & $-1$ & $0$ \\ \hline
$z=c$ & $-1$ & $1$ & $-1$ \\ \hline
\end{tabular}%
$ & $%
\begin{tabular}{|c|c|c|c|}
\hline
$u_{i}\left( z,\theta _{3}\right) $ & $i=1$ & $=2$ & $=3$ \\ \hline
$z=a$ & $1$ & $1$ & $1$ \\ \hline
$z=b$ & $0$ & $0$ & $0$ \\ \hline
$z=c$ & $-1$ & $-1$ & $-1$ \\ \hline
\end{tabular}%
$%
\end{tabular}%
$}%
\end{tabular}
\\
\begin{tabular}{c}
$f\left( \theta _{4}\right) =c$. \\ 
$%
\begin{tabular}{|c|c|c|c|}
\hline
$u_{i}\left( z,\theta _{4}\right) $ & $i=1$ & $=2$ & $=3$ \\ \hline
$z=a$ & $-1$ & $0$ & $1$ \\ \hline
$z=b$ & $0$ & $-1$ & $-1$ \\ \hline
$z=c$ & $1$ & $1$ & $0$ \\ \hline
\end{tabular}%
$%
\end{tabular}%
\end{gather*}

In this example, $f$ satisfies strict Maskin monotonicity$^{\ast \ast }$ but
not strict Maskin monotonicity$^{\ast }$. To see this, note%
\begin{equation*}
\mathcal{L}_{i}^{\circ }\left[ a,\text{ }\theta _{1}\right] \cup \mathcal{L}%
_{i}^{\circ }\left[ a,\text{ }\theta _{2}\right] \subset \triangle \left(
Z\right) =\mathcal{L}_{i}\left[ a,\text{ }\theta _{3}\right] \text{, }%
\forall i\in \left\{ 1,2,3\right\} \text{,}
\end{equation*}%
i.e., at the true state $\theta _{3}$, when all agents report $\theta _{1}$
(or $\theta _{2}$), no whistle blower exists. As a result, the only possible
partition in both strict Maskin monotonicity$^{\ast }$ and strict Maskin
monotonicity$^{\ast \ast }$ is $\mathcal{P}\left( \theta \right) \equiv 
\mathcal{P}_{f}\left( \theta \right) $, i.e.,%
\begin{equation*}
\mathcal{P}\left( \theta _{1}\right) =\mathcal{P}\left( \theta _{2}\right) =%
\mathcal{P}\left( \theta _{3}\right) =\left\{ \theta _{1},\theta _{2},\theta
_{3}\right\} \text{ and }\mathcal{P}\left( \theta _{4}\right) =\left\{
\theta _{4}\right\} \text{.}
\end{equation*}%
To see why strict Maskin monotonicity$^{\ast }$ fails, suppose that the true
state is $\theta _{4}$ and that all agents report $\theta _{1}$. Then, agent
3 cannot be a whistle-blower because $f\left[ \mathcal{P}\left( \theta
_{1}\right) \right] =\left\{ a\right\} $ and $a$ is the unique best outcome
for agent $3$ at $\theta _{4}$, i.e., no blocking plan exists for agent $3$.
Furthermore, neither agent 1 nor agent 2 can be a whistle blower as required
by strict Maskin monotonicity$^{\ast }$ because%
\begin{gather*}
\mathcal{L}_{1}^{\circ }\left[ f\left( \theta _{1}\right) =a,\text{ }\theta
_{1}\right] =\left\{ \gamma ^{1}a+\gamma ^{2}b+\gamma ^{3}c\in \triangle
\left( Z\right) :\gamma ^{2}>\gamma ^{3}\text{ and }\gamma ^{1}<1\right\} 
\text{,} \\
\mathcal{L}_{2}^{\circ }\left[ f\left( \theta _{1}\right) =a,\text{ }\theta
_{1}\right] =\left\{ \gamma ^{1}a+\gamma ^{2}b+\gamma ^{3}c\in \triangle
\left( Z\right) :\gamma ^{2}<\gamma ^{3}\text{ and }\gamma ^{1}<1\right\} 
\text{,} \\
\mathcal{L}_{1}^{\circ }\left[ f\left( \theta _{2}\right) =a,\text{ }\theta
_{2}\right] =\left\{ \gamma ^{1}a+\gamma ^{2}b+\gamma ^{3}c\in \triangle
\left( Z\right) :\gamma ^{2}<\gamma ^{3}\text{ and }\gamma ^{1}<1\right\} 
\text{,} \\
\mathcal{L}_{2}^{\circ }\left[ f\left( \theta _{2}\right) =a,\text{ }\theta
_{2}\right] =\left\{ \gamma ^{1}a+\gamma ^{2}b+\gamma ^{3}c\in \triangle
\left( Z\right) :\gamma ^{2}>\gamma ^{3}\text{ and }\gamma ^{1}<1\right\} 
\text{,} \\
\text{and hence, }\mathcal{L}_{1}^{\circ }\left[ a,\text{ }\theta _{1}\right]
\cap \mathcal{L}_{1}^{\circ }\left[ a,\text{ }\theta _{2}\right] =\mathcal{L}%
_{2}^{\circ }\left[ a,\text{ }\theta _{1}\right] \cap \mathcal{L}_{2}^{\circ
}\left[ a,\text{ }\theta _{2}\right] =\varnothing \text{.}
\end{gather*}

Nevertheless, strict Maskin monotonicity$^{\ast \ast }$ holds. At true state 
$\theta _{4}$, if all agents report states in $\mathcal{P}\left( \theta
_{1}\right) $, agent 1 can act as a whistle-blower with the following
state-contingent blocking plan $\left[ \theta _{1}\text{: }b\text{; \ }%
\theta _{2}\text{: }c\text{; \ }\theta _{3}\text{: }c\right] $ because%
\begin{eqnarray*}
u_{1}\left( a,\theta _{1}\right) &>&u_{1}\left( b,\theta _{1}\right) \text{
and }u_{1}\left( b,\theta _{4}\right) >u_{1}\left( a,\theta _{4}\right) 
\text{,} \\
u_{1}\left( a,\theta _{2}\right) &>&u_{1}\left( c,\theta _{2}\right) \text{
and }u_{1}\left( c,\theta _{4}\right) >u_{1}\left( a,\theta _{4}\right) 
\text{,} \\
u_{1}\left( a,\theta _{3}\right) &>&u_{1}\left( c,\theta _{3}\right) \text{
and }u_{1}\left( c,\theta _{4}\right) >u_{1}\left( a,\theta _{4}\right) 
\text{.}
\end{eqnarray*}%
Furthermore, at true state $\theta _{1}$ or $\theta _{2}$ or $\theta _{3}$,
if all agents report $\theta _{4}$, agent 3 can act as a whistle-blower and
block false reporting by the blocking plan of "$b$," because%
\begin{eqnarray*}
u_{3}\left( c,\theta ^{IV}\right) &>&u_{3}\left( b,\theta ^{IV}\right) \text{
and }u_{3}\left( b,\theta ^{\prime }\right) >u_{3}\left( c,\theta ^{\prime
}\right) \text{,} \\
u_{3}\left( c,\theta ^{IV}\right) &>&u_{3}\left( b,\theta ^{IV}\right) \text{
and }u_{3}\left( b,\theta ^{\prime \prime }\right) >u_{3}\left( c,\theta
^{\prime \prime }\right) \text{,} \\
u_{3}\left( c,\theta ^{IV}\right) &>&u_{3}\left( b,\theta ^{IV}\right) \text{
and }u_{3}\left( b,\theta ^{\prime \prime \prime }\right) >u_{3}\left(
c,\theta ^{\prime \prime \prime }\right) \text{.}
\end{eqnarray*}%
By Theorem \ref{theorem:non-responsive}, we can achieve rationalizable
implementation in this example, even though strict Maskin monotonicity$%
^{\ast }$ fails.

\subsection{Rationalizable implementation Vs Nash implementation}

\label{sec:Nash-rationalizability}

Our results imply that rationalizable implementation is stronger than Nash
implementation. In this section, we use examples to illustrate their
difference. To simplify the problem, we impose only responsiveness in
Appendix \ref{sec:responsive-appendix} and only NWA in Appendix \ref%
{sec:NWA-appendix}.

\subsubsection{Responsiveness}

\label{sec:responsive-appendix}

In this subsection, we impose resonsiveness, but not NWA.

\paragraph{A simple example}

We first provide a very simple example in which we can achieve Nash
implementation but not rationalizable implementation.

\begin{equation*}
\begin{tabular}{c}
Example 5: $\mathcal{I}=\left\{ i_{1},i_{2},i_{3},i_{4}\right\} $, $\Theta
=\left\{ \theta _{1},\theta _{2}\right\} $, $Z=\left\{ a,b\right\} $, \\ 
\multicolumn{1}{l}{$%
\begin{tabular}{cc}
$f\left( \theta _{1}\right) =a$, & $f\left( \theta _{2}\right) =b$, \\ 
$%
\begin{tabular}{|c|c|c|c|c|}
\hline
$u_{i}\left( z,\theta _{1}\right) $ & $i_{1}$ & $i_{2}$ & $i_{3}$ & $i_{4}$
\\ \hline
$z=a$ & $0$ & $0$ & $1$ & $1$ \\ \hline
$z=b$ & $1$ & $1$ & $0$ & $0$ \\ \hline
\end{tabular}%
$ & $%
\begin{tabular}{|c|c|c|c|c|}
\hline
$u_{i}\left( z,\theta _{2}\right) $ & $i_{1}$ & $i_{2}$ & $i_{3}$ & $i_{4}$
\\ \hline
$z=a$ & $1$ & $1$ & $0$ & $0$ \\ \hline
$z=b$ & $0$ & $0$ & $1$ & $1$ \\ \hline
\end{tabular}%
$%
\end{tabular}%
$}%
\end{tabular}%
\end{equation*}%
In Example 1, Maskin monotonicity and no-veto power hold, and hence, we can
achive Nash implementation. However, 
\begin{equation*}
\mathcal{I}^{\theta _{1}}=\left\{ i_{3},i_{4}\right\} \text{, }\mathcal{I}%
^{\theta _{2}}=\left\{ i_{1},i_{2}\right\} \text{, }\mathcal{I}^{\Theta }=%
\mathcal{I}^{\theta _{1}}\cap \mathcal{I}^{\theta _{2}}=\varnothing \text{.}
\end{equation*}%
Our results imply that $\mathcal{I}^{\Theta }\neq \varnothing $ is a
necessary condition for rationalizable implementation. Therefore, we cannot
achieve rationalizable implementation in Example 5.

\paragraph{An economic example}

We consider the Condorcet function which is defined as follows.

\begin{define}
$z\in Z$ is Condorcet winner at $\theta \in \Theta $ if 
\begin{equation*}
\left\vert \left\{ i\in \mathcal{I}:u_{i}\left( z,\text{ }\theta \right)
>u_{i}\left( z^{\prime },\text{ }\theta \right) \right\} \right\vert
>\left\vert \left\{ i\in \mathcal{I}:u_{i}\left( z,\text{ }\theta \right)
<u_{i}\left( z^{\prime },\text{ }\theta \right) \right\} \right\vert \text{, 
}\forall z^{\prime }\in Z\diagdown \left\{ z\right\} \text{.}
\end{equation*}%
An SCF $f:\Theta \longrightarrow Z$ is a Condorcet function if for every $%
\theta \in \Theta $, $f\left( \theta \right) $ is a condorcet winner at $%
\theta $.
\end{define}

In the following example, $f^{C}$ is the Condorcet function.

\begin{equation*}
\begin{tabular}{c}
Example 6: $\mathcal{I}=\left\{ i_{1},i_{2},i_{3}\right\} $, $\Theta
=\left\{ \theta _{1},\theta _{2},\theta _{3}\right\} $, $Z=\left\{
a,b,c\right\} $, \\ 
\begin{tabular}{c}
state $\theta _{1}$ with $f^{C}\left( \theta _{1}\right) =a$, \\ 
$%
\begin{tabular}{|c|c|c|c|}
\hline
$u_{i}\left( z,\theta _{1}\right) $ & $i_{1}$ & $i_{2}$ & $i_{3}$ \\ \hline
$z=a$ & $0$ & $2$ & $2$ \\ \hline
$z=b$ & $1$ & $1$ & $1$ \\ \hline
$z=c$ & $2$ & $0$ & $0$ \\ \hline
\end{tabular}%
$%
\end{tabular}%
\begin{tabular}{c}
state $\theta _{2}$ with $f^{C}\left( \theta _{2}\right) =b$, \\ 
$%
\begin{tabular}{|c|c|c|c|}
\hline
$u_{i}\left( z,\theta _{2}\right) $ & $i_{1}$ & $i_{2}$ & $i_{3}$ \\ \hline
$z=a$ & $0$ & $2$ & $0$ \\ \hline
$z=b$ & $2$ & $0$ & $2$ \\ \hline
$z=c$ & $1$ & $1$ & $1$ \\ \hline
\end{tabular}%
$%
\end{tabular}%
\begin{tabular}{c}
state $\theta _{3}$ with $f^{C}\left( \theta _{3}\right) =c$, \\ 
$%
\begin{tabular}{|c|c|c|c|}
\hline
$u_{i}\left( z,\theta _{3}\right) $ & $i_{1}$ & $i_{2}$ & $i_{3}$ \\ \hline
$z=a$ & $1$ & $1$ & $1$ \\ \hline
$z=b$ & $0$ & $0$ & $2$ \\ \hline
$z=c$ & $2$ & $2$ & $0$ \\ \hline
\end{tabular}%
$%
\end{tabular}%
\end{tabular}%
\end{equation*}

In Example 6, Maskin monotonicity and no-veto power hold, and hence, we can
achive Nash implementation. However, 
\begin{equation*}
\mathcal{I}^{\theta _{1}}=\left\{ i_{2},i_{3}\right\} \text{, }\mathcal{I}%
^{\theta _{2}}=\left\{ i_{1},i_{3}\right\} \text{, }\mathcal{I}^{\theta
_{3}}=\left\{ i_{1},i_{2}\right\} \text{, }\mathcal{I}^{\Theta }=\mathcal{I}%
^{\theta _{1}}\cap \mathcal{I}^{\theta _{2}}\cap \mathcal{I}^{\theta
_{3}}=\varnothing \text{.}
\end{equation*}%
Since $\mathcal{I}^{\Theta }\neq \varnothing $ is a necessary condition for
rationalizable implementation, we cannot achieve rationalizable
implementation in Example 6.

\subsubsection{NWA: an economic example}

\label{sec:NWA-appendix}

In this subsection, we impose NWA, but resonsiveness. In the following
example, $f^{C}$ is the Condorcet function.%
\begin{equation*}
\begin{tabular}{c}
Example 7: $\mathcal{I}=\left\{ i_{1},i_{2},i_{3}\right\} $, $\Theta
=\left\{ \theta _{1},\theta _{2},\theta _{3},\theta _{4}\right\} $, $%
Z=\left\{ a,b,c\right\} $, \\ 
\multicolumn{1}{l}{$%
\begin{tabular}{ccc}
$f^{C}\left( \theta _{1}\right) =a$, & $f^{C}\left( \theta _{2}\right) =a$,
& $f^{C}\left( \theta _{3}\right) =a$. \\ 
$%
\begin{tabular}{|c|c|c|c|}
\hline
$u_{i}\left( z,\theta _{1}\right) $ & $i_{1}$ & $i_{2}$ & $i_{3}$ \\ \hline
$z=a$ & $1$ & $1$ & $2$ \\ \hline
$z=b$ & $0$ & $2$ & $1$ \\ \hline
$z=c$ & $2$ & $0$ & $0$ \\ \hline
\end{tabular}%
$ & $%
\begin{tabular}{|c|c|c|c|}
\hline
$u_{i}\left( z,\theta _{2}\right) $ & $i_{1}$ & $i_{2}$ & $i_{3}$ \\ \hline
$z=a$ & $2$ & $1$ & $1$ \\ \hline
$z=b$ & $1$ & $0$ & $2$ \\ \hline
$z=c$ & $0$ & $2$ & $0$ \\ \hline
\end{tabular}%
$ & $%
\begin{tabular}{|c|c|c|c|}
\hline
$u_{i}\left( z,\theta _{3}\right) $ & $i_{1}$ & $i_{2}$ & $i_{3}$ \\ \hline
$z=a$ & $2$ & $2$ & $2$ \\ \hline
$z=b$ & $0$ & $0$ & $0$ \\ \hline
$z=c$ & $1$ & $1$ & $1$ \\ \hline
\end{tabular}%
$%
\end{tabular}%
$} \\ 
$%
\begin{tabular}{c}
$f^{C}\left( \theta _{4}\right) =c$. \\ 
$%
\begin{tabular}{|c|c|c|c|}
\hline
$u_{i}\left( z,\theta _{4}\right) $ & $i_{1}$ & $i_{2}$ & $i_{3}$ \\ \hline
$z=a$ & $1$ & $1$ & $2$ \\ \hline
$z=b$ & $0$ & $0$ & $0$ \\ \hline
$z=c$ & $2$ & $2$ & $1$ \\ \hline
\end{tabular}%
$%
\end{tabular}%
$%
\end{tabular}%
\end{equation*}

In Example 7, Maskin monotonicity and no-veto power hold, and hence, we can
achive Nash implementation. However, strict Maskin monotonicity$^{\ast \ast
} $ fails, and by Theorem 1, we cannot achieve rationalizable implemenation
in Example 3.

To see the failure of strict Maskin monotonicity$^{\ast \ast }$, let $%
\mathcal{L}_{i}^{\circ }\left[ a,\text{ }\theta \right] $ denote the strict
lower contour set of $a$ for agent $i$ at state $\theta $, and $\mathcal{L}%
_{i}\left[ a,\text{ }\theta \right] $ the weak lower contour set. Note that%
\begin{equation*}
\mathcal{L}_{i}^{\circ }\left[ a,\text{ }\theta ^{\prime }\right] \cup 
\mathcal{L}_{i}^{\circ }\left[ a,\text{ }\theta ^{\prime \prime }\right]
\subset \triangle \left( Z\right) =\mathcal{L}_{i}\left[ a,\text{ }\theta
^{\prime \prime \prime }\right] \text{, }\forall i\in \left\{ 1,2,3\right\} 
\text{,}
\end{equation*}%
i.e., at the true state $\theta _{3}$, when all agents report $\theta _{1}$
(or $\theta _{2}$), no whistle blower exists. As a result, the only possible
partition in strict Maskin monotonicity$^{\ast \ast }$ is $\mathcal{P}\left(
\theta \right) \equiv \mathcal{P}_{f}\left( \theta \right) $, i.e.,%
\begin{equation*}
\mathcal{P}\left( \theta _{1}\right) =\mathcal{P}\left( \theta _{2}\right) =%
\mathcal{P}\left( \theta _{3}\right) =\left\{ \theta _{1},\theta _{2},\theta
_{3}\right\} \text{ and }\mathcal{P}\left( \theta _{4}\right) =\left\{
\theta _{4}\right\} \text{.}
\end{equation*}

Suppose that the true state is $\theta _{4}$ and that all agents report
either $\theta _{1}$, or $\theta _{2}$, or $\theta _{3}$. Strict Maskin
monotonicity$^{\ast \ast }$ requires a whistle-blower $j$ such that%
\begin{equation*}
\mathcal{L}_{j}^{\circ }\left[ a,\text{ }\theta \right] \diagdown \mathcal{L}%
_{j}\left[ a,\text{ }\theta _{4}\right] \neq \varnothing \text{, }\forall
\theta \in \left\{ \theta _{1},\theta _{2},\theta _{3}\right\} \text{.}
\end{equation*}%
Since%
\begin{equation*}
\mathcal{L}_{i_{1}}^{\circ }\left[ a,\text{ }\theta _{1}\right] \subset 
\mathcal{L}_{j}\left[ a,\text{ }\theta _{4}\right] \text{,}
\end{equation*}%
the whistle-blower $j$ cannot be $i_{1}$. Similarly, we have%
\begin{equation*}
\mathcal{L}_{i_{2}}^{\circ }\left[ a,\text{ }\theta _{2}\right] \subset 
\mathcal{L}_{i_{2}}\left[ a,\text{ }\theta _{4}\right] \text{,}
\end{equation*}%
\begin{equation*}
\mathcal{L}_{i_{3}}^{\circ }\left[ a,\text{ }\theta _{3}\right] \subset 
\mathcal{L}_{i_{3}}\left[ a,\text{ }\theta _{4}\right] \text{,}
\end{equation*}%
and as a result, the whistle-blower $j$ can neither be $i_{2}$ nor be $i_{3}$%
. Therefore, there is no such a whistle-blower, i.e., strict Maskin
monotonicity$^{\ast \ast }$ fails.

\bibliographystyle{econometrica}
\bibliography{RI}

\end{document}